\documentclass{article}

\usepackage{PRIMEarxiv}

\usepackage[utf8]{inputenc} 
\usepackage[T1]{fontenc}    
\usepackage{hyperref}       
\usepackage{url}            
\usepackage{booktabs}       
\usepackage{amsfonts}       
\usepackage{nicefrac}       
\usepackage{microtype}      
\usepackage{lipsum}
\usepackage{fancyhdr}       
\usepackage{graphicx}       
\graphicspath{{media/}}     
\usepackage{amsmath}
\usepackage{doi}
\usepackage{bbold}

\usepackage{rotating}
\usepackage{float}
\usepackage{placeins}
\usepackage{wrapfig}

\title{ Assessing the impact of external factors on the occurrence of emergencies
}

\author{
  Félicien Hêche \thanks{These authors contributed equally to this work.} \\
  School of Engineering and Management \\
  University of Applied Sciences and \\
  Arts Western Switzerland (HES-SO) \\
  Yverdon-les-Bains, Switzerland \\
  \texttt{felicien.heche@gmail.com} \\
   \And
  Philipp Schiller \footnotemark[1] \\
  School of Engineering and Management \\
  University of Applied Sciences and  \\
  Arts Western Switzerland (HES-SO) \\
  Yverdon-les-Bains, Switzerland \\
  \texttt{philipp.schiller@hotmail.com} \\
  \And
  Oussama Barakat \\
  SINERGIES Laboratory \\
  University of Bourgogne-Franche-Comté \\
  Besançon, France \\
  \And
  Thibaut Desmettre \\
  Emergency Departement \\
  Hôpitaux Universitaires de Genève (HUG) \\
  Genève, Switzerland \\
  \And
  Stephan Robert-Nicoud \\
  School of Engineering and Management \\
  University of Applied Sciences and \\
  Arts Western Switzerland (HES-SO) \\
  Yverdon-les-Bains, Switzerland \\
}

\begin{document}
\maketitle

\begin{abstract}
This study investigates the impact of 19 external factors, related to weather, road traffic conditions, air quality, and time, on the occurrence of emergencies using historical data provided by the dispatch center of the Centre Hospitalier Universitaire Vaudois (CHUV). This center is responsible for managing Emergency Medical Service (EMS) resources in the majority of the French-speaking part of Switzerland. First, classical statistical methods, such as correlation, Chi-squared test, Student's $t$-test, and information value, are employed to identify dependencies between the occurrence of emergencies and the considered parameters. Additionally, SHapley Additive exPlanations (SHAP) values and permutation importance are computed using eXtreme Gradient Boosting (XGBoost) and Multilayer Perceptron (MLP) models. The results indicate that the hour of the day, along with correlated parameters, plays a crucial role in the occurrence of emergencies. Conversely, other factors do not significantly influence emergency occurrences. Subsequently, a simplified model that considers only the hour of the day is compared with our XGBoost and MLP models. These comparisons reveal no significant difference between the three models in terms of performance, supporting the use of the basic model in this context. These observations provide valuable insights for EMS resource relocation strategies, benefit predictive modeling efforts, and inform decision-making in the context of EMS. The implications extend to enhancing EMS quality, making this research essential.
\end{abstract}


\section{Introduction}\label{intro}
\begin{figure}[h]
\centering
\includegraphics[width=5cm]{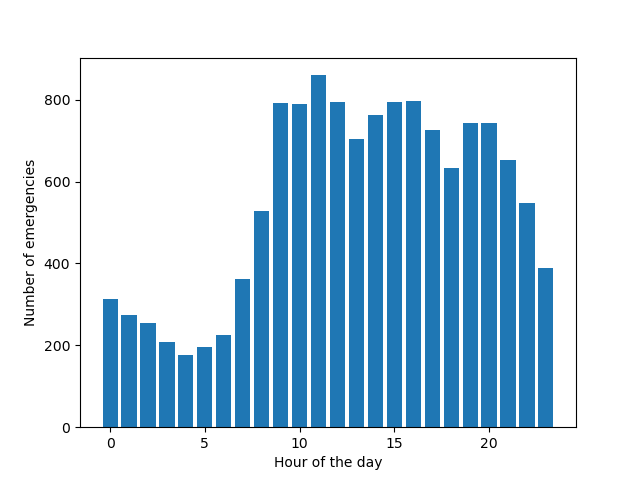}\hfill
\includegraphics[width=5cm]{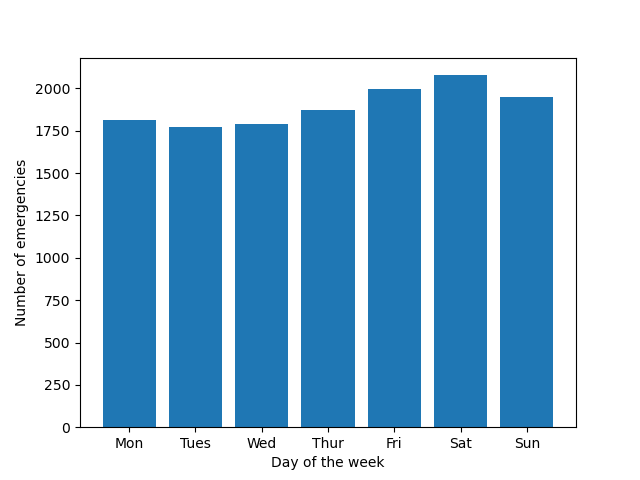}\hfill
\includegraphics[width=5cm]{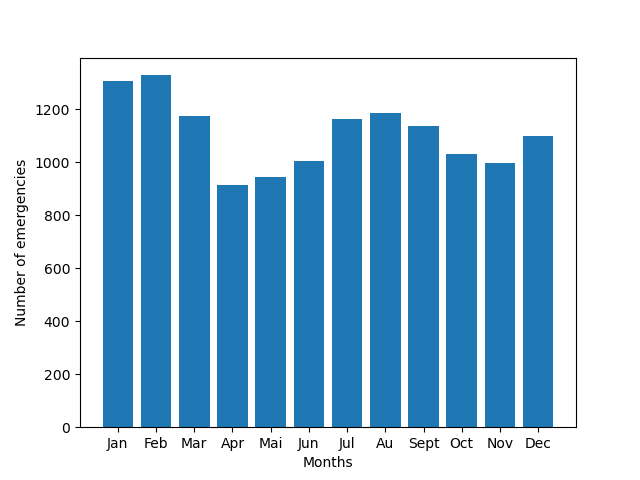}
\caption{Number of emergencies related to Aigle's station regarding different time intervals.}
\label{hists}
\end{figure}

Emergency Medical Service (EMS) plays an important role in healthcare, garnering the attention of researchers over the past three decades \cite{savas1969simulation, furuta2013minisum, swalehe2016dynamic, bruglieri2016optimizing, laatz2019developing, black2004appropriate, heche2023inhomogeneous, heche2024dispatch}. Numerous aspects of EMS have been subject to study, including the triage system \cite{bandara2014priority}, the effect of fatigue training
\cite{barger2018effect}, and crisis management \cite{maudet2020emergency}.
In particular, resource management has received a lot of attention. EMS resources are managed at strategic and operational levels.
At the strategic level, the optimal locations for ambulance stations are determined \cite{BERALDI2009323, mackle2020data, bandara2014priority}.  At the operational level, resources can be relocated among different stations to reduce response time or enhance coverage \cite{BROTCORNE2003451, lee2011role, LEKNES2017122, lee2012role}. In practice, resource relocation can be decided by humans or algorithms. However, the aim remains the same in both cases: relocate resources for optimal management of future emergencies. 
\medskip
\newline
Previous studies have identified several external factors that may contribute to generating emergencies. For instance, road traffic is recognized to cause a high amount of accidents \cite{bachani2017road, zhang2011road, mohan2008road}, which can sometimes lead to premature death \cite{hamzeh2016epidemiology, spoerri2011mortality}. Furthermore, seasonal and weather conditions have also been shown to impact physical activities \cite{turrisi2021seasons, shih2009impact}, thereby influencing the risk of injuries \cite{habelt2011sport, maisonneuve2020epidemiology, de2017severe, neumair2022influence}. Additionally, seasonal factors have been found to affect the suicide rate \cite{ruuhela2009climate, fountoulakis2016relationship}. Some studies also indicate that air quality can have a significant short-term effect on mortality rates \cite{pascal2014short, samoli2013associations}. Finally, as illustrated in Fig. \ref{hists}, it is evident that temporal factors influence the number of emergencies.
\medskip
\newline
EMS resource relocation strategies heavily rely on estimating the likelihood of future emergencies occurring within particular regions. While the above-mentioned studies suggest the potential benefits of integrating external factors into these estimations to enhance EMS resource relocation methods, the actual advantages of such incorporation remain unclear. Moreover, there is uncertainty about which parameters should be prioritized for building these methods. In this paper, we aim to tackle these fundamental questions through a series of experiments. To the best of our knowledge, these aspects have not been studied in prior research.
\medskip
\newline
For this purpose we investigate the impact of 19 different factors related to road traffic conditions, weather, air quality, and time on the occurrence of emergencies, using historical data provided by the Centre Hospitalier Universitaire Vaudois (CHUV) dispatch center. This center is responsible for managing EMS resources in a region that encompasses the majority of the French-speaking part of Switzerland. Considered parameters are listed and described in Table \ref{list parameters}. First, classical statistical methods, such as correlation \cite{ly2018analytic}, Chi-squared test \cite{plackett1983karl}, Student's t-test \cite{daniel2018biostatistics}, and information value \cite{siddiqi2012credit}, are employed to identify dependencies between the occurrence of emergencies and the considered parameters. Additionally, SHapley Additive exPlanations (SHAP) values \cite{lundberg2017unified} and permutation importance \cite{huang2016permutation} are computed using eXtreme Gradient Boosting (XGBoost) \cite{chen2016xgboost} and Multilayer Perceptron (MLP) \cite{Goodfellow-et-al-2016} models. Finally, the performances of a simplified model considering only the hour of the day are compared with our XGBoost and MLP models.
\medskip
\newline
Our experiments highlight the significant role of the hour of the day, along with correlated parameters, on the occurrence of emergencies. Conversely, our results indicate that other factors have minimal influence on emergency occurrences. Furthermore, our findings suggest that, among the examined factors, there are limited advantages to integrating additional external variables into EMS resource relocation methods, apart from the Hour parameter. Therefore, our results argue for focusing only on the hour of the day when developing EMS resource relocation methods. These insights provide valuable guidance for practitioners seeking to refine EMS resource relocation methods, thereby enhancing the quality of EMS.
\medskip
\newline
This paper is organized as follows. Section \ref{data} introduces the dataset and outlines the pre-processing methods employed. The conducted experiments are detailed in Section \ref{experiments}. Finally, in the last section, we draw our conclusions and discuss future research opportunities.

\begin{table}
\centering
\caption{Parameters description.}\label{list parameters}
\begin{tabular}{lll}
\toprule
Parameter & Description & Unity \\
\midrule
Temperature (min) & Air temperature at 2 m above ground; hourly minimum & °C  \\
Precipitations (max) &  Precipitations; 10 minutes summation, hourly maximum & mm \\
Precipitations (sum) & Precipitations, hourly sum & mm \\
Thunders (3 km) & Close thunders (distance 0-3 km); hourly sum & - \\
Thunders (3km to 30km) & Distant thunders (distance 3-30 km); hourly sum & - \\
Air humidity & Relative air humidity at 2 m above ground; hourly average & \% \\
Wind speed & Scalar wind speed; hourly average & m/s \\
Snow height & Snow height (automatically measured); hourly instantaneous value  & cm \\
Wind speed (max) &  Wind speed; average over 10 minutes, maximum over the last 2 hours & m/s \\
Sunlight & Sunshine duration; hourly sum & h \\
Sun rays & Global radiation; hourly average & W/$\text{m}^{2}$  \\
Road 1 & Number of vehicles in the first direction; hourly sum & -   \\
Road 2 & Number of vehicles in the other direction; hourly sum & -   \\
PM10 & Particulate matter less than 10 µm in diameter &  $\mu$g/$\text{m}^{3}$  \\
O$_{3}$ & O$_{3}$ concentration & $\mu$g/$\text{m}^{3}$ \\
NO$_{2}$ & NO$_{2}$ concentration & $\mu$g/$\text{m}^{3}$ \\
Hour & Hour of the day; between 0 and 23 & -   \\
Weekday & Day of the week; between 0 and 6 & - \\
Month & Month of the year; between 0 and 11 & -  \\
\bottomrule
\end{tabular}
\end{table}

\section{Data}\label{data}
This section provides a detailed description of the data used in this study and the pre-processing methods employed.  It begins with an overview of the data concerning emergencies, followed by the presentation of data related to road traffic conditions. Subsequently, an introduction to the weather data is provided. Section \ref{air quality} delves into the specifics of the data concerning air quality. Furthermore, the incorporation of temporal information into the analysis is discussed. Lastly, the pre-processing methods applied in this study are explained.

\subsection{Emergencies}

\begin{figure}[t]
\begin{minipage}{.5\linewidth}
    \begin{center}
    \includegraphics[width=7cm]{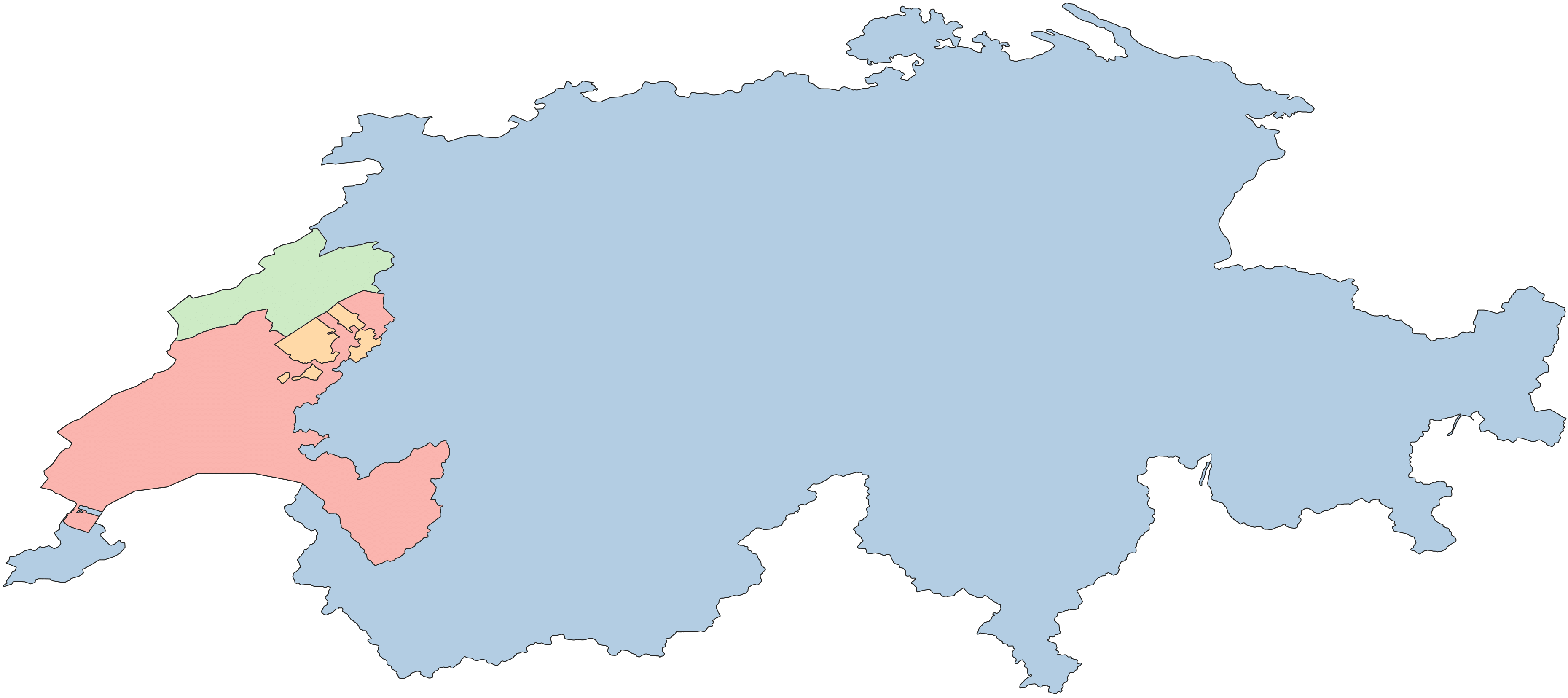}
    \caption{Region of interest.}
    \label{region of interest}
    \end{center}
\end{minipage}
\qquad
\begin{minipage}{.45\linewidth}
    \begin{center}
    \includegraphics[width=7cm]{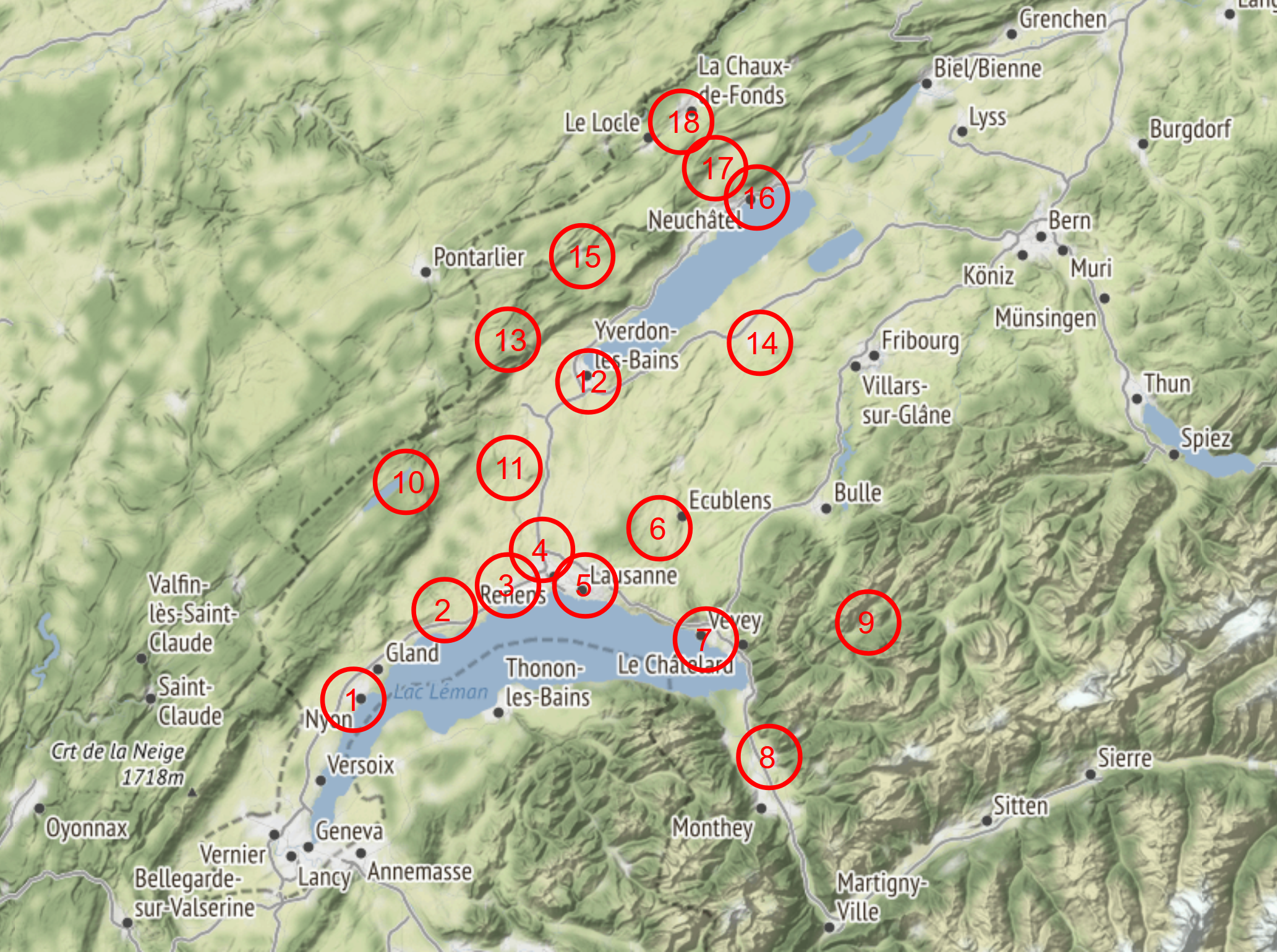}
    \caption{Ambulances stations location. Correspondence between station number and name can be found in Table \ref{stations stats}.}
    \label{station location}
    \end{center}
\end{minipage}
\end{figure}

As mentioned in the Introduction, historical data provided by the CHUV dispatch center has been used. In this center, and as a common practice, emergency calls are classified into three priority levels: Priority 1 for the most urgent life-threatening calls, Priority 2 for urgent but not life-threatening, and Priority 3 for non-urgent calls. In this paper, incidents and emergencies refer to Priority 1 and 2 calls.
\medskip
\newline
The provided data spans a time interval from January 2015 to the end of December 2021. The region of interest, as shown in Fig. \ref{region of interest}, covers the majority of the French-speaking part of Switzerland. Within this region, there are a total of $18$ ambulance stations, as illustrated in Fig. \ref{station location}. Each emergency has been assigned to its closest station using the traveling time estimated using Openrouteservice API \footnote{\url{https://openrouteservice.org/}}. In cases where, we were unable to find the closest station, such as due to GPS location errors, we excluded these incidents from our analysis. Some of the considered areas show a major difference in the number of incidents. Indeed, while $70'796$ emergencies were registered in the Lausanne station (by far the biggest city), in L'Abbaye (a rather rural area), only $5'511$ incidents occurred in the same period. Statistics highlighting differences between these stations can be found in Table \ref{stations stats}.

\begin{table}
\centering
\caption{Some statistics regarding emergencies related to each station. The column labeled "emergencies" corresponds to the number of incidents assigned to this station, while the column called "zeros emergencies" represents the percentage of one-hour timesteps with zero incidents. "More than one emergency" indicates the percentage of timesteps with more than one emergency. Stations with a value of less than $3\%$ in the last column are bolded.}\label{stations stats}
\setlength{\tabcolsep}{15pt} 
\begin{tabular}{lllll}
\toprule
Id &  Name & Emergencies & Zeros emergencies & More than one emergency  \\
\midrule
1 & Nyon & 20'884 & 72.06\% &  5.24\% \\
\textbf{2} & \textbf{Aubonne} & \textbf{12'555} & \textbf{83.24\%} & \textbf{2.92\%}  \\
\textbf{3} & \textbf{Morges} & \textbf{11'138} & \textbf{83.82\%} & \textbf{1.79\%}  \\
4 & Villars-Sainte-Croix & 18'500 & 74.63\% & 4.14\% \\
5 & Lausanne & 70'923 & 35.49\% & 31.39\% \\
\textbf{6} & \textbf{Mézières} & \textbf{9'444} & \textbf{86.05\%} & \textbf{1.32\%} \\
7 & Tour-de-Peilz & 28'321 & 65.07\% & 8.59\% \\
\textbf{8} & \textbf{Aigle} & \textbf{13'265} & \textbf{81.84\%}  & \textbf{2.83\%} \\
\textbf{9} & \textbf{Château-d'Oex} & \textbf{2'898} & \textbf{95.53\%} & \textbf{0.24\%} \\
\textbf{10} & \textbf{L'Abbaye} & \textbf{5'509} & \textbf{92.07\%} & \textbf{0.95\%} \\
\textbf{11} & \textbf{Pompales} & \textbf{8'443} & \textbf{87.7\%} & \textbf{1.3\%} \\
12 & Yverdon-les-Bains & 15'914 & 77.97\% & 3.32\% \\
\textbf{13} & \textbf{Sainte-Croix} & \textbf{4'614} & \textbf{93.03\%} & \textbf{0.45\%} \\
\textbf{14} & \textbf{Payerne} & \textbf{10'563} & \textbf{84.75\%} & \textbf{1.77\%} \\
\textbf{15} & \textbf{Val-de-Travers} & \textbf{6'901} & \textbf{89.59\%} & \textbf{0.77\%} \\
16 & Neuchâtel & 22'965 & 71.04\% & 6.39\% \\
\textbf{17} & \textbf{Malviliers} & \textbf{9'879} & \textbf{85.79\%} & \textbf{1.62\%}  \\
18 & La Chaux-de-Fonds & 21'563 & 73.29\% & 6.22\% \\
\bottomrule
\end{tabular}
\end{table}

\subsection{Road Traffic}

The road traffic data has been provided by the Federal Roads Office \href{https://www.astra.admin.ch/astra/en/home.html}{(FEDRO)} of Switzerland, which owns counting points located throughout Switzerland's road network. Each of them measures the number of vehicles passing through a specific road in one hour and each direction of the flow. There are $58$ counting points located in our region of interest. Most of these stations are on highways, and thus, highly correlated. To address unavoidable missing data, two different approaches are used. First, if there exists a road where the correlation between the route with the missing data and this road is higher than $95\%$, we just re-scale data provided by this road. If such a high correlation does not exist, we use the Prophet model \cite{taylor2018forecasting} due to the high seasonality. Finally, since a high correlation between close counting points has been observed, for each station we consider only the closest road according to the Euclidean distance.

\subsection{Weather}
The data portal for teaching and research of MeteoSwiss \href{https://www.meteoswiss.admin.ch/services-and-publications/service/weather-and-climate-products/data-portal-for-teaching-and-research.html}{(IDAweb)}, has been consulted to collect a total of $11$ different parameters, such as temperature, snow height or precipitation. The list of complete weather parameters can be found in Table \ref{list parameters}. This portal grants access to raw data collected by weather stations located throughout Switzerland. Instead of measuring all possible features, weather station usually measures only a few. All together, there are $125$ weather stations located within our region of interest. Each of them has been assigned to its closest ambulance station using the Euclidean distance.
\medskip
\newline
To address missing data, we use a weighted average approach using data from stations that have recorded the feature of interest. The considered weights are proportional to the Euclidean distance between the weather station and the ambulance station. Specifically, for the weather station $i$ and ambulance station $j$, we calculate the weight with the following formula
$$
w_{ij} = \left( \frac{d(i, j)^{s}}{\sum_{k} d(k, j)^{s}} \right),
$$
where $d(i, j)$ denotes the Euclidean distance between $i$ and $j$. Note that $\sum_{k} w_{kj} = 1$.
The hyperparameter $s$ controls the importance of the closest stations.  Assigning a higher value to $s$ increases the significance of the nearest stations. This approach eliminates the need for generating synthetic data. In our study, we set $s$ to $3$.

\subsection{Air quality}\label{air quality}
Another aspect we take into account in this work is air quality. For this purpose, we use data provided by the Federal Office for the Environment \href{https://www.bafu.admin.ch/bafu/en/home.html}{(FOEN)} of Switzerland. Specifically, we consider the concentration of fine particles with a diameter smaller than $10$ $\mu m$ (PM10), the concentration of ozone (O$_3$), and nitrogen dioxide (NO$_2$). We use four different FOEN stations which are located in Sion, Lausanne, Chaumont, and Payerne. For each ambulance station, we exclusively use data from the nearest FOEN station. Since we have only one measurement per day, we duplicate this value for every hour of the day. Additionally, to address missing data, we duplicate the most recently measured value.

\begin{figure}[h]
    \centering
    \includegraphics[width = 12cm]{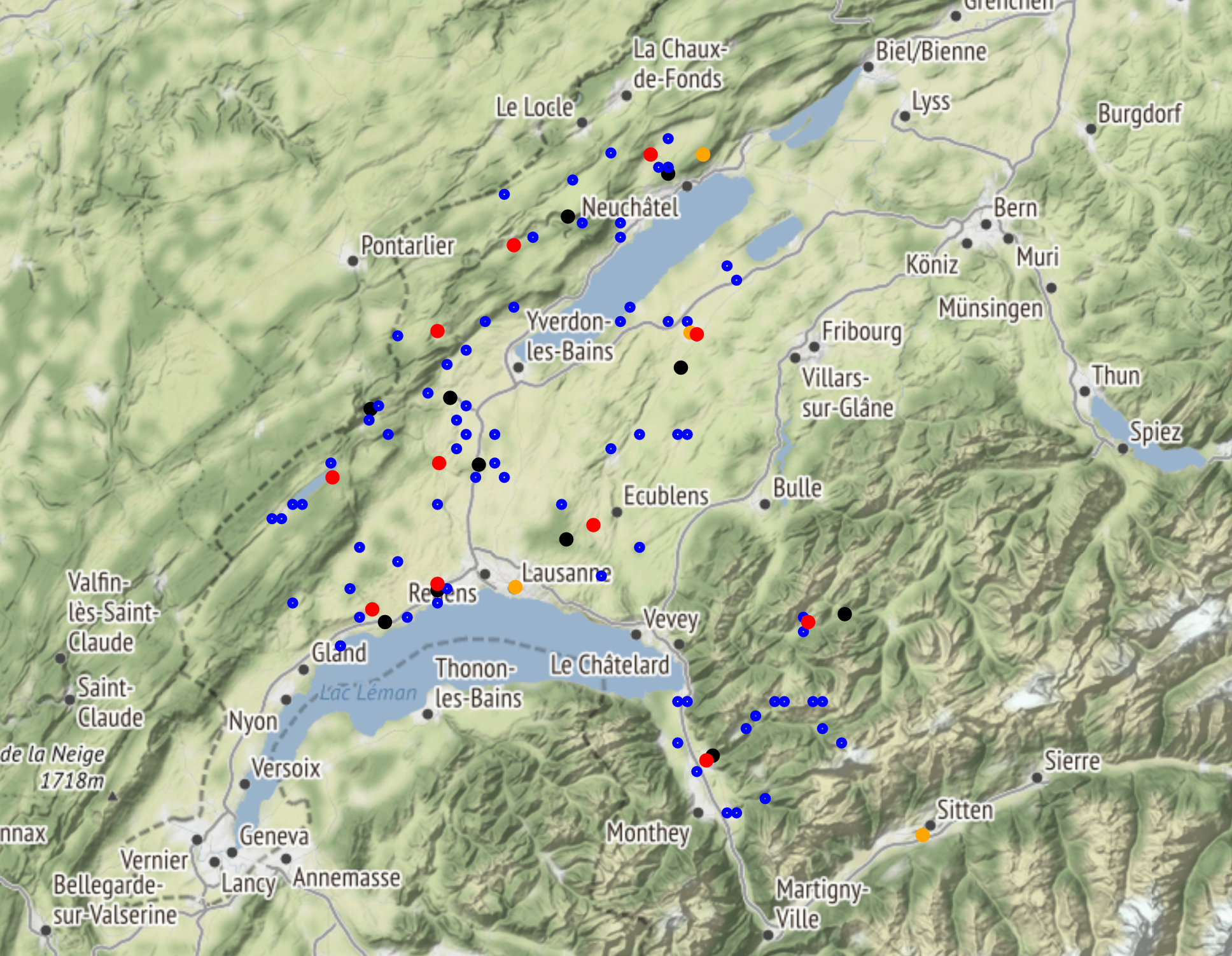} 
    \caption{Location of all different considered stations and traffic points. Ambulance stations are in red, weather stations in blue, counting points in black, and air quality stations in orange.}
    \label{all stns}
\end{figure}

\subsection{Time}
Finally, temporal parameters are also considered. Specifically, we analyze temporal variables encompassing the hour of the day, weekdays, and months, which are incorporated as components of a normalized vector. For example, Tuesday corresponds to the value of $1/6$.

\subsection{Data pre-processing}
Before conducting our experiments, we normalize each parameter by rescaling it to the interval $[0,1]$. To mitigate potential issues related to multi-class scenarios, we treat hours with multiple emergencies as if there was only one emergency. To ensure that this approximation does not significantly impact our results, we consider only those stations where hours with more than one incident occur in less than $3\%$ of the cases. Consequently, our analysis includes only 11 ambulance stations (see Table \ref{stations stats}), 11 traffic counting points, 77 weather stations, and 4 air quality stations. The locations of all these stations are shown in Fig. \ref{all stns}.

\section{Experiments}\label{experiments}
In this paper, we employ several approaches to conduct our experiments. First, we apply classical statistical methods to assess the influence of different parameters on the occurrence of emergencies. Specifically, we use 
correlation \cite{ly2018analytic}, 
Chi-squared test \cite{plackett1983karl}, 
Student's $t$-test and information value \cite{siddiqi2012credit}.
\medskip
\newline
Subsequently, we proceed to conduct additional experiments using XGBoost \cite{chen2016xgboost} and MLP \cite{Goodfellow-et-al-2016} models. While we initially explored alternative models such as Support Vector Machine (SVM) \cite{cortes1995support} and Long-Short Term Memory (LSTM) \cite{hochreiter1997long}, our results revealed no significant performance discrepancies among them. Therefore, we focus our experimentation solely on XGBoost and MLP. Interestingly, the combination of XGBoost and feature importance through model interpretation has been previously applied in the context of car accident predictions \cite{parsa2020toward, meng2018expressway, mousa2019extreme}.
\medskip
\newline
To evaluate the performance of each model, we employ classical metrics such as the Area Under the receiver operating characteristic Curve (AUC) and precision. Subsequently, utilizing the XGBoost and MLP models, we delve into assessing the significance of individual factors by computing SHAP value and permutation importance for each parameter. Finally, we compare the outcome of XGBoost and MLP models with a simple baseline called Prior model which takes only the hour of the day into account.
\medskip
\newline
Since the figures do not show notable differences between stations, we present only figures related to Morges here. Figures corresponding to other stations can be found in the Appendix.

\subsection{Correlation}
\begin{figure}[h]
    \centering
    \includegraphics[width =\textwidth, height=5in]{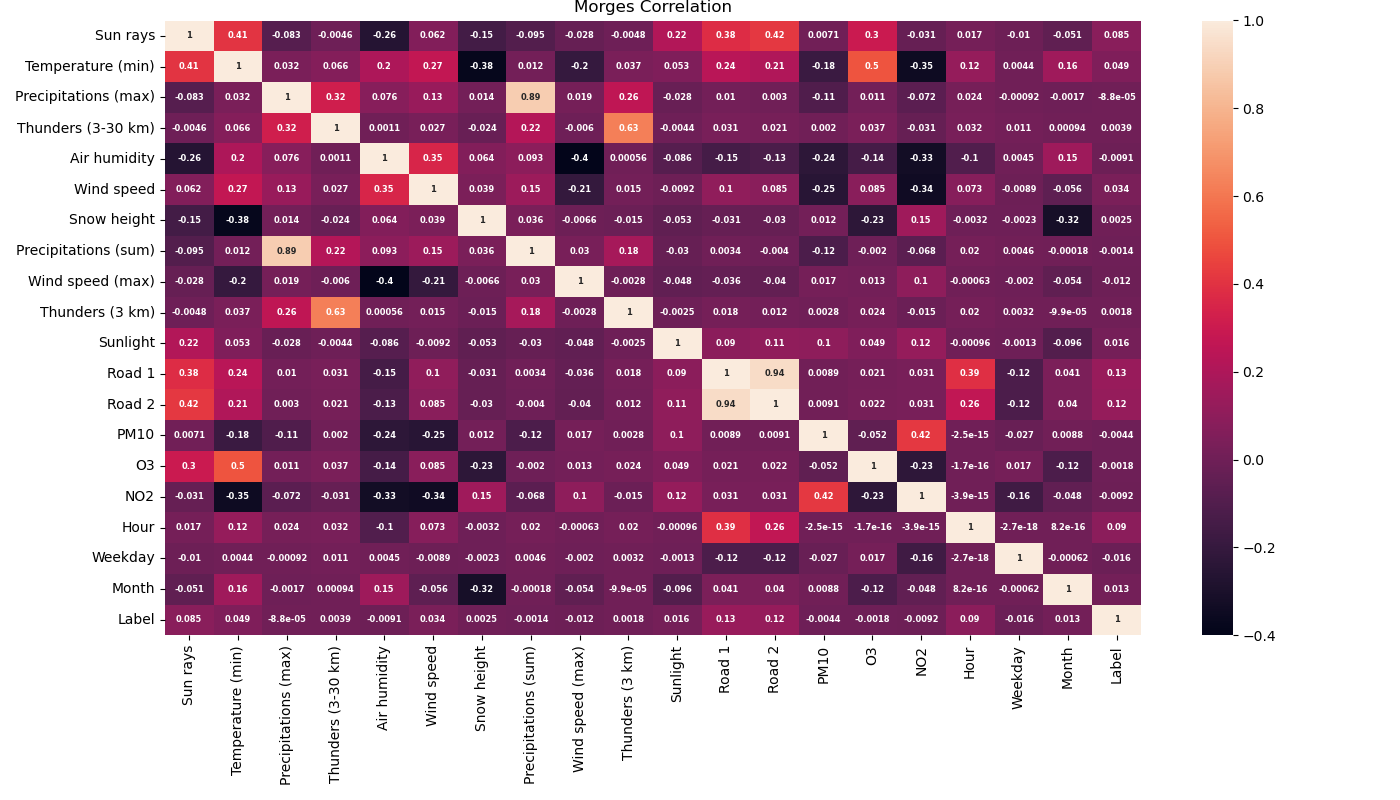}
    \caption{Correlation matrix, Morges.}
    \label{morges_corr}
\end{figure}

First, we begin by computing the correlation coefficient, also known as Pearson correlation coefficient \cite{ly2018analytic}, between each parameter and label but also between each pair of parameters. As it can be observed in Fig. \ref{morges_corr}, the parameters that exhibit the highest correlation with the label are the Hour and parameters related to traffic conditions. However, note that Road 1 and Road 2 are also correlated with the hour of the day. Furthermore, weather parameters do not show important correlations with the target, apart from Sun rays, which only appear during daytime.

\subsection{Chi-squared tests}
Then, we use Chi-squared tests \cite{plackett1983karl} to investigate the influence of categorical parameters on the occurrence of emergencies. Specifically, these parameters correspond to the hour of the day, the day of the week, and the month. The null hypothesis $H_{0}$ states that the factor of interest and labels are independent. 
\medskip
\newline
As shown in Table \ref{p-value}, all time parameters seem to have some influence on the occurrence of emergencies for at least one station. Nevertheless, it is worth noting that, unlike the day of the week or the month, the hour has a significant influence (assuming we use a threshold of $1\%$) on the occurrence of emergencies for all stations.

\begin{table}
\centering
\caption{$p$-values related to Chi-squared test.}\label{p-value}
\setlength{\tabcolsep}{15pt} 
\begin{tabular}{llll}
\toprule
 & Hour  & Weekday  & Month  \\
\midrule
Aubonne & $0.0$ & $0.0227$ & $0.0053$  \\
Morges & $0.0$ & $0.0000$ & $0.0160$   \\
Mézières & $0.0$ & $0.0017$ & $0.0706$  \\
Aigle & $0.0$ & $0.0000$ & $0.0000$   \\
Château-d'Oex & $0.0$ & $0.0000$ & $0.0000$   \\
L'Abbaye & $0.0$ & $0.9358$ & $0.0159$   \\
Pompales & $0.0$ & $0.3195$ & $0.0009$   \\
Sainte-Croix & $0.0$ & $0.0603$ & $0.0032$   \\
Payerne & $0.0$ & $0.6750$ & $0.0026$   \\
Val-de-Travers & $0.0$ & $0.8757$ & $0.0019$  \\
Malviliers & $0.0$ & $0.0431$ & $0.0007$   \\
\bottomrule
\end{tabular}
\end{table}

\subsection{Student's $t$-test}
As mentioned above, a Chi-squared test is
used for categorical parameters. However, in our case, we also need to handle continuous parameters. Therefore, we perform a Student's $t$-test \cite{daniel2018biostatistics} to measure the influence of these parameters on the occurrence of emergencies. The results related to these tests are presented in Table \ref{t-test}. 
\medskip
\newline
First, using a statistical threshold of $1\%$, it can be observed that Sun rays, Road 1 and Road 2 are statistically not independent of the occurrence of emergencies in all considered stations. Furthermore, Temperature (min), Air humidity, Wind speed, and Thunders (3-30km) also have a $p$-value smaller than $1\%$ in almost all stations. Conversely, Precipitations (max), Precipitations (sum), Thunders (3 km) and O$_{3}$ rarely obtain $p$-values lower than $1\%$. For the remaining parameters (Snow height, Sunlight, PM10, NO$_{2}$, Wind speed (max)) a general pattern cannot be observed.

\begin{sidewaystable}[ph!]
\footnotesize
\caption{$p$-values $t$-test.}\label{t-test}
\begin{tabular}{llllllllllll}
\toprule
 & Aubonne  & Morges   & Mézières & Aigle & Château-d'Oex & L'Abbaye & Pompales & Sainte-Croix & Payerne & Val-de-Travers & Malviliers \\
\midrule
Sun rays & $0.0$  & $0.0$ & $0.0$  & $0.0$ & $0.0$ & $0.0$ & $0.0$ & $0.0$ & $0.0$  & $0.0$ & $0.0$ \\
Temperature (min) & $0.0$ & $0.0$ & $0.0$  & $0.0$ & $0.2185$ & $0.0$ & $0.0$ & $0.0$ & $0.0$ & $0.0$ & $0.0$ \\
Precipitations (max) & $0.0252$ & $0.9826$ & $0.3861$ & $0.6121$ & $0.0725$ & $0.6136$ & $0.0071$ & $0.3452$ & $0.7623$ & $0.4642$ & $0.5708$ \\
Thunders (3-30km) & $0.0024$ & $0.3368$ & $0.0324$ & $0.9700$ & $0.3041 $ & $0.0265$  & $0.1534$ & $0.0042$ & $0.7155$ & $0.0027$ & $0.1569$ \\
Air humidity & $0.0$ & $0.0247$ & $0.0063$ & $0.0$ & $0.0$ & $0.0$ & $0.0$ & $0.0$ & $0.0$ & $0.0$ & $0.0$ \\
Wind speed & $0.3407$ & $0.0$ & $0.0$ & $0.0$ & $0.0026$ & $ 0.0$ & $0.0$ & $0.0$ & $0.0$ & $0.0$ & $0.0$ \\
Snow height & $0.4701$  & $0.5311$  & $0.2274$ & $0.0357$ & $0.0$ & $0.0050$ & $0.5144$ & $0.0005$ & $0.3010$ & $0.5910$ & $0.0004$ \\
Precipitations (sum) & $0.0001$  & $0.7367$  & $0.7244$ & $0.7591$ & $ 0.0316$ & $0.7528$ & $0.0$ & $0.5363$ & $0.8988$ & $0.4597$ & $0.2320$ \\
Wind speed (max) & $0.0$  & $0.0038$  & $0.0$ & $0.1375$ & $0.9256$ & $0.8708$ & $0.0003$ & $0.0$ & $0.0339$ & $0.1526$ & $0.0$ \\
Thunders (3 km) & $0.0158$ & $0.6495$ & $0.0638$ & $0.9914$ & $0.3865$ & $0.1227$  & $0.6561$ & $0.3047$ & $0.8300$ & $0.0$ & $0.6123$ \\
Sunlight & $0.0835$ & $0.0$ & $0.6162$  & $0.0007$  & $0.0$ & $0.1317$ & $0.9556$ & $0.0009$ & $0.0652$ & $0.7091$ & $0.9242$ \\
Road 1 & $0.0$ & $0.0$  & $0.0$ & $0.0$ & $0.0$ & $0.0$ & $0.0$ & $0.0$ & $0.0$ & $0.0$ & $0.0$ \\
Road 2 & $0.0$ & $0.0$ & $0.0$ & $0.0$ & $0.0$ & $0.0$ & $0.0$ & $0.0$ & $0.0$ & $0.0$ & $0.0$ \\
PM10 & $0.0$  & $0.2797$  & $0.0$ & $0.0033$ & $0.0972$ & $0.0018$ & $0.0008$ & $0.0762$ & $0.0009$ & $0.1912$ & $0.0028$  \\
03 & $0.0948$ & $0.6520$  & $0.1750$ & $0.0350$ & $0.0$ & $0.4856$ & $0.6616$ & $0.6216$ & $0.0005$ & $0.2117$ & $0.0291$ \\
NO2 & $0.0$  & $0.0233$ & $0.0$ & $0.0225$  & $0.0$ & $0.0$ & $0.0$ & $0.1362$ & $0.0$ & $0.0312$ & $0.2709$ \\
\bottomrule
\end{tabular}
\vspace{0.8cm}
\caption{Information values. We bold information values higher than $0.1$.}\label{Table IV}
\centering
\footnotesize
\begin{tabular}{llllllllllll}
\toprule
 & Aubonne  & Morges  & Mézières & Aigle & Château-d'Oex & L'Abbaye & Pompales & Sainte-Croix & Payerne & Val-de-Travers & Malviliers \\
\midrule
Sun rays & $ \textbf{0.1444}$ & $0.0858$ & $0.0692$ & $0.0873$ & $ \textbf{0.2351}$ & $0.0931$ & $0.0820$ & $0.0968$ & $0.0800$ & $ \textbf{0.1120}$ & $0.0629$ \\
Temperature (min) & $0.0298$ & $0.0236$ & $0.0382$ & $0.0271$ & $0.0319$ & $0.0279$ & $0.0490$ & $0.0530$ & $0.0350$ & $0.0287$ & $0.0143$ \\
Precipitations (max) & $0.0012$ & $0.0009$ & $0.0002$ & $0.0003$ & $0.0018$ & $0.0010$ & $0.0020$ & $0.0010$ & $0.0003$ & $0.0015$ & $0.0006$ \\
Thunders (3-30km) & $0.0$ & $0.0$ & $0.0$ & $0.0$ & $0.0$ & $0.0$ & $0.0$ & $0.0$ & $0.0$ & $0.0$ & $0.0$ \\
Air humidity & $0.0159$ & $0.0254$ & $0.0082$ & $0.0390$ & $0.0713$ & $0.0309$ & $0.0295$ & $0.0406$ & $0.0275$ & $0.0133$ & $0.0166$ \\
Wind speed & $0.0061$ & $0.0152$ & $0.0214$ & $0.0089$ & $0.0103$ & $0.0211$ & $0.0161$ & $0.0457$ & $0.0236$ & $0.0154$ & $0.0228$ \\
Snow height & $0.0040$ & $0.0013$ & $0.0064$ & $0.0064$ & $0.0429$ & $0.0032$ & $0.0047$ & $0.0075$ & $0.0058$ & $0.0023$ & $0.0080$ \\
Precipitations (sum) & $0.0014$ & $0.0004$ & $0.0002$ & $0.0006$ & $0.0019$ & $0.0004$ & $0.0025$ & $0.0007$ & $0.0002$ & $0.0017$ & $0.0010$ \\
Wind speed (max) & $0.0130$ & $0.0056$ & $0.0080$ & $0.0061$ & $0.0126$ & $0.0095$ & $0.0097$ & $0.0152$ & $0.0018$ & $0.0003$ & $0.0029$ \\
Thunders (3 km) & $0.0$ & $0.0$ & $0.0$ & $0.0$ & $0.0$ & $0.0$ & $0.0$ & $0.0$ & $0.0$ & $0.0$ & $0.0$ \\
Sunlight & $0.0$ & $0.0$ & $0.0$ & $0.0$ & $0.0$ & $0.0$ & $0.0$ & $0.0$ & $0.0$ & $0.0$ & $0.0$ \\
Road 1 & $\textbf{0.2424}$ & $\textbf{0.1802}$ & $\textbf{0.1618}$ & $\textbf{0.2026}$ & $\textbf{0.2644}$ & $\textbf{0.1912}$ & $\textbf{0.2022}$ & $\textbf{0.2203}$ & $\textbf{0.1663}$ & $\textbf{0.2002}$ & $\textbf{0.1521}$ \\
Road 2 & $\textbf{0.2443}$ & $\textbf{0.1713}$ & $\textbf{0.1561}$ & $\textbf{0.1770}$ & $\textbf{0.3057}$ & $\textbf{0.1600}$ & $\textbf{0.1673}$ & $\textbf{0.1945}$ & $\textbf{0.1463}$ & $\textbf{0.1698}$ & $\textbf{0.1239}$ \\
PM10 & $0.0081$ & $0.0040$ & $0.0082$ & $0.0026$ & $0.0136$ & $0.0090$ & $0.0080$ & $0.0040$ & $0.0030$ & $0.0035$ & $0.0049$ \\
03 & $0.0080$ & $0.0020$ & $0.0041$ & $0.0039$ & $0.0256$ & $0.0081$ & $0.0077$ & $0.0067$ & $0.0054$ & $0.0041$ & $0.0058$ \\
NO2 & $0.0158$ & $0.0027$ & $0.0108$ & $0.0056$ & $0.0277$ & $0.0084$ & $0.0164$ & $0.0057$ & $0.0055$ & $0.0057$ & $0.0017$  \\
Hour & $\textbf{0.3280}$ & $\textbf{0.2225}$ & $\textbf{0.2339}$ & $\textbf{0.2553}$ & $\textbf{0.3937}$  & $\textbf{0.2242}$ & $\textbf{0.2429}$ & $\textbf{0.2497}$ & $\textbf{0.2036}$ & $\textbf{0.2304}$ & $\textbf{0.1921}$ \\
Weekday & $0.00205$ & $0.0044$ & $0.0034$ & $0.0041$ & $0.0128$  & $0.0004$ & $0.0012$ & $0.0033$ & $0.0006$ & $0.0005$ & $0.0020$ \\
Month & $0.0042$ & $0.0027$ & $0.0020$ & $0.0170$ & $0.0956$ & $0.0063$ & $0.0043$ & $0.0070$ & $0.0036$ & $0.0049$ & $0.0061$  \\
\bottomrule
\end{tabular}
\end{sidewaystable}

\subsection{Information values}
Other metrics used to measure the influence of features include information value (IV) \cite{siddiqi2012credit}. This tool offers an estimation of the parameter's ability to distinguish between classes. A parameter that exhibits higher separation between classes is expected to have a stronger influence on the label. Consequently, a higher IV is desirable.

\begin{wrapfigure}{r}{.5\linewidth}
\includegraphics[width=7cm]{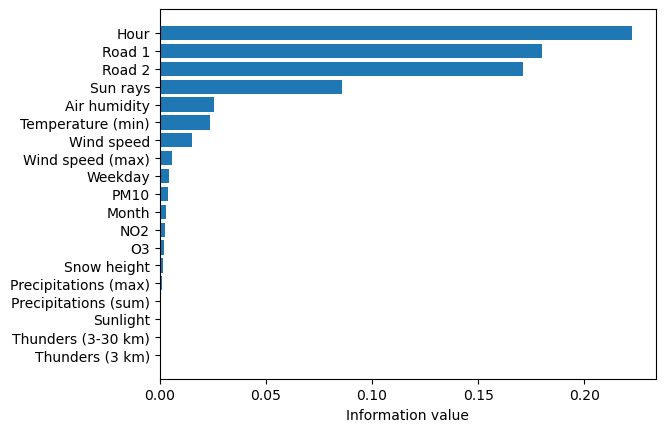}
\caption{Information values, Morges.}
\label{Morges IV}
\end{wrapfigure}
\medskip
To compute this score for each parameter, we begin by dividing the data into $m$ bins. For categorical parameters (Hour, Weekday, Month) $m$ is equal to the number of classes. For continuous parameters, we set $m$ to $24$. $X$ denotes the set of observed data and $Y$ to the corresponding labels. Then, we compute the Weight-of-Evidence \cite{wod1985weight} for each bin $i$ (WoE$_{i}$) using
    \begin{equation}
    \text{WoE}_i = \log \left( \frac{\mathbb{P}[X = i|Y=1]}{\mathbb{P}[X=i|Y=0]} \right).
    \end{equation}
Finally, IV is defined as
\begin{equation*}\label{IV}
    \text{IV} := \sum_{i=1}^m \left( \mathbb{P}[X=i|Y=0]- \mathbb{P}[X=i|Y=1] \right) \text{WoE}_{i}.
\end{equation*}
Using this equation it is possible to compute information value related to each parameter and each station.
The results of these computations can be found in Table \ref{Table IV}. A histogram concerning the information values related to Morges station can be found in Fig. \ref{Morges IV}. Histograms regarding other stations are presented in the Appendix.
\medskip
\newline
A rule of thumb asserts that up to $0.02$, a parameter can be considered as insignificant; up to $0.1$,  as weak; from $0.1$ to $0.3$ as a medium predictor; and finally, anything beyond that is considered as strong \cite{siddiqi2012credit}. Thus, according to this rule, none of the parameters is strong. While Road 1, Road 2, and Hour seem to be medium predictors for all stations, Sun rays appear to be a medium predictor only for three stations. All remaining parameters can be considered weak or insignificant.

\subsection{Models interpretation}
As explained at the beginning of this section, XGBoost, MLP are used to investigate factor influence from another perspective. First, we describe these models and their performance using AUC score and precision curves. Then, using SHAP values and permutation importance, we examine how parameters affect their predictions. Finally, we compare the performance of these two models with a simple baseline called Prior model.

\subsubsection{XGBoost} \label{XGBoost description}
XGBoost \cite{chen2016xgboost} is an implementation of the gradient boosting framework  \cite{friedman2000additive, friedman2001greedy}. Specifically, XGBoost is a boosting tree ensemble trained to minimize the sum of a convex loss function that depends on the specific problem and a regularization term.
\medskip
\newline
We utilized data from the years $2015$ to $2020$ to fine-tune and train our model, while data from the year $2021$ was used to perform the experiments presented in this paper. XGBoost provides a range of hyperparameters that require careful selection. In this study, we employed  HyperOpt \cite{bergstra2013making}, a Bayesian optimization method, to tune these hyperparameters. The chosen evaluation metric was the mean of a 5-fold cross-validation using the AUC score. We conducted a total of $100$ attempts. Each XGBoost was trained using the classical binary cross-entropy loss function.
\begin{table}
\caption{AUC scores related to each XGBoost model.}\label{AUC}
\setlength{\tabcolsep}{15pt} 
\centering
\begin{tabular}{llll}
\toprule
Station & XGBoost  & MLP & Prior \\
\midrule
Aubonne & $0.6525$ & $0.6481$ & $0.6632$  \\
Morges &  $0.6179$ & $0.6218$ & $0.6211$   \\
Mézières & $0.6160$ & $0.6128$ & $0.6264$   \\
Aigle &  $0.6397$ & $0.6373$ & $0.6383$   \\
Château-d'Oex &  $0.6522$ & $0.6532$ & $0.6444$  \\
L'Abbaye &  $0.6129$ & $0.6160$ & $0.6271$   \\
Pompales & $0.6549$ & $0.6299$ & $0.6448$   \\
Sainte-Croix &  $0.6324$ & $0.6294$ & $0.6313$   \\
Payerne & $0.6048$ & $0.6075$ & $0.6091$   \\
Val-de-Travers & $0.6290$ & $0.6225$ & $0.6326$  \\
Malviliers &  $0.6051$ & $0.6064$ & $0.6120$  \\
\bottomrule
\end{tabular}
\end{table}

AUC refers to the area under the receiver operating characteristic (ROC) curve \cite{fawcett2006introduction}. ROC curve is a method used to visualize the performance of a binary classifier. It specifically focuses on the tradeoff between the true positive rate (TPR) (i.e. the ratio between correct positive predictions and all positive predictions) and the false positive rate (FPR), (i.e. the ratio between correct negative predictions and all negative predictions). A ROC graph contains the FPR on the $x$-axis and the TPR on the $y$-axis. This method offers a two-dimensional representation of the classifier's performance, where the diagonal line represents complete randomness. The further the curve deviates from the diagonal, the better the classifier performs. An AUC score close to 1 indicates good performance, while a score near 0.5 suggests random classifications. 
\medskip
\newline
As a standard practice, we first measure the performance of each XGBoost model using the AUC score. These scores can be found in Table \ref{AUC}. These models do not show any significant difference between each other in terms of AUC score, presenting a performance varying from $0.6048$ to $0.6549$.

\subsubsection{MLP}
MLP constitutes a class of neural networks characterized by multiple layers of interconnected nodes \cite{Goodfellow-et-al-2016}. These networks excel in learning complex patterns inherent in data by iteratively adjusting weights during training to minimize a defined loss function. Leveraging nonlinear activation functions, MLP possesses the capability to capture intricate relationships within data, rendering them highly adaptable for a wide range of tasks, including classification and regression.
\medskip
\newline
In alignment with our protocol used for optimizing XGBoost models, we employed a similar approach for training our MLP models. Specifically, we utilized data spanning from $2015$ to $2020$ for training and fine-tuning purposes, reserving data from the year $2021$ for conducting experiments presented in this study. Each model has been trained using the binary cross-entropy loss function and Adam optimizer \cite{kingma2014adam}. Furthermore, our architectures were enriched with contemporary deep learning techniques like batch normalization \cite{ioffe2015batch} and dropout regularization \cite{srivastava2014dropout}. Hyperparameters have been tuned using the Bayesian optimization method provided by Keras Tuner \cite{omalley2019kerastuner}, employing the AUC score as the evaluation metric on a dedicated evaluation dataset, comprising $20\%$ of the data from years $2015$ to $2020$. Our hyperparameters optimization process encompassed a total of $75$ attempts for each station.
\medskip
\newline
As detailed in Table \ref{AUC}, the AUC scores of MLP models closely mirror the performance of XGBoost, ranging between $0.6064$ and $0.6532$. Moreover, no significant differences can be observed across stations.

\subsubsection{Precision}
For incorporating these models into EMS resource relocation methods, obtaining a high precision of class zeros is crucial. Recall that the precision of class zero is calculated as the ratio of non-emergency instances correctly predicted to the total number of predictions for class zero.
\medskip
\newline
When the model predicts that no emergency will occur at a specific station, there may be a temptation to redistribute EMS resources from that station to others. However, relocating resources prematurely can result in insufficient coverage, leaving certain areas vulnerable to potential emergencies and possibly leading to catastrophic outcomes. Therefore, ensuring high precision for class zeros is essential.
\medskip
\newline
Since the output of XGBoost and MLP consists of probabilities, it is possible to control their precision by adapting a threshold that will be denoted $c$. Specifically, if $\hat{p}_{i}$ corresponds to the predicted probability of belonging to class zero, we can define a classification as follows
\begin{equation}
    \hat{y}_i = 
  \begin{cases}
    0 & \text{for } \hat{p}_{i} \geq c \\
    1 & \text{otherwise}
  \end{cases}
\end{equation}
where $c \in [0,1]$. Selecting a high value for $c$ will result in the model predicting class zero only when it is highly confident. Since the precision depends on the threshold $c$, we will denote it as $P(c)$.  
\medskip
\newline 
For the case where $c$ is equal to one, and since there usually is not any prediction for the zero class, precision is not well defined. However, in this paper, we define it as $1$. It is worth noting that precision does not necessarily increase monotonically with $c$, as both the number of true positives and the predicted positives depend on threshold value $c$.

\begin{figure}[h]
\begin{minipage}{.45\linewidth}
    \includegraphics[width = 7cm]{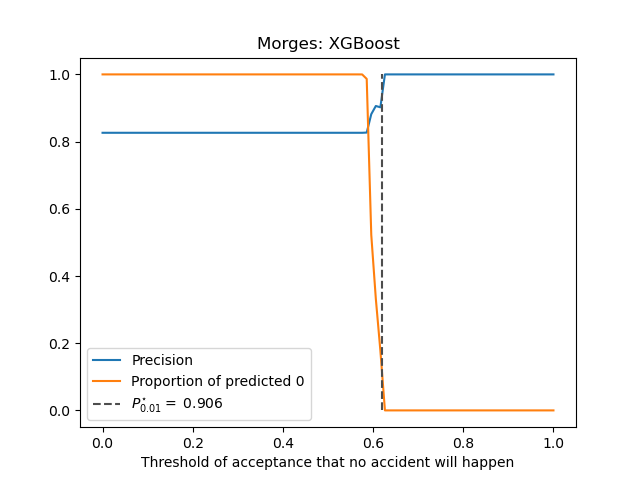}
    \caption{Precision curve, XGBoost, Morges.}
    \label{morges_precision}
\end{minipage}
\qquad
\begin{minipage}{.45\linewidth}
    \includegraphics[width = 7cm]{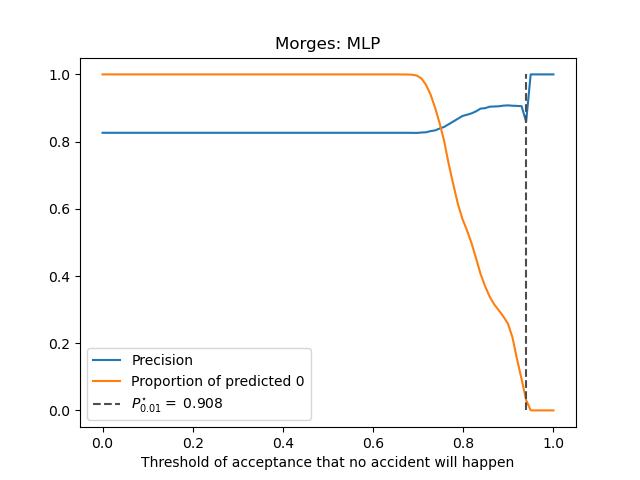}
    \caption{Precision curve, MLP, Morges.}
    \label{precision_mlp}
\end{minipage}
\end{figure}
\medskip

Now, let's consider a scenario where we have a model that consistently predicts class one with only one correct exception for class zero. Although such a model would achieve perfect precision, it would be practically useless. To prevent such cases, we can require that the model predicts a minimum number of class zero instances. For this purpose, we introduce the $\gamma$-precision, which is defined as follows
\begin{equation}
    P^{\star}_\gamma = \max\{P(c)\text{ } : \text{ }c \in [0,1], \text{ } \# \text{Predicted }0 > n\gamma\} , 
    \label{precision_star}
\end{equation}
where $n$ is the size of the dataset. 
\medskip
\newline
Thus, in our experiments, we compute the precision related to each XGBoost and MLP model according to the threshold $c$. We also measure the proportion of predicted $0$ and the value of $P^{\star}_{0.01}$. The result of these computations regarding Morges station can be observed in Fig. \ref{morges_precision} and \ref{precision_mlp}.

\subsubsection{SHAP values} \label{shap method}
\begin{figure}[h]
\begin{minipage}{.45\linewidth}
    \includegraphics[width = 7cm]{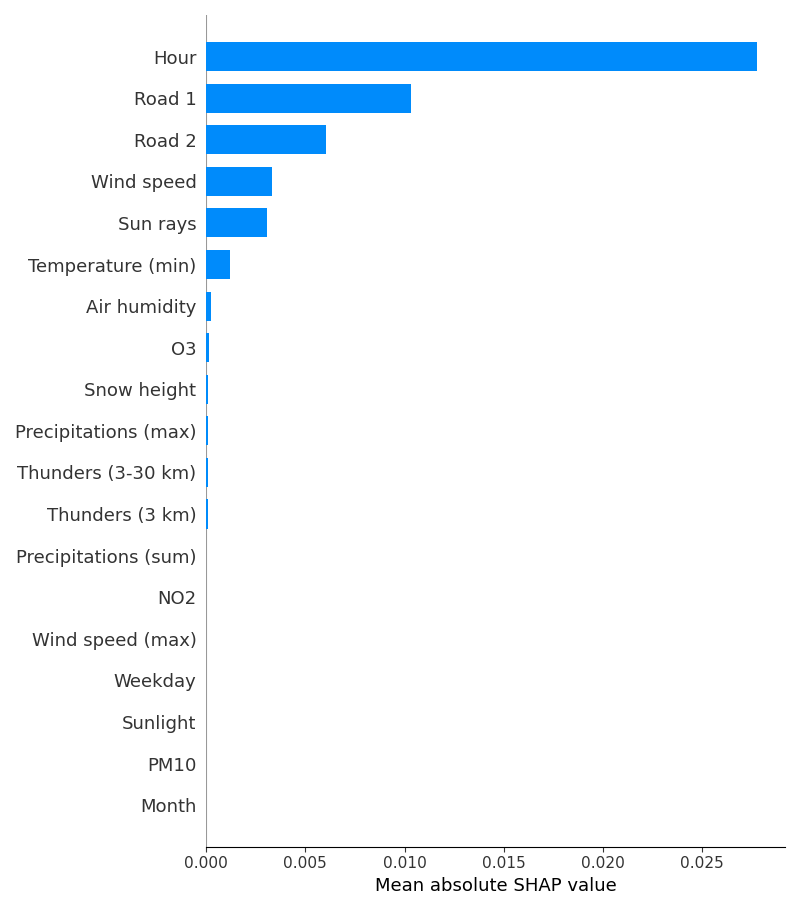}
    \caption{SHAP values, XGBoost, Morges.}
    \label{shap_morges}
\end{minipage}
\qquad
\begin{minipage}{.45\linewidth}
    \includegraphics[width = 7cm]{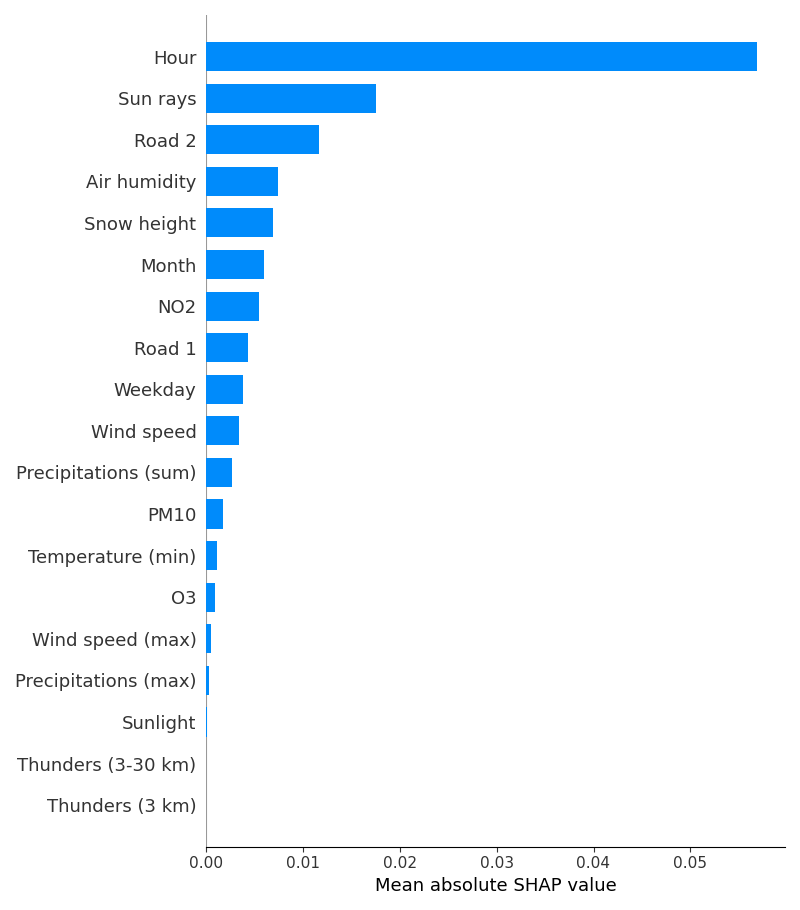}
    \caption{SHAP values, MLP, Morges.}
    \label{shap_mlp_morges}
\end{minipage}
\end{figure}

SHAP value \cite{lundberg2017unified} measures the contribution of each parameter to the prediction of a specific input $x$, given a model $f$. The key idea is to measure how the base value $\mathbb{E}[f(z)]$ that would be predicted if we did not know any parameter, will change by adding some factors. Specifically, given a potentially simplified input $x'$, the SHAP value associated to the parameter $i$, denoted $\phi_{i}(f, x)$ is given by
\begin{equation} \label{shap def}
    \phi_i(f, x) = 
   \sum_{z' \subseteq x'}  \frac{|z'|!(M-|z'|-1)!}{M!} 
   \left(  \mathbb{E}[f(z) | z_{S}] - \mathbb{E}[f(z) | z_{S \setminus\{i\} }] \right),
\end{equation}
where $z' \subseteq x' $ represents all vectors where the non-zeros entries are a subset of the non-zeros entries in $x'$. The set $S$ refers to the non-zero indices in $z'$, while $z_{S}$ represents the input values corresponding to these indices. It might not be immediately clear how to compute $\mathbb{E}[f(z) | z_{S}]$, but assuming feature independence and model linearity, we can approximate it as
$$
\mathbb{E}[f(z) | z_{S}]  \approx f([z_{S}, \mathbb{E}[z_{\overline{S}}]]),
$$
where $z_{\overline{S}}$ denotes the values of the input which are not in $S$. Using this last approximation, Equation \ref{shap def} and taking the mean over the entire dataset, the mean SHAP value for each parameter $\phi_{i}$ can be calculated.
\medskip
\newline
SHAP values related to each parameter and each station have been computed using data from the year 2021. SHAP values corresponding to Morges station and related to XGBoost and MLP model can be observed in Fig. \ref{shap_morges} and Fig. \ref{shap_mlp_morges}, respectively. SHAP values corresponding to other stations can be found in the Appendix. 
\medskip
\newline
Regarding SHAP values computed with the XGBoost model, it can be noticed that Hour provides the higher SHAP value for almost all stations. The only exception is for the station called Val-de-Travers. Then, in the three following parameters, we find almost every time Road 1 and Road 2. The station Château-d'Oex is the only exception to this rule. Finally, in the trio following the parameter Hour, several parameters such as Sun rays, Wind speed, Temperature (min), or even Month can be met.
\medskip
\newline
While results concerning MLP models exhibit slight deviations from those obtained with XGBoost, they generally align in the same direction. First, Hour emerges as the most important parameter since it provides the highest SHAP values across $6$ stations and the second-highest across $4$ stations. Then,  parameters Road $1$ and/or Road $2$ usually yield high SHAP values. Interestingly, one of these two parameters consistently appears in the trio of parameters with the highest SHAP values. The composition of this trio varies across stations but may include factors such as Sun rays, Air humidity, NO$_{2}$, or O$_{3}$.

\subsubsection{Permutation importance}
\begin{figure}[h]
\begin{minipage}{.45\linewidth}
    \includegraphics[width = 7cm]{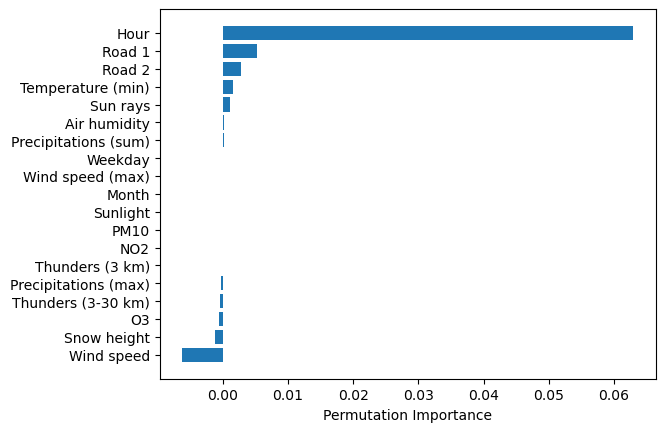}
    \caption{Permutation importance, XGBoost, Morges.}
    \label{permutation_imp}
\end{minipage}
\qquad
\begin{minipage}{.45\linewidth}
    \includegraphics[width = 7cm]{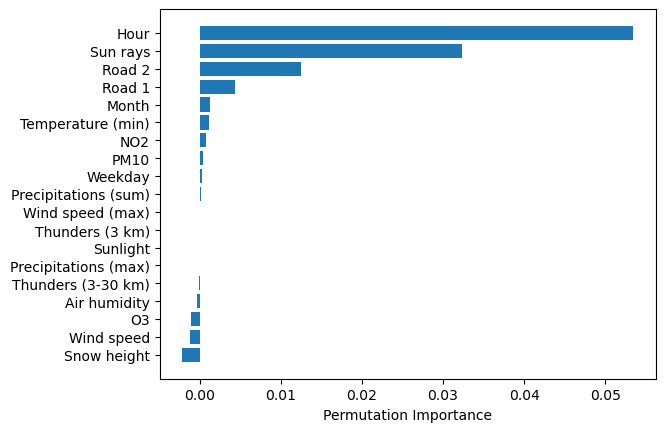}
    \caption{Permutation importance, MLP, Morges.}
    \label{permutation_MLP_morges}
\end{minipage}
\end{figure}

Another method used to measure the influence of each parameter is permutation importance \cite{huang2016permutation}. Permutation importance measures the error between the case where the model has been trained using all parameters and the scenario where one parameter of interest has been replaced with random noise.
\medskip
\newline
More formally, for any model $f$, dataset $X$, target vector $Y$ and an arbitrary loss function $\mathcal{L}$, the original error is given by $E_{0} = \mathcal{L}(Y,f(X))$ . Then, denoting $X_{\bar{j}}$  as the dataset $X$ with random values for the parameter $j$, we can compute the error in this case $E_{j} := \mathcal{L}(Y,(f(X_{\bar{j}}))$. Finally, permutation importance related to the parameter $j$ is given by the difference between these two values:
\begin{equation}
     E_{j} - E_{0} = \mathcal{L}(Y,(f(X_{\bar{j}})) - \mathcal{L}(Y,f(X)).
\end{equation}
Permutation importance values corresponding to Morges station and related to XGBoost and MLP model are shown in Fig. \ref{permutation_imp} and \ref{permutation_MLP_morges} respectively. Permutation importance values concerning other stations can be found in the Appendix.
\medskip
\newline

Regarding permutation importance values computed with XGBoost, it can be noticed that Hour consistently produces higher results across all stations. However, the gap between Hour and the other parameters is more pronounced when considering permutation importance compared to SHAP values. Additionally, we find that Road 1, Road 2, Sun rays, and Temperature (min) are among the parameters that exhibit notable permutation importance values.
\medskip
\newline
In analyzing results concerning MLP models, it can be observed that Hour, Road 1, and Road 2 yield high permutation importance values across all stations. Notably, Hour and either Road 1 or Road 2 are included in the three parameters that exhibit higher permutation importance values. Additionally, it is noteworthy that this couple is complemented with Sun rays for 9 stations.

\subsubsection{Prior model}
\begin{wrapfigure}{r}{.5\linewidth}
\includegraphics[width=7cm]{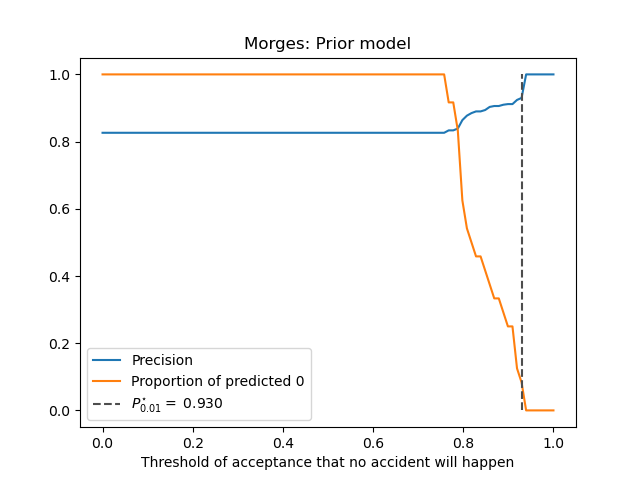}
\caption{Precision curves, Prior model, Morges.}
\label{prior precision}
\end{wrapfigure}

To compare the performance of the XGBoost and MLP models with a basic classifier, we build the following model, called the Prior model. This model solely utilizes the hour of the day as input and is defined as follows
\begin{equation}\label{prior eq}
    \tilde{f}(t)  := \mathbb{1}_{ \{ \mathbb{P}[Y=1|t \geq c] \} },
\end{equation}
where $t$ stands for the hour of the day and $c$ is a hyperparameter. In other words, if the probability for a specific hour of the day exceeds a certain threshold, the Prior model assumes that an emergency will occur.
\medskip
\newline
Similar to XGBoost and MLP, the Prior model was constructed using data spanning from 2015 to 2020. Data from the year 2021 has exclusively been utilized for computing the  AUC score and the precision curve. As it can be seen in Table \ref{AUC}, Prior models achieve comparable results to XGBoost and MLP models in terms of AUC score, ranging between $0.6091$ and $0.6632$. Furthermore, by adjusting the threshold $c$ of Equation \ref{prior eq}, it is also possible to compute the precision curve of the Prior model. As shown in Fig. \ref{prior precision}, XGBoost and MLP do not obtain a better precision or a better $P_{0.01}^{\star}$ value than the Prior model for the Morges station.  The same pattern can be observed in almost all stations, as presented in the Appendix. Therefore, XGBoost and MLP do not seem to provide better results than our basic model, which is surprising since Prior models take only the hour of the day into account.

\subsection{Discussion}
Initial analyses using Chi-squared and Student's $t$-tests reveal that several parameters show dependencies on the occurrence of emergencies across all stations (Hour, Sun rays, Road 1, Road 2). These tests also suggest that a significant portion of parameters exhibit dependence on the occurrence of emergencies within specific stations. 
Only a few parameters appear to be independent of the targets. However, information values provide more nuanced insights, suggesting that only Hour, Road 1, and Road 2 could significantly influence the occurrence of emergencies. Subsequently, additional experiments were conducted using XGBoost and MLP models. Specifically, SHAP values and permutation importance have been computed. The results of these computations align with the patterns observed for information values. Only Hour, Road 1, and Road 2 appear to yield noteworthy results. Nevertheless, the correlation matrices indicate that Road 1 and Road 2 exhibit some correlation with the Hour parameter. Consequently, it remains unclear whether these parameters yield interesting results due to their content of relevant information or solely because of their temporal relation.
\medskip
\newline
In an attempt to shed light on this issue, we built a basic model called the Prior model, which only takes the hour of the day into account, and compared its performance with our sophisticated XGBoost and MLP models. Surprisingly, the performances of these three models showed no significant differences. This observation raises several questions, given that the Prior model only has access to one parameter (Hour), while XGBoost and MLP have access to the $19$ parameters described in Table \ref{list parameters}. We hypothesize that this observation could be attributed to the fact that only the hour of the day contains a genuine causal factor for the occurrence of emergencies. Other parameters that appear to influence the occurrence of emergencies, such as the traffic road conditions, are interesting solely because they are correlated with the hour. This hypothesis is supported by the observation that parameters that are independent of the time of day do not show a significant impact on the occurrence of emergencies. Specifically, there seems to be no real evidence that parameters related to weather or air quality influence the occurrence of emergencies.
\medskip
\newline
Our findings suggest that, aside from the parameter Hour, there are limited benefits to integrating other studied external factors into EMS resource relocation strategies. While impressive results obtained from machine learning methods may lead to the temptation of developing EMS relocation methods using machine learning models that consider a multitude of parameters, our work argues against this idea.  Given that only the hour of the day appears to have a genuine influence on emergency occurrences, models should consider solely this parameter. In such cases, there is no imperative to use machine learning models specifically. For instance, it would be feasible to model the occurrence of emergencies using stochastic processes and build EMS resource relocation methods based on these processes. This approach, which relies on step-by-step procedures, offers a level of transparency that is lacking in machine learning methods. Such transparency is valuable in a context where every decision can significantly impact human lives. Additionally, since our basic Prior model performs as well as our XGBoost and MLP models, there is optimism that a less naive stochastic process could more accurately model emergency occurrences.
\medskip
\newline
From a broader perspective, the results obtained in this paper underline the limitations of statistical tests. While these tests suggest several features are not independent of emergency occurrences, further experiments indicate that these parameters are, in fact, not useful in predicting such occurrences. These results highlight the nuanced nature of the relationship between predictor variables and the target outcome. While statistical tests offer an initial understanding of the association between parameters and the target, they alone are insufficient to ensure predictive capability.

\section{Conclusion}
In this paper, we investigated the impact of $19$ different parameters on the occurrence of emergencies within hourly intervals using historical data provided by the CHUV dispatch center and spanning six years. Initially, basic statistical tests were conducted to examine any relationships between emergency occurrences and these parameters. The results indicated that almost all factors showed some degree of dependence on the occurrence of emergencies. However, further analysis revealed that only the hour of the day exhibited a significant influence on emergency occurrences.
\medskip
\newline
Based on these findings, it appears that among the studied parameters, only the hour of the day warrants consideration when devising EMS resource relocation strategies. Therefore, and as discussed in the previous section, exploring stochastic methods for resource allocation may be pertinent. Additionally, since our experiments demonstrated that machine learning models did not outperform a basic stochastic model, there is potential for less naive stochastic modeling to yield superior performance. Consequently, future research could focus on developing novel methods for EMS resource allocation and relocation using a more sophisticated model, such as an inhomogeneous Poisson process.
\medskip
\newline
Future research could also involve conducting additional experiments to explore the impact of various other factors on emergency occurrences. For instance, investigating the potential influence of epidemiological data on the occurrence of emergencies could offer valuable insights. Finally, while we do not anticipate significant deviations from the results presented in this paper, broadening the scope of experiments to encompass other machine learning models, such as SVM or LSTM networks, could provide a more comprehensive assessment of factor influence.

\section*{Acknowledgement} 
Research presented in this paper has been funded by \href{https://www.interreg-francesuisse.eu}{Interreg France-Suisse (project SIA-REMU)}. Furthermore, we would like to thank the CHUV dispatch center and more specifically Dr. Fabrice Dami, the Swiss Federal Roads Office, MeteoSwiss, and the Swiss Federal Office for the Environment for providing historical data used in this study.


\appendix

\section{Appendix}

\subsection{Aubonne}
\begin{figure*}[!htbp]
    \centering
    \includegraphics[width =0.9\textwidth, height=4in]{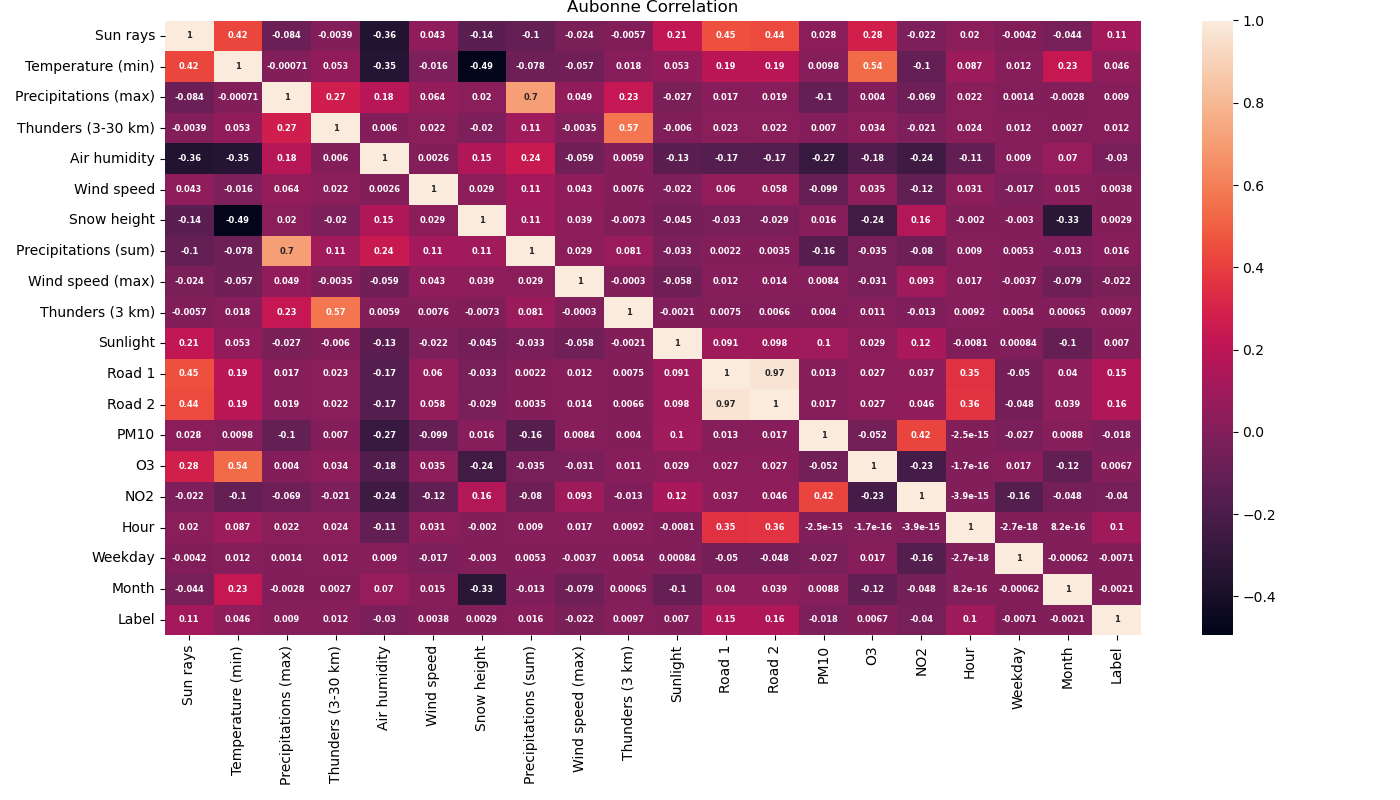}
    \caption{Correlation matrix, Aubonne}
    \label{aubonne_corr}
\end{figure*}

\begin{figure}[!h]
\begin{minipage}{.45\linewidth}
    \centering
    \includegraphics[width =7cm]{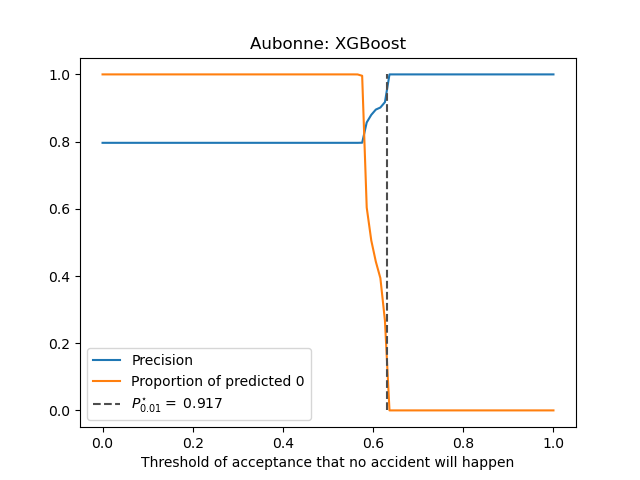}
    \caption{Precision curve, XGBoost, Aubonne}
\end{minipage}
\quad
\begin{minipage}{.45\linewidth}
    \centering
    \includegraphics[width = 7cm]{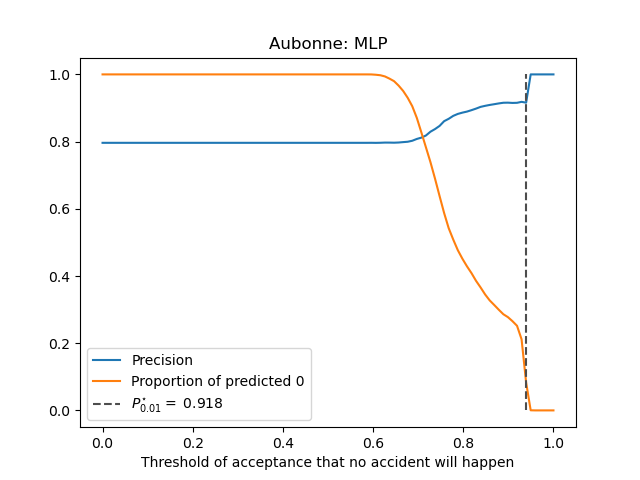}
    \caption{Precision curve, MLP, Aubonne}
\end{minipage}
\end{figure}

\begin{figure}[!h]
\begin{minipage}{.45\linewidth}
    \centering
    \includegraphics[width =7cm]{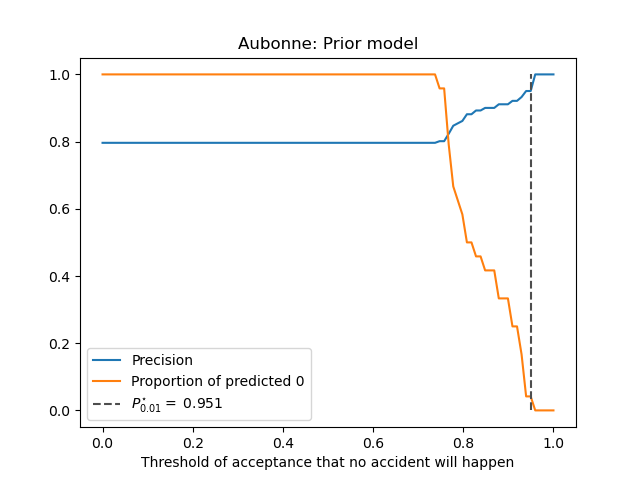}
    \caption{Precision curve, Prior model, Aubonne}
\end{minipage}
\quad
\begin{minipage}{.45\linewidth}
    \centering
    \includegraphics[width = 7cm]{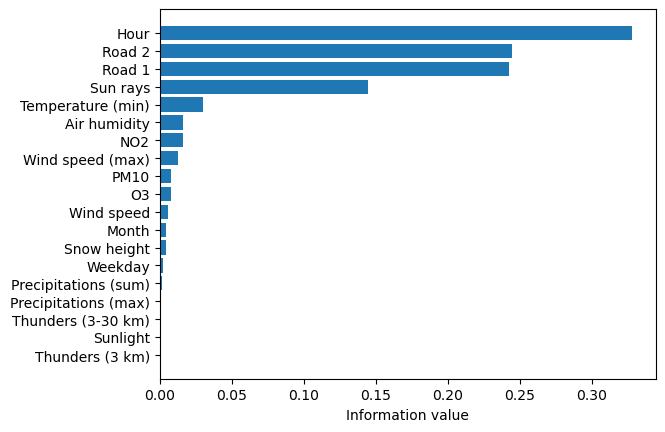}
    \caption{Information values, Aubonne}
\end{minipage}
\end{figure}

\begin{figure}[!h]
\begin{minipage}{.45\linewidth}
    \centering
    \includegraphics[width =7cm]{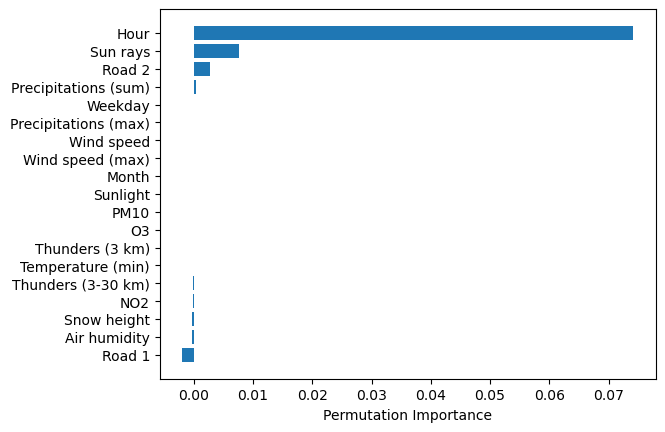}
    \caption{Permutation importance, XGBoost, Aubonne}
\end{minipage}
\quad
\begin{minipage}{.45\linewidth}
    \centering
    \includegraphics[width = 7cm]{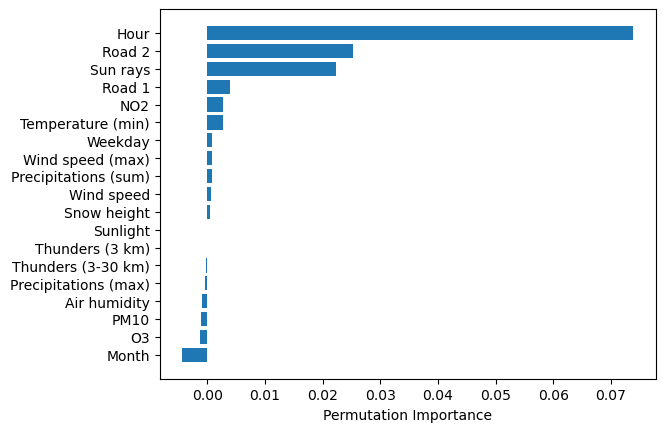}
    \caption{Permutation importance, MLP, Aubonne}
\end{minipage}
\end{figure}

\begin{figure}[!h]
\begin{minipage}{.45\linewidth}
    \centering
    \includegraphics[width =7cm]{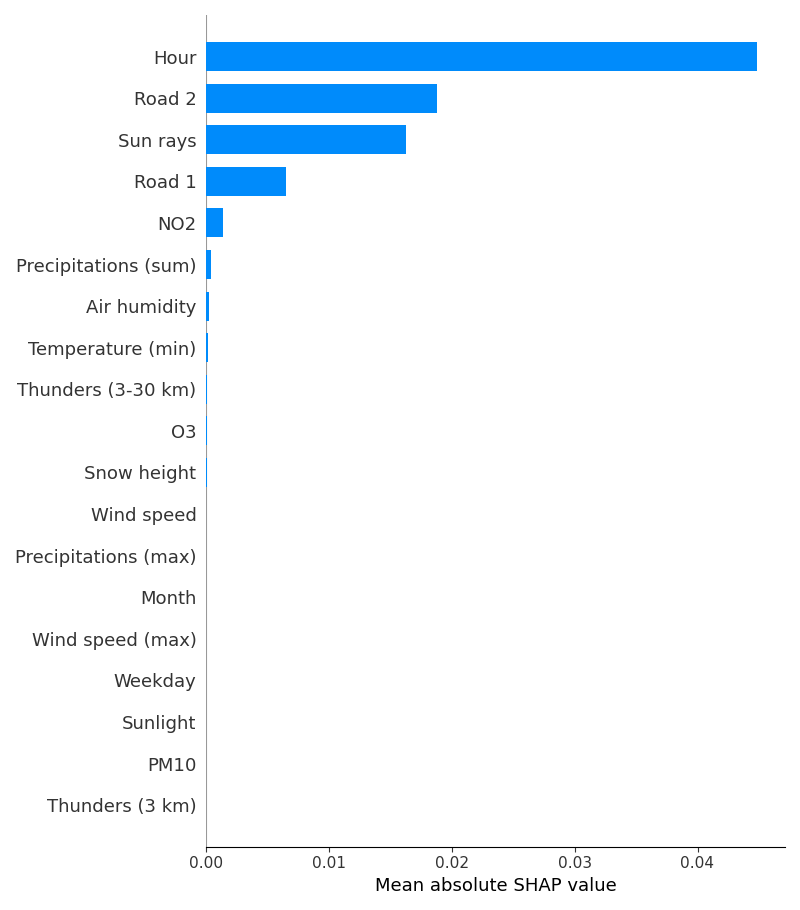}
    \caption{SHAP values, XGBoost, Aubonne}
\end{minipage}
\quad
\begin{minipage}{.45\linewidth}
    \centering
    \includegraphics[width = 7cm]{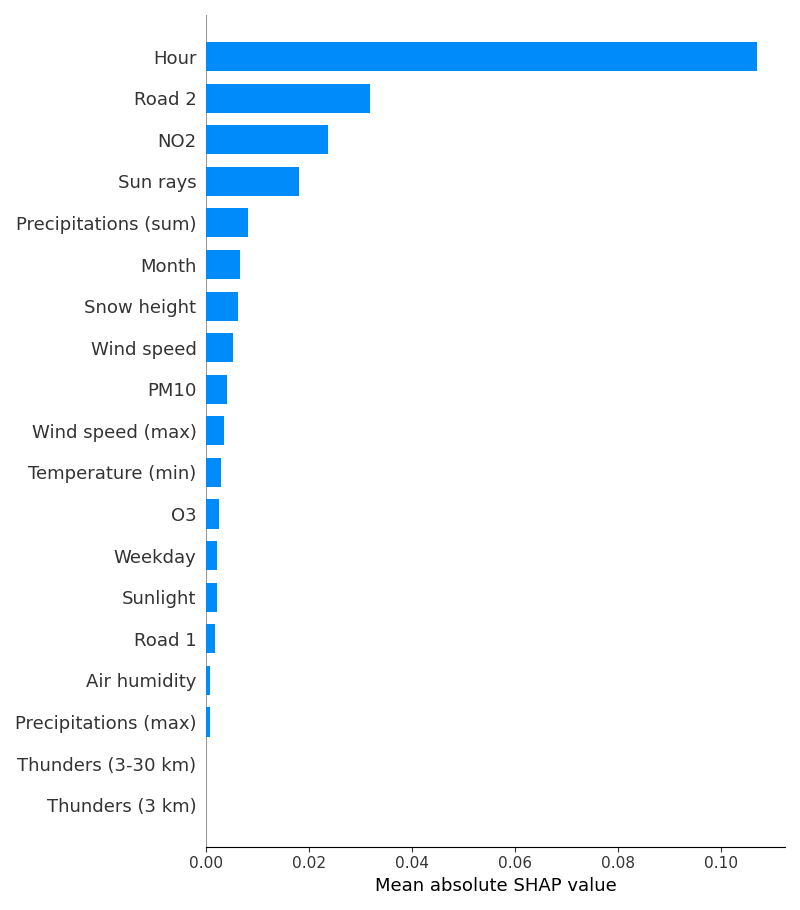}
    \caption{SHAP values, MLP, Aubonne}
\end{minipage}
\end{figure}

\FloatBarrier

\vspace{1pts}
\subsection{Mézières}
\begin{figure*}[!h]
    \centering
    \includegraphics[width =0.9\textwidth, height=4in]{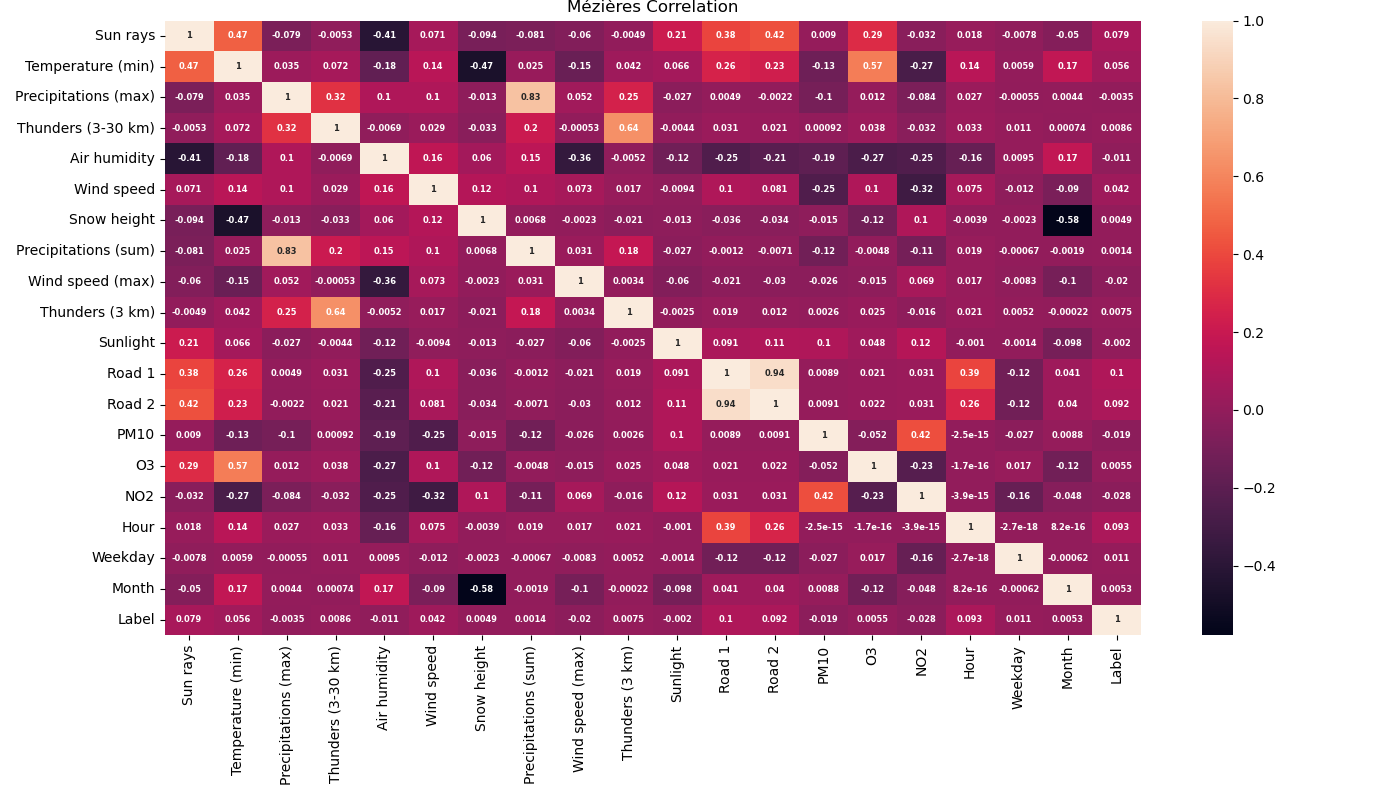}
    \caption{Correlation matrix, Mézières}
    \label{mezieres_corr}
\end{figure*}

\begin{figure}[!h]
\begin{minipage}{.45\linewidth}
    \centering
    \includegraphics[width =7cm]{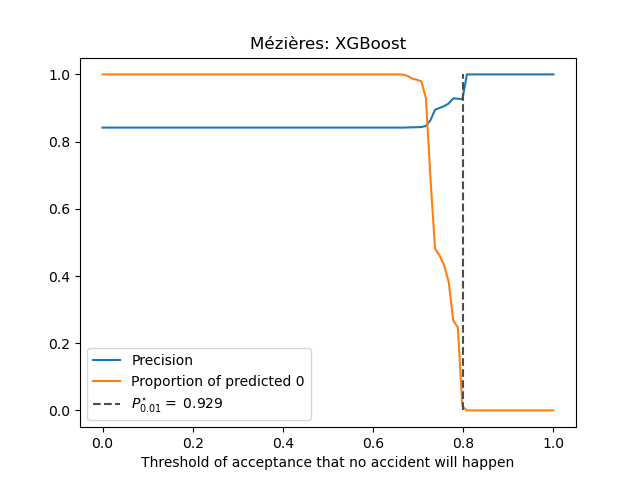}
    \caption{Precision curve, XGBoost, Mézières}
\end{minipage}
\quad
\begin{minipage}{.45\linewidth}
    \centering
    \includegraphics[width = 7cm]{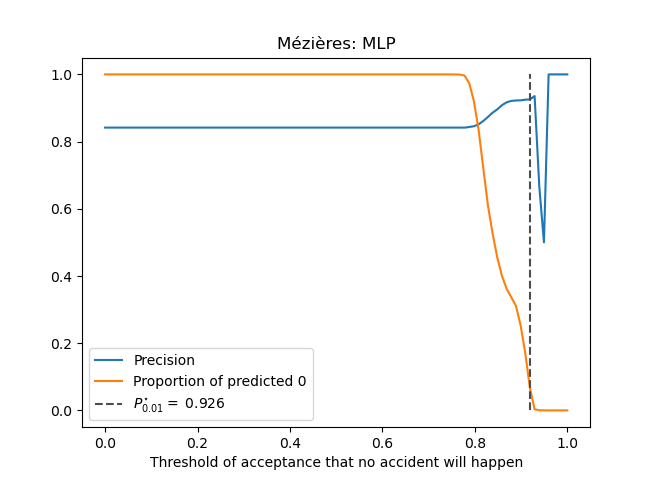}
    \caption{Precision curve, MLP, Mézières}
\end{minipage}
\end{figure}

\begin{figure}[!h]
\begin{minipage}{.45\linewidth}
    \centering
    \includegraphics[width =7cm]{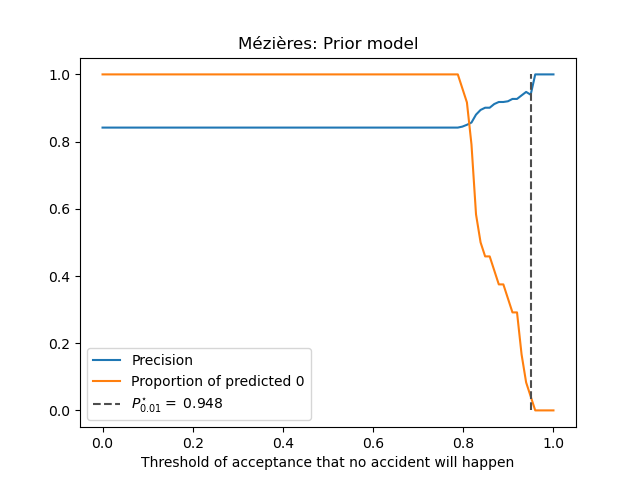}
    \caption{Precision curve, Prior model, Mézières}
\end{minipage}
\quad
\begin{minipage}{.45\linewidth}
    \centering
    \includegraphics[width = 7cm]{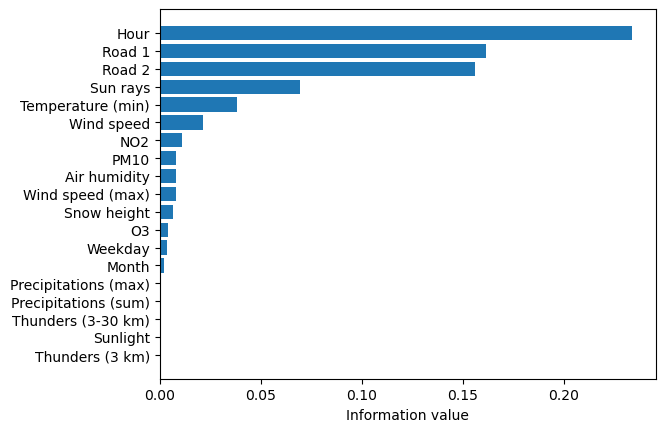}
    \caption{Information values, Mézières}
\end{minipage}
\end{figure}

\begin{figure}[!h]
\begin{minipage}{.45\linewidth}
    \centering
    \includegraphics[width =7cm]{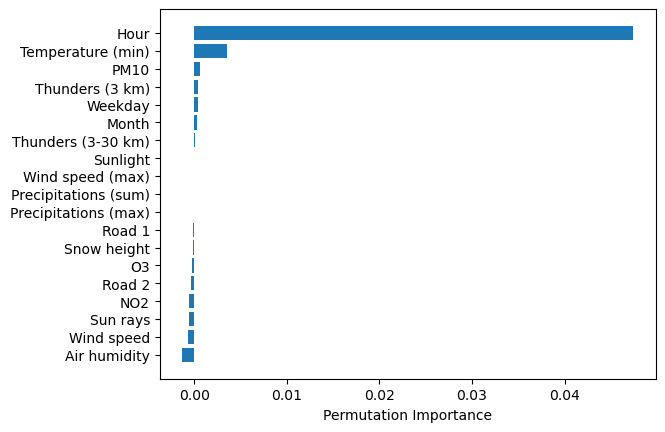}
    \caption{Permutation importance, XGBoost, Mézières}
\end{minipage}
\quad
\begin{minipage}{.45\linewidth}
    \centering
    \includegraphics[width = 7cm]{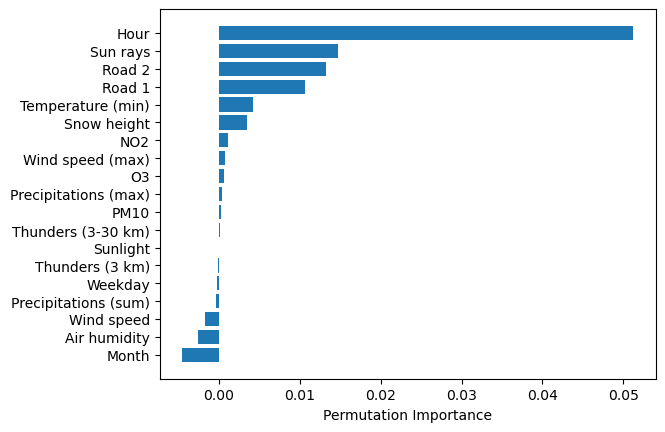}
    \caption{Permutation importance, MLP, Mézières}
\end{minipage}
\end{figure}

\begin{figure}[!h]
\begin{minipage}{.45\linewidth}
    \centering
    \includegraphics[width =7cm]{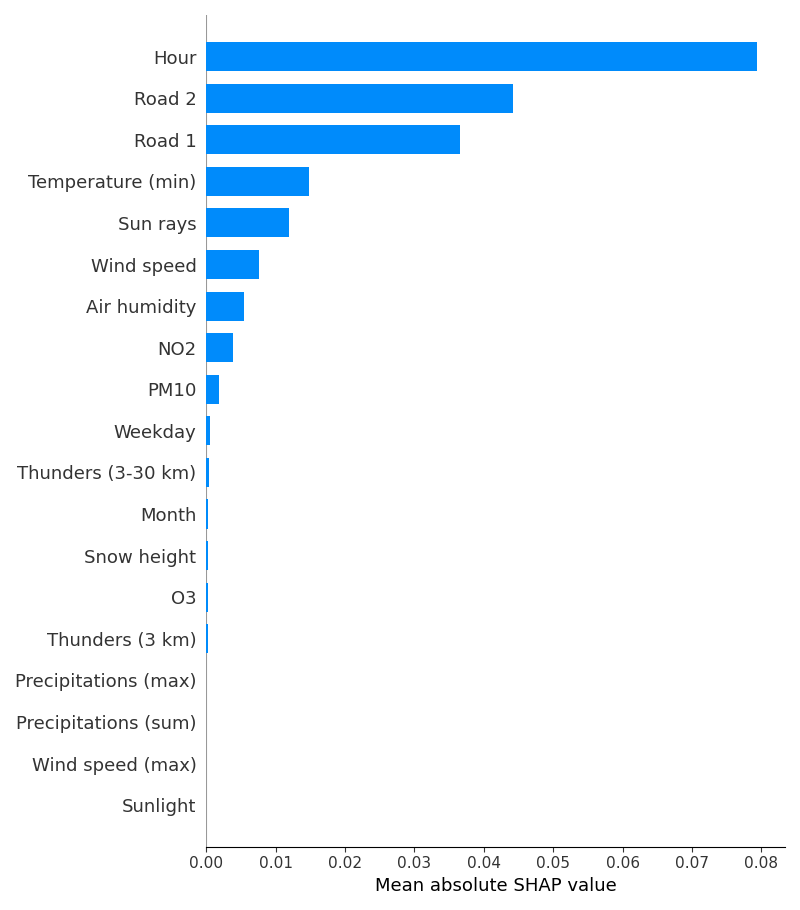}
    \caption{SHAP values, XGBoost, Mézières}
\end{minipage}
\quad
\begin{minipage}{.45\linewidth}
    \centering
    \includegraphics[width = 7cm]{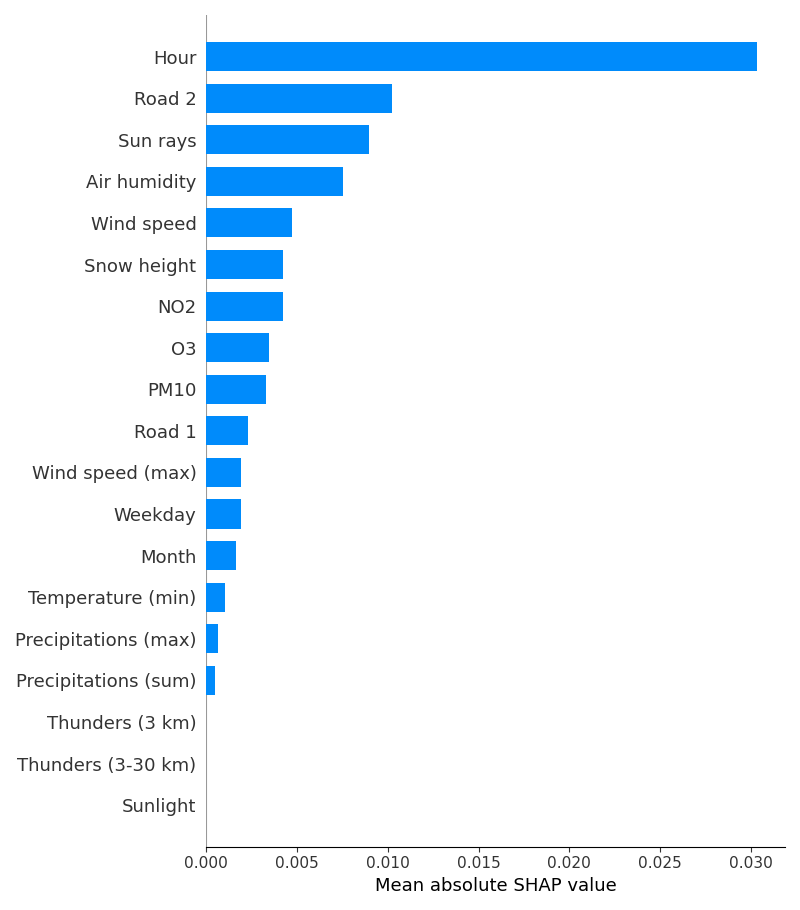}
    \caption{SHAP values, MLP, Mézières}
\end{minipage}
\end{figure}

\FloatBarrier

\vspace{1pts}

\subsection{Aigle}
\begin{figure*}[!h]
    \centering
    \includegraphics[width =0.9\textwidth, height=4in]{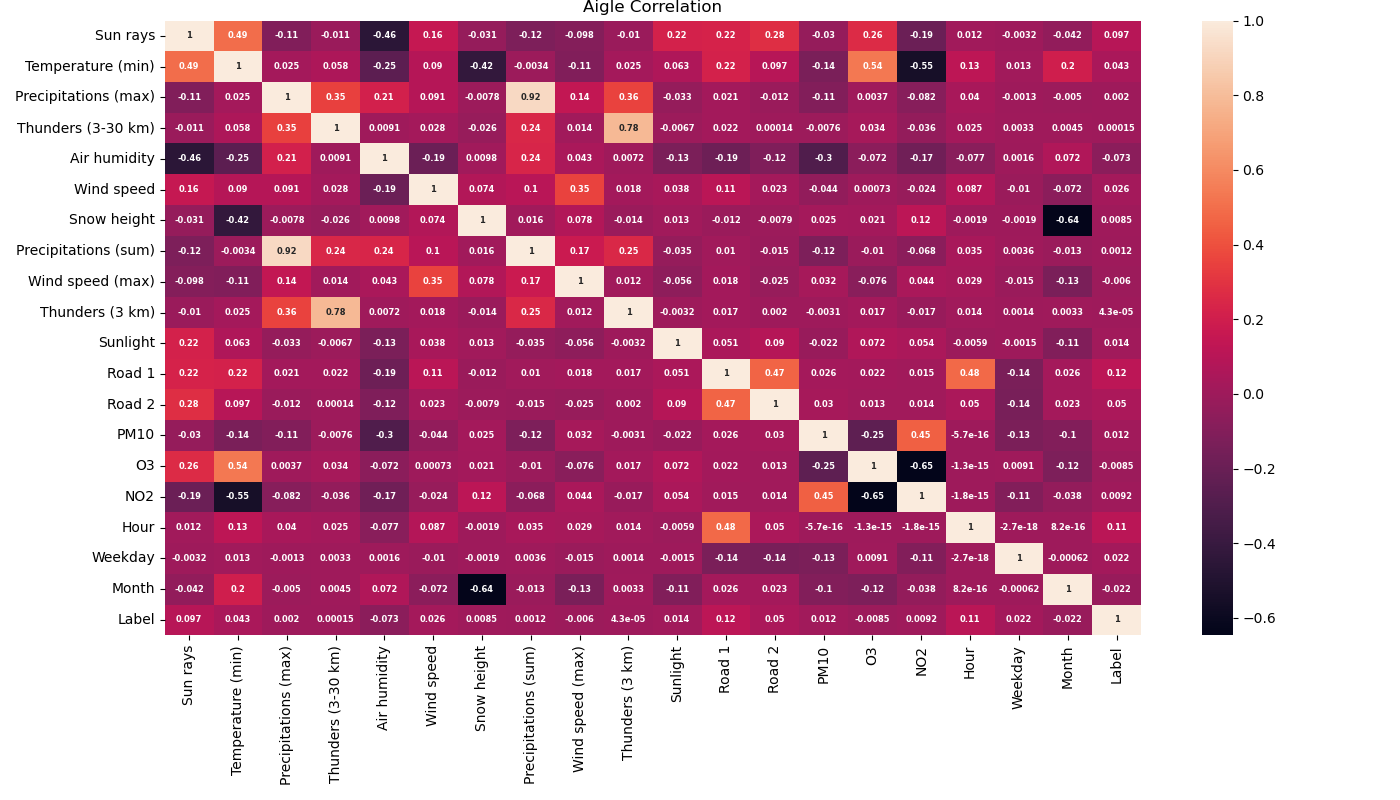}
    \caption{Correlation matrix, Aigle}
\end{figure*}

\begin{figure}[!h]
\begin{minipage}{.45\linewidth}
    \centering
    \includegraphics[width =7cm]{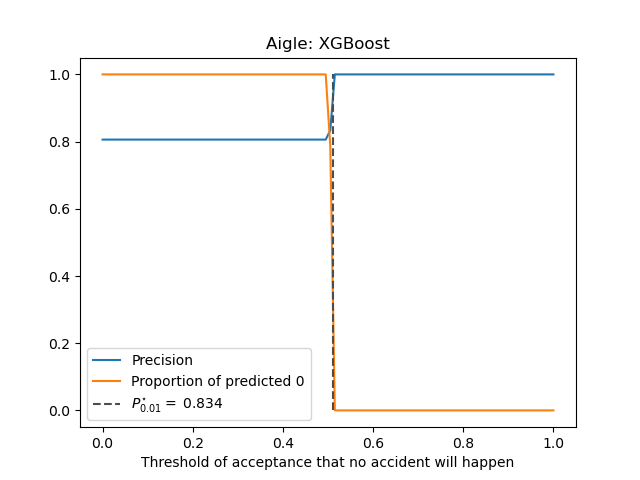}
    \caption{Precision curve, XGBoost, Aigle}
\end{minipage}
\quad
\begin{minipage}{.45\linewidth}
    \centering
    \includegraphics[width = 7cm]{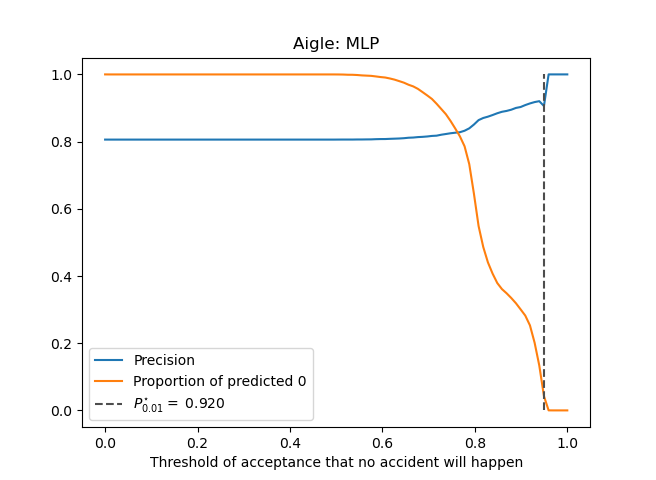}
    \caption{Precision curve, MLP, Aigle}
\end{minipage}
\end{figure}

\begin{figure}[!h]
\begin{minipage}{.45\linewidth}
    \centering
    \includegraphics[width =7cm]{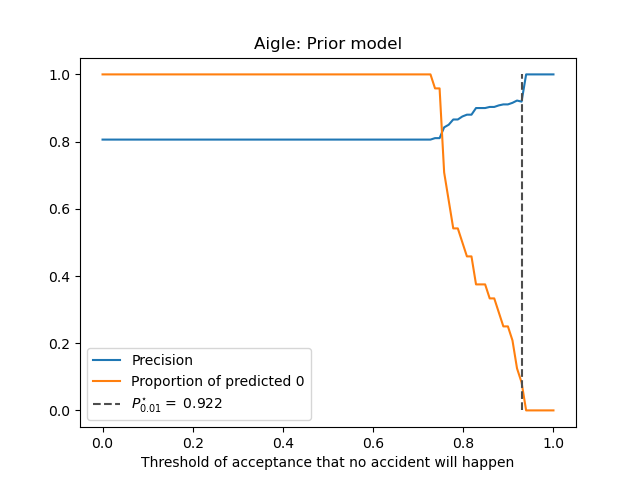}
    \caption{Precision curve, Prior model, Aigle}
\end{minipage}
\quad
\begin{minipage}{.45\linewidth}
    \centering
    \includegraphics[width = 7cm]{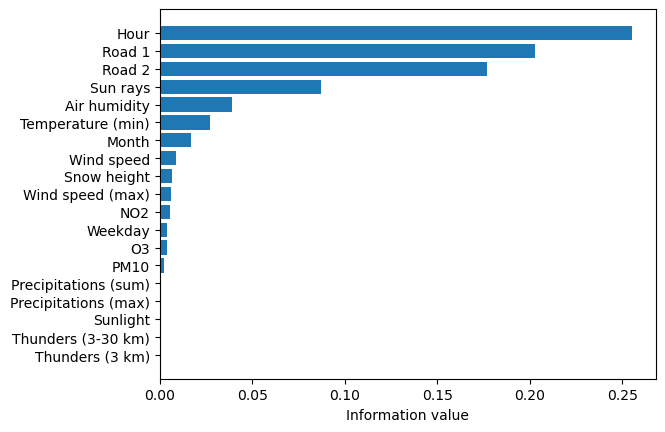}
    \caption{Information values, Aigle}
\end{minipage}
\end{figure}

\begin{figure}[!h]
\begin{minipage}{.45\linewidth}
    \centering
    \includegraphics[width =7cm]{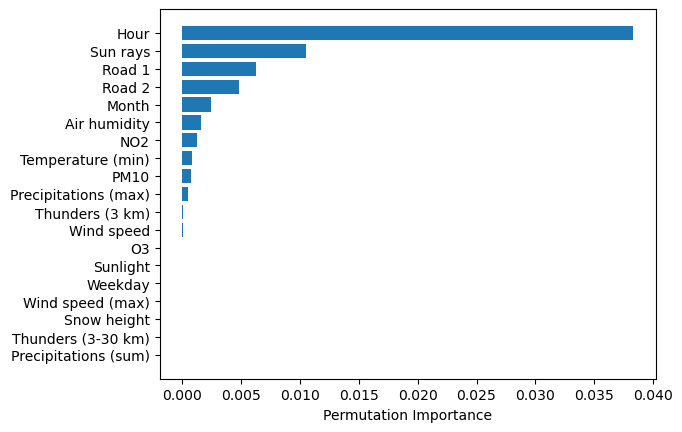}
    \caption{Permutation importance, XGBoost, Aigle}
\end{minipage}
\quad
\begin{minipage}{.45\linewidth}
    \centering
    \includegraphics[width = 7cm]{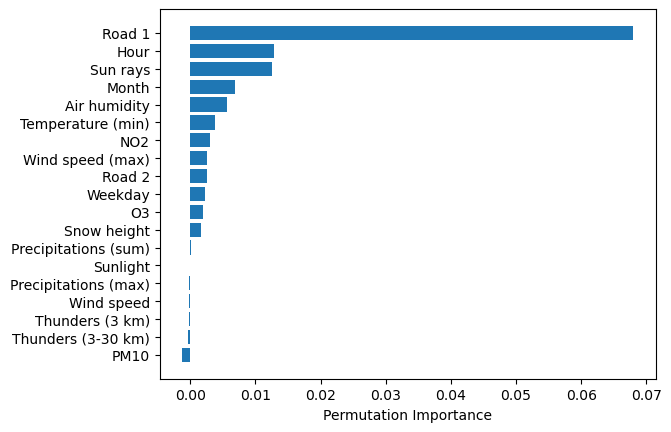}
    \caption{Permutation importance, MLP, Aigle}
\end{minipage}
\end{figure}

\begin{figure}[!h]
\begin{minipage}{.45\linewidth}
    \centering
    \includegraphics[width =7cm]{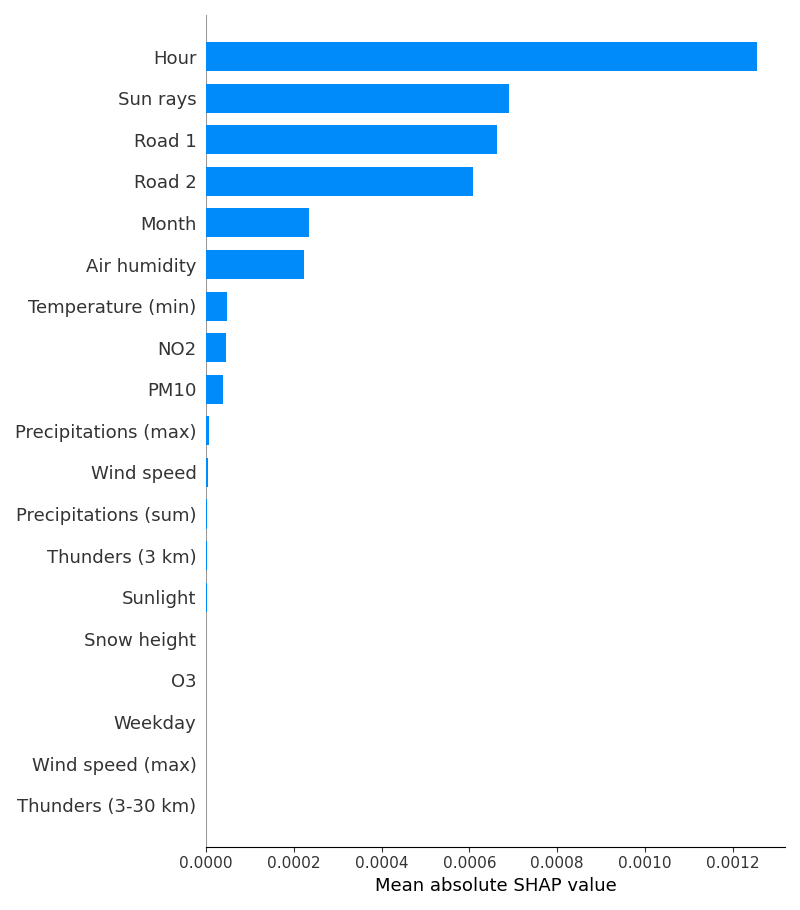}
    \caption{SHAP values, XGBoost, Aigle}
\end{minipage}
\quad
\begin{minipage}{.45\linewidth}
    \centering
    \includegraphics[width = 7cm]{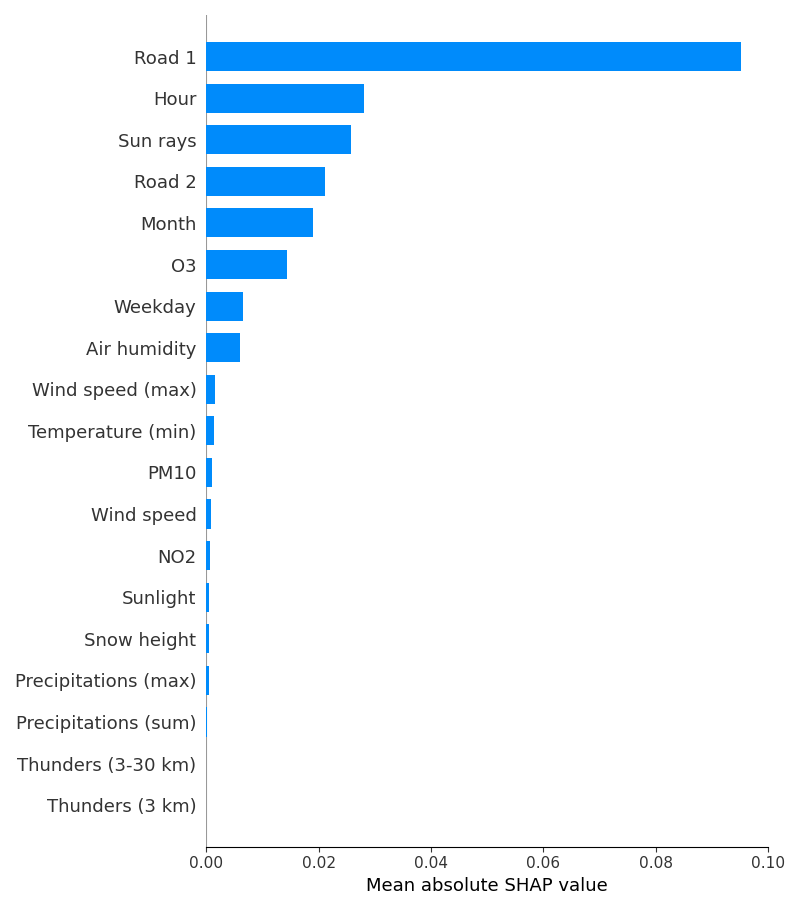}
    \caption{SHAP values, MLP, Aigle}
\end{minipage}
\end{figure}

\FloatBarrier

\vspace{10pts}
\subsection{Château-d'Oex}
\begin{figure*}[!h]
    \centering
    \includegraphics[width =0.9\textwidth, height=4in]{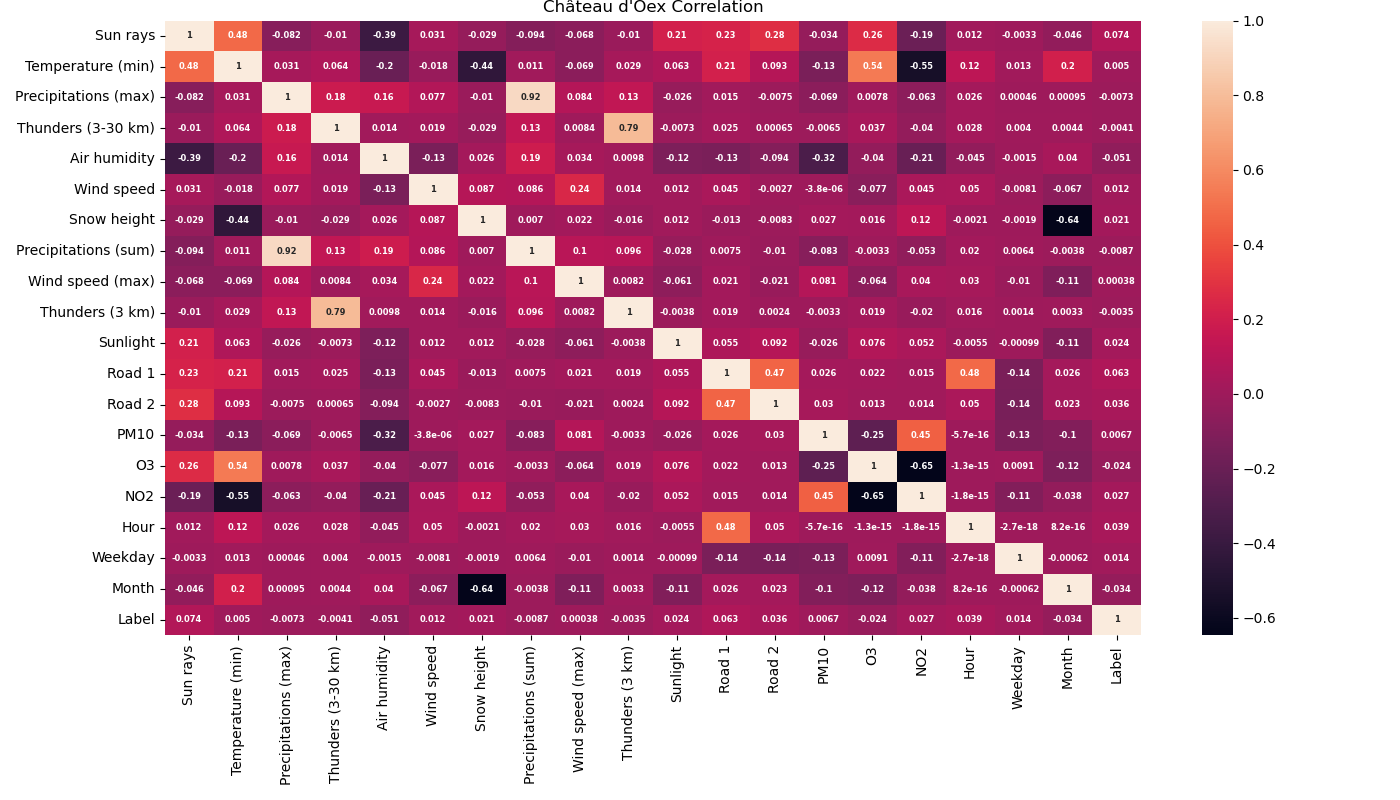}
    \caption{Correlation matrix, Château-d'Oex}
    \label{chateau_corr}
\end{figure*}

\begin{figure}[!h]
\begin{minipage}{.45\linewidth}
    \centering
    \includegraphics[width =7cm]{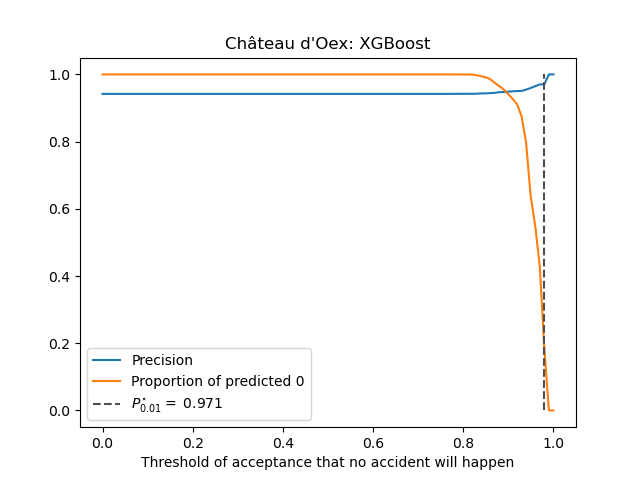}
    \caption{Precision curve, XGBoost, Château-d'Oex}
\end{minipage}
\quad
\begin{minipage}{.45\linewidth}
    \centering
    \includegraphics[width = 7cm]{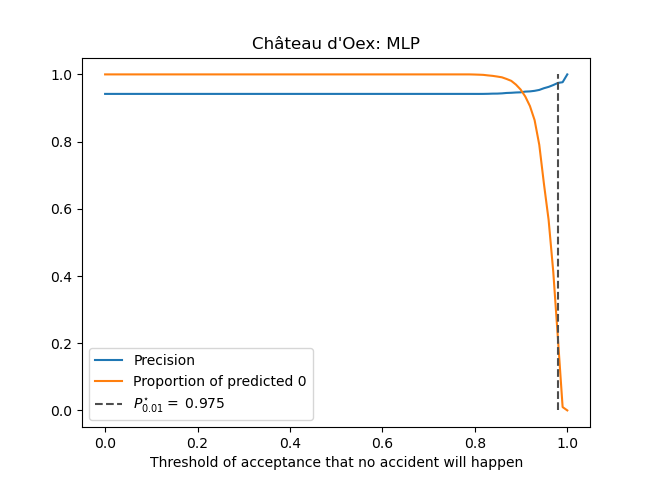}
    \caption{Precision curve, MLP, Château-d'Oex}
\end{minipage}
\end{figure}

\begin{figure}[!h]
\begin{minipage}{.45\linewidth}
    \centering
    \includegraphics[width =7cm]{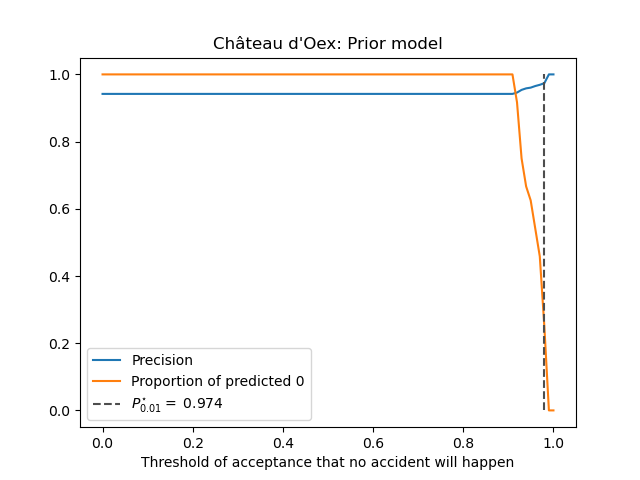}
    \caption{Precision curve, Prior model, Château-d'Oex}
\end{minipage}
\quad
\begin{minipage}{.45\linewidth}
    \centering
    \includegraphics[width = 7cm]{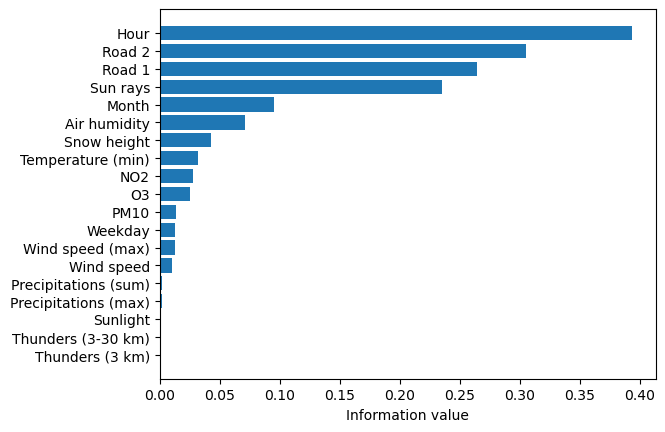}
    \caption{Information values, Château-d'Oex}
\end{minipage}
\end{figure}

\begin{figure}[!h]
\begin{minipage}{.45\linewidth}
    \centering
    \includegraphics[width =7cm]{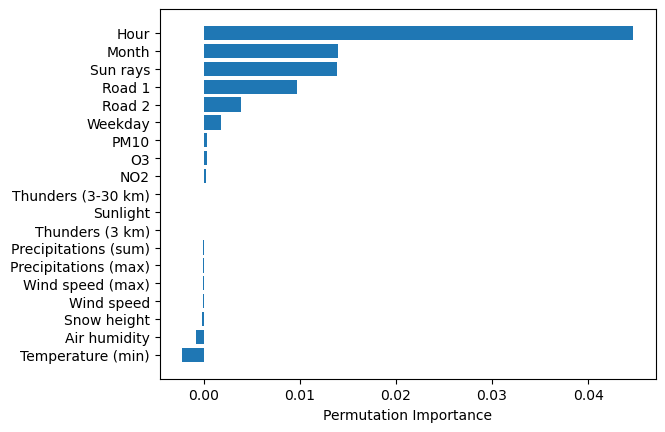}
    \caption{Permutation importance, XGBoost, Château-d'Oex}
\end{minipage}
\quad
\begin{minipage}{.45\linewidth}
    \centering
    \includegraphics[width = 7cm]{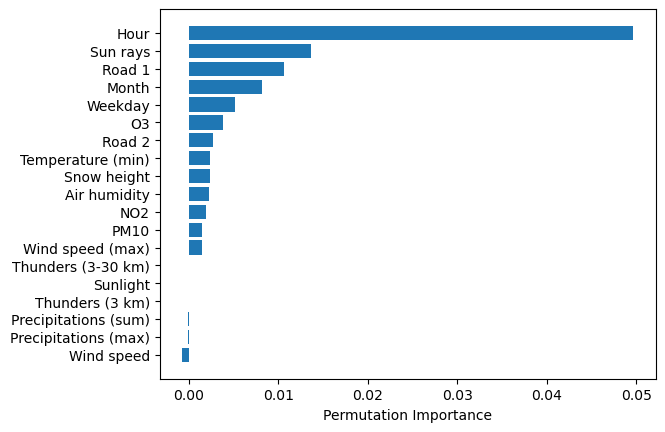}
    \caption{Permutation importance, MLP, Château-d'Oex}
\end{minipage}
\end{figure}

\begin{figure}[!h]
\begin{minipage}{.45\linewidth}
    \centering
    \includegraphics[width =7cm]{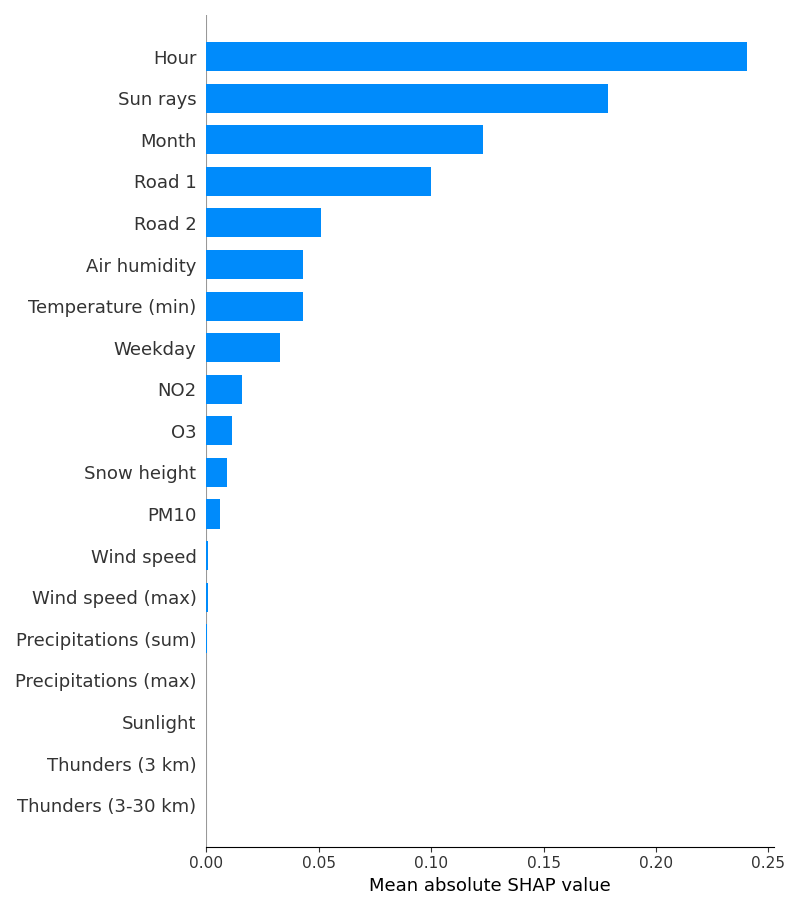}
    \caption{SHAP values, XGBoost, Château-d'Oex}
\end{minipage}
\quad
\begin{minipage}{.45\linewidth}
    \centering
    \includegraphics[width = 7cm]{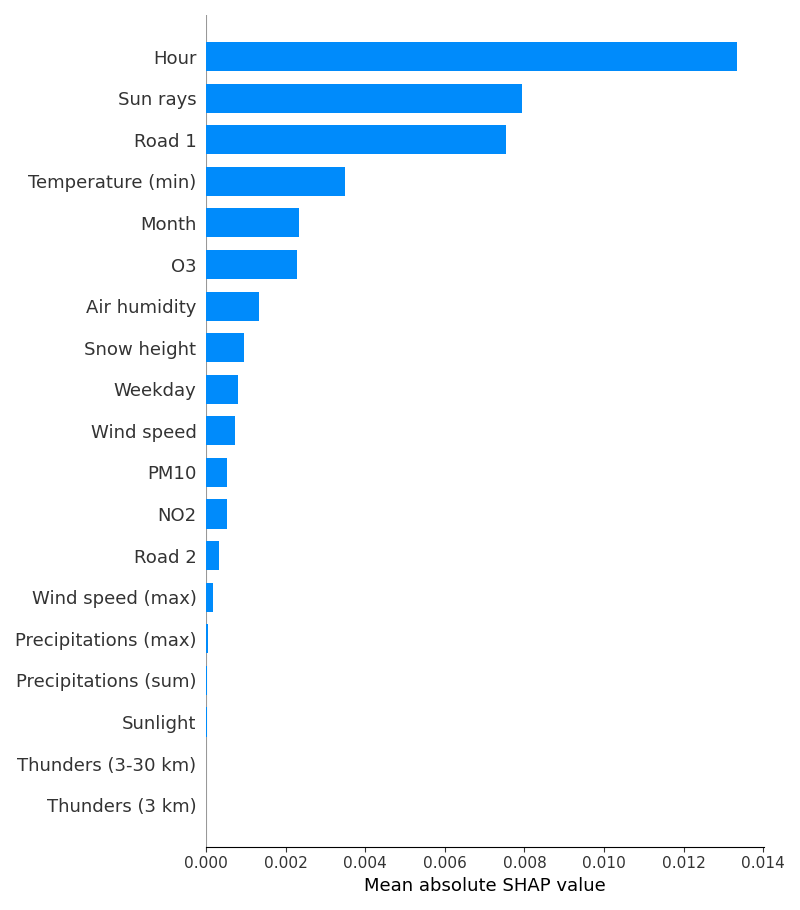}
    \caption{SHAP values, MLP, Château-d'Oex}
\end{minipage}
\end{figure}

\FloatBarrier

\vspace{1pts}
\subsection{L'Abbaye}
\begin{figure*}[!h]
    \centering
    \includegraphics[width =0.9\textwidth, height=4in]{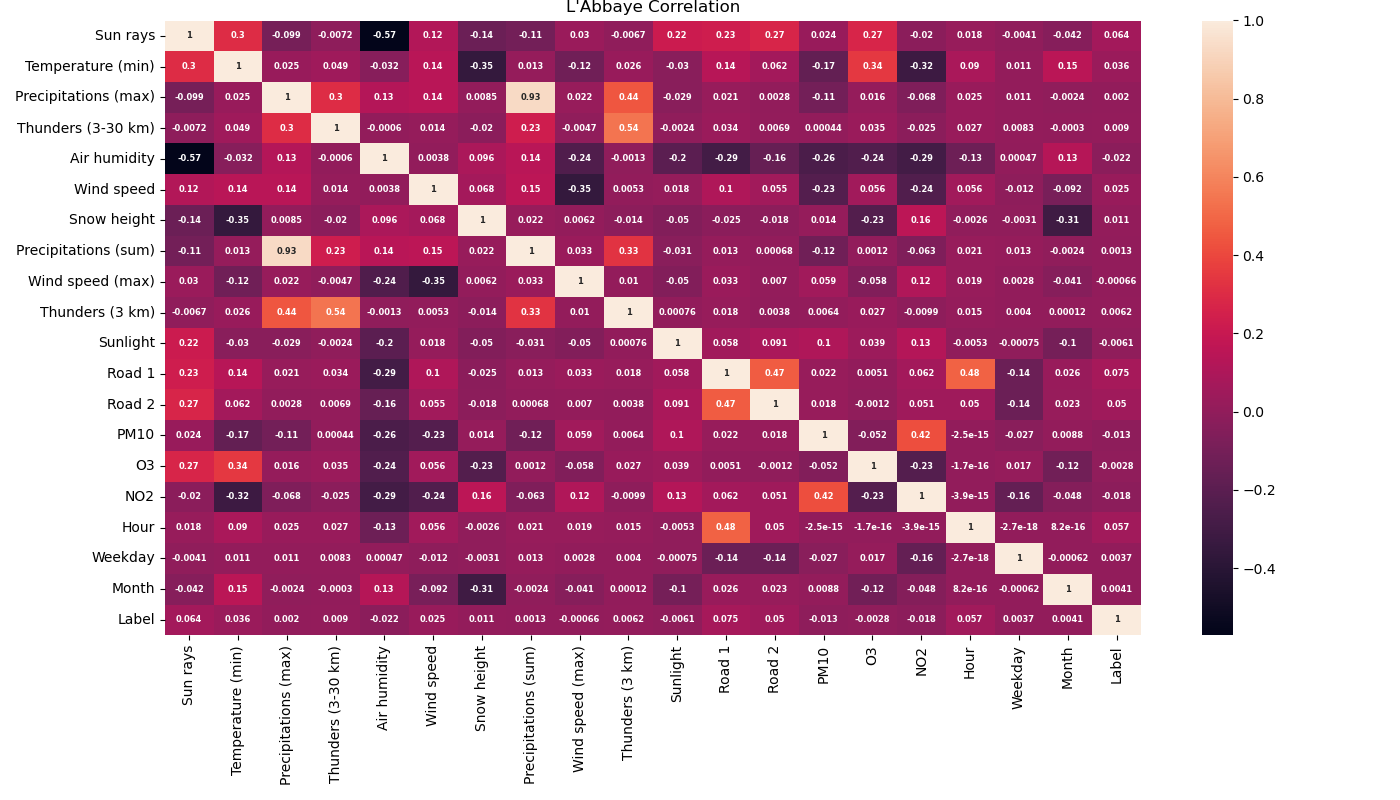}
    \caption{Correlation matrix, L'Abbaye}
    \label{abbaye_corr}
\end{figure*}

\begin{figure}[!h]
\begin{minipage}{.45\linewidth}
    \centering
    \includegraphics[width =7cm]{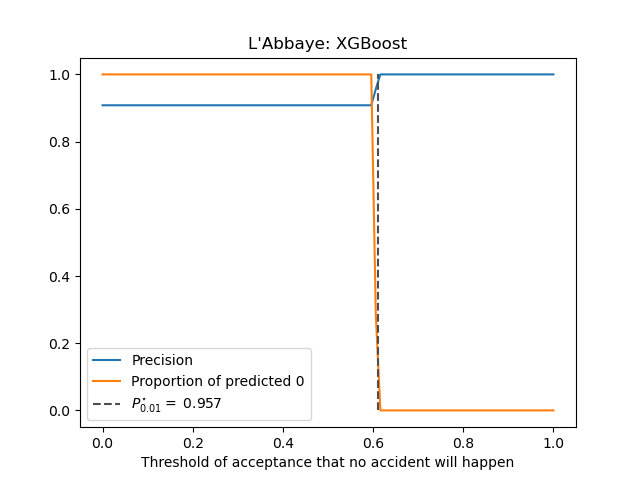}
    \caption{Precision curve, XGBoost, L'Abbaye}
\end{minipage}
\quad
\begin{minipage}{.45\linewidth}
    \centering
    \includegraphics[width = 7cm]{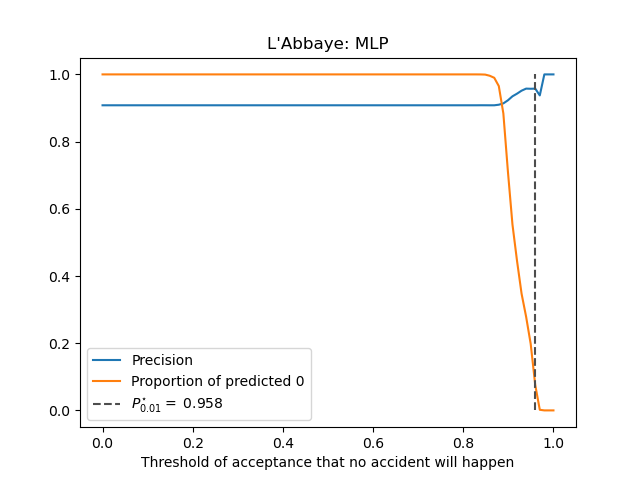}
    \caption{Precision curve, MLP, L'Abbaye}
\end{minipage}
\end{figure}

\begin{figure}[!h]
\begin{minipage}{.45\linewidth}
    \centering
    \includegraphics[width =7cm]{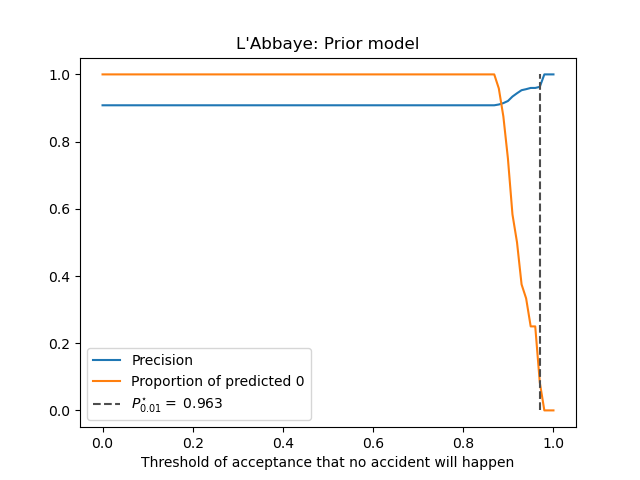}
    \caption{Precision curve, Prior model, L'Abbaye}
\end{minipage}
\quad
\begin{minipage}{.45\linewidth}
    \centering
    \includegraphics[width = 7cm]{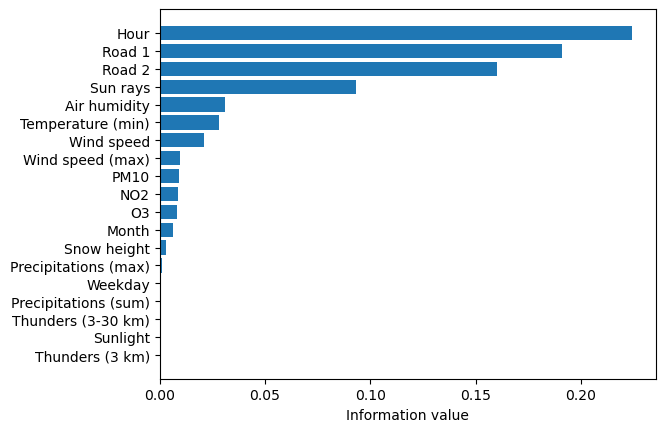}
    \caption{Information values, L'Abbaye}
\end{minipage}
\end{figure}

\begin{figure}[!h]
\begin{minipage}{.45\linewidth}
    \centering
    \includegraphics[width =7cm]{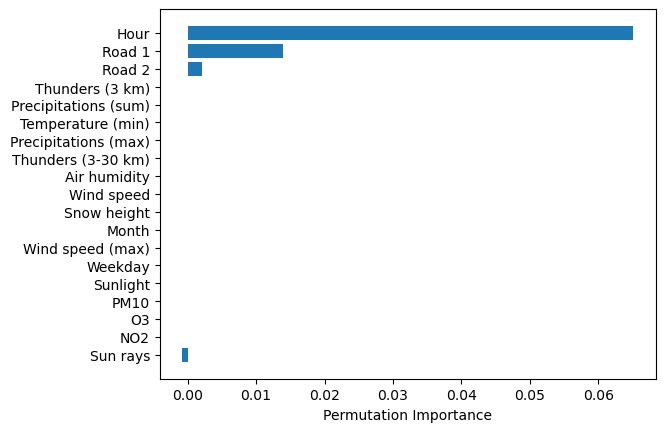}
    \caption{Permutation importance, XGBoost, L'Abbaye}
\end{minipage}
\quad
\begin{minipage}{.45\linewidth}
    \centering
    \includegraphics[width = 7cm]{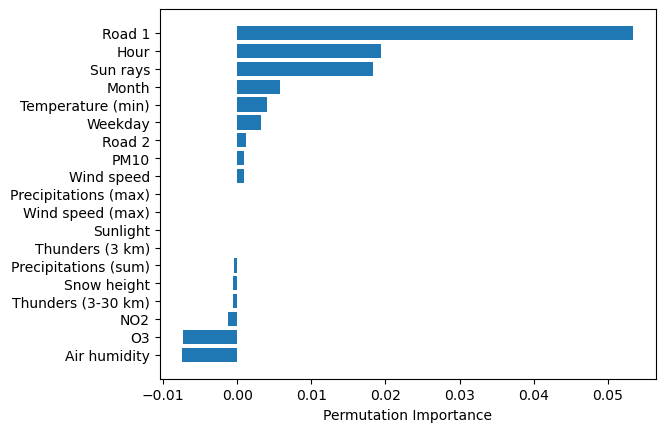}
    \caption{Permutation importance, MLP, L'Abbaye}
\end{minipage}
\end{figure}

\begin{figure}[!h]
\begin{minipage}{.45\linewidth}
    \centering
    \includegraphics[width =7cm]{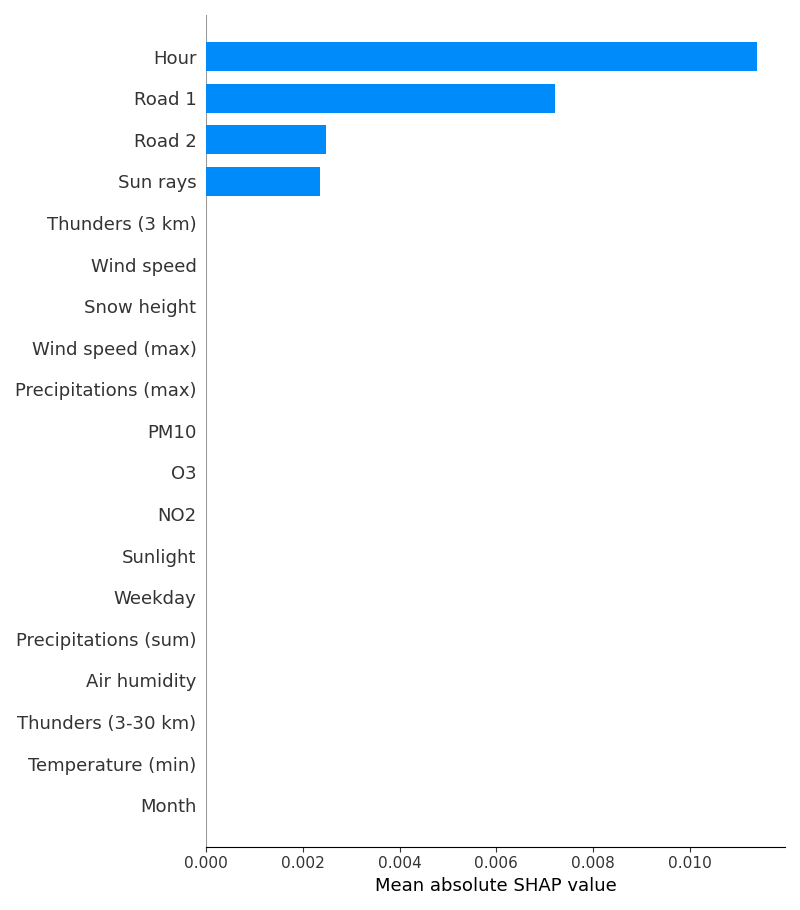}
    \caption{SHAP values, XGBoost, L'Abbaye}
\end{minipage}
\quad
\begin{minipage}{.45\linewidth}
    \centering
    \includegraphics[width = 7cm]{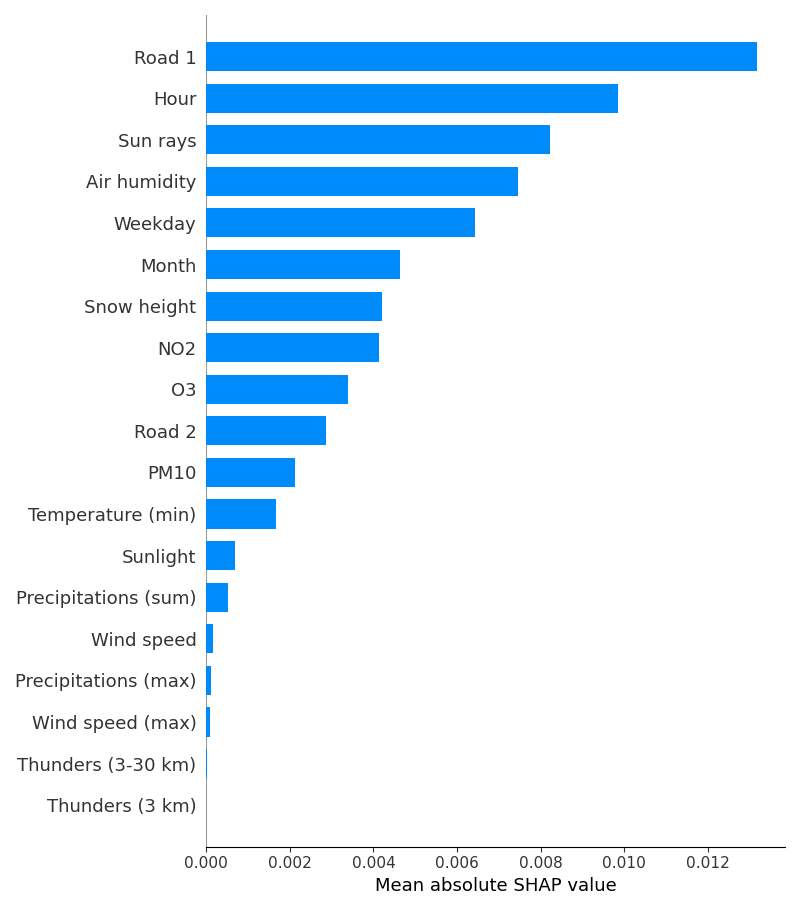}
    \caption{SHAP values, MLP, L'Abbaye}
\end{minipage}
\end{figure}

\FloatBarrier
\vspace{1pts}

\subsection{Pompales}
\begin{figure*}[!h]
    \centering
    \includegraphics[width =0.9\textwidth, height=4in]{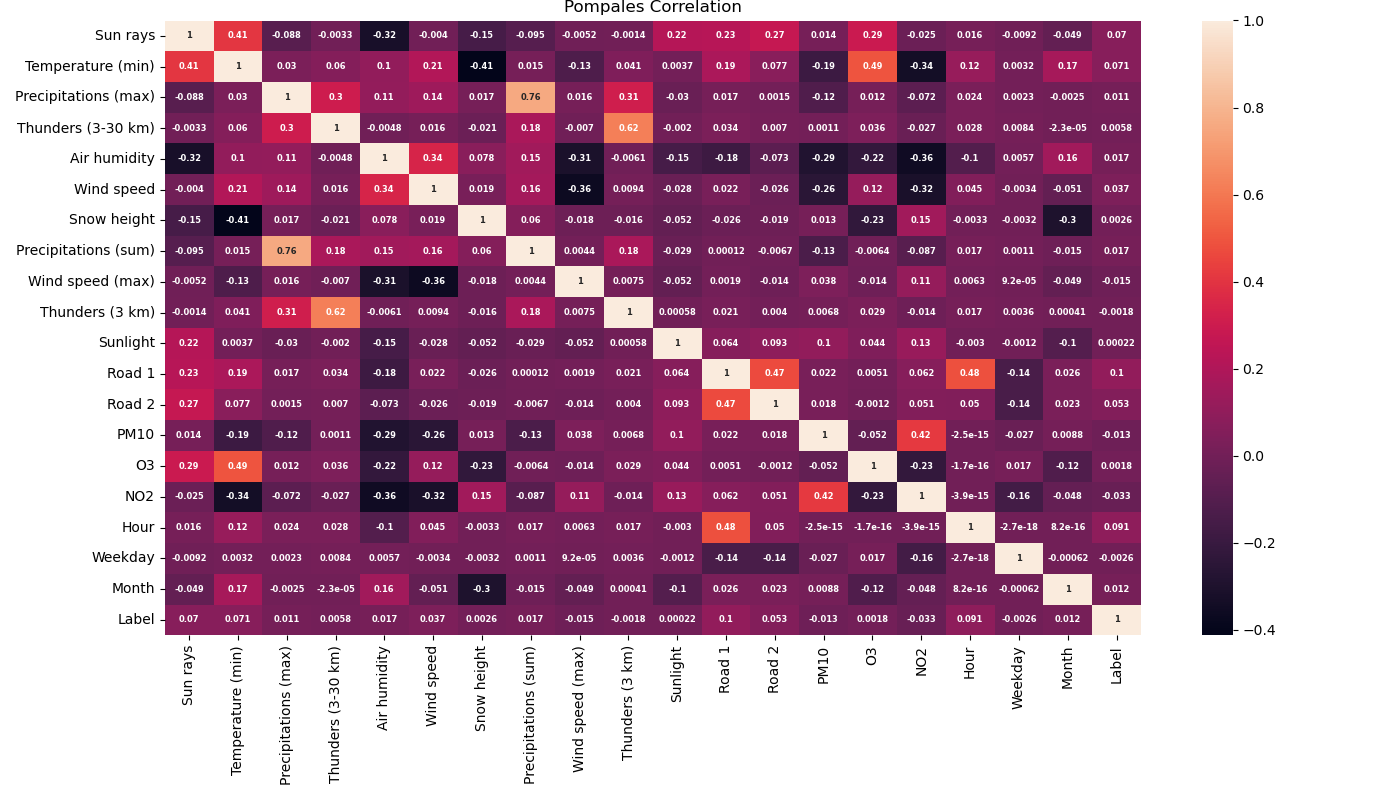}
    \caption{Correlation matrix, Pompales}
    \label{pompales_corr}
\end{figure*}

\begin{figure}[!h]
\begin{minipage}{.45\linewidth}
    \centering
    \includegraphics[width =7cm]{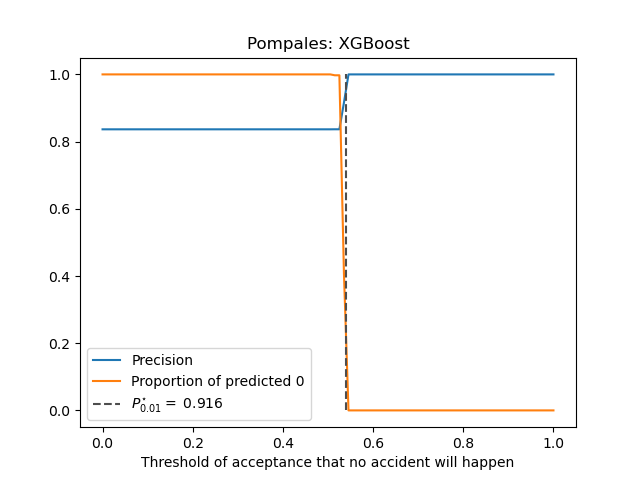}
    \caption{Precision curve, XGBoost, Pompales}
\end{minipage}
\quad
\begin{minipage}{.45\linewidth}
    \centering
    \includegraphics[width = 7cm]{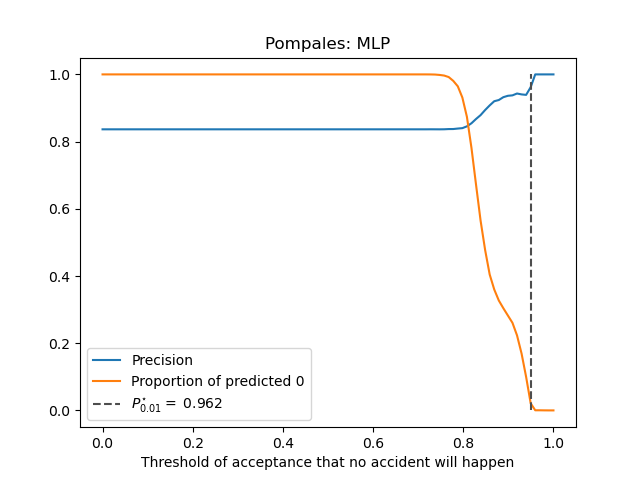}
    \caption{Precision curve, MLP, Pompales}
\end{minipage}
\end{figure}

\begin{figure}[!h]
\begin{minipage}{.45\linewidth}
    \centering
    \includegraphics[width =7cm]{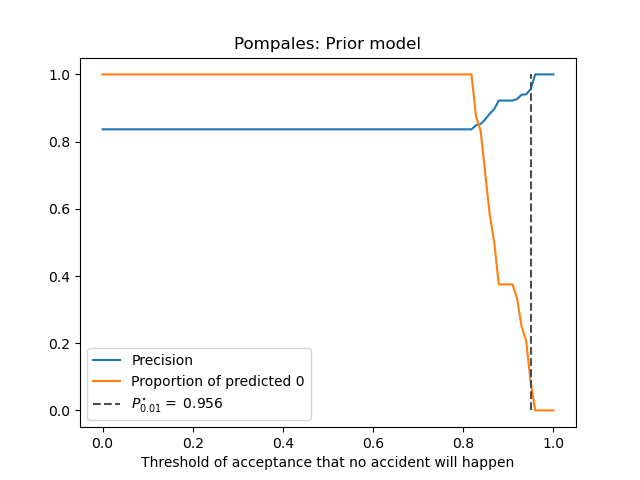}
    \caption{Precision curve, Prior model, Pompales}
\end{minipage}
\quad
\begin{minipage}{.45\linewidth}
    \centering
    \includegraphics[width = 7cm]{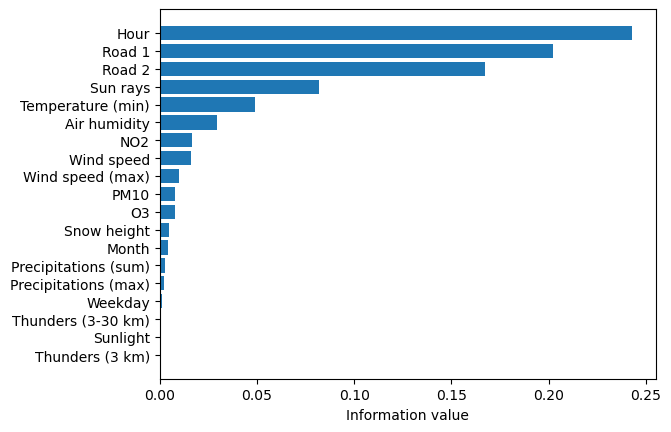}
    \caption{Information values, Pompales}
\end{minipage}
\end{figure}

\begin{figure}[!h]
\begin{minipage}{.45\linewidth}
    \centering
    \includegraphics[width =7cm]{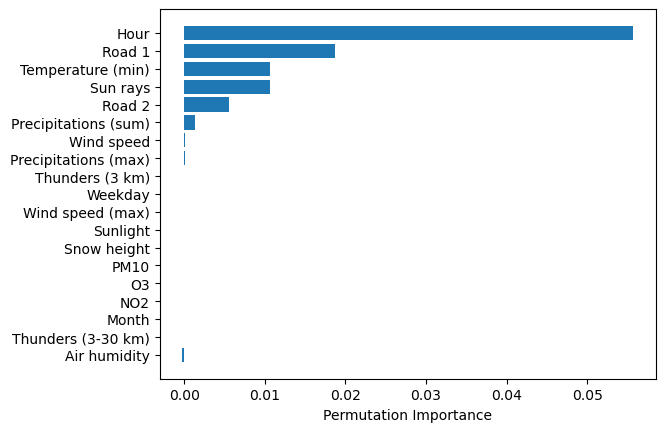}
    \caption{Permutation importance, XGBoost, Pompales}
\end{minipage}
\quad
\begin{minipage}{.45\linewidth}
    \centering
    \includegraphics[width = 7cm]{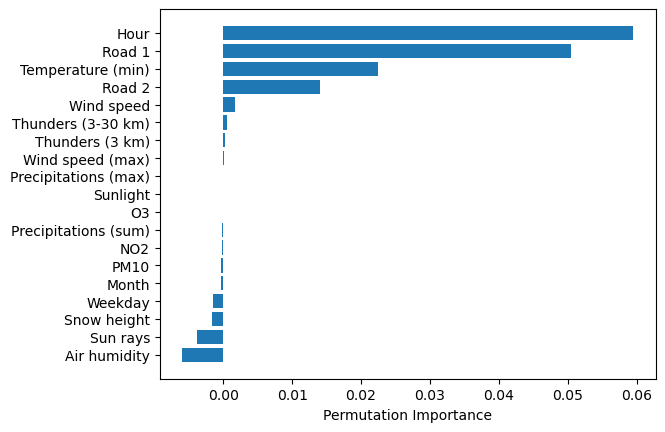}
    \caption{Permutation importance, MLP, Pompales}
\end{minipage}
\end{figure}

\begin{figure}[!h]
\begin{minipage}{.45\linewidth}
    \centering
    \includegraphics[width =7cm]{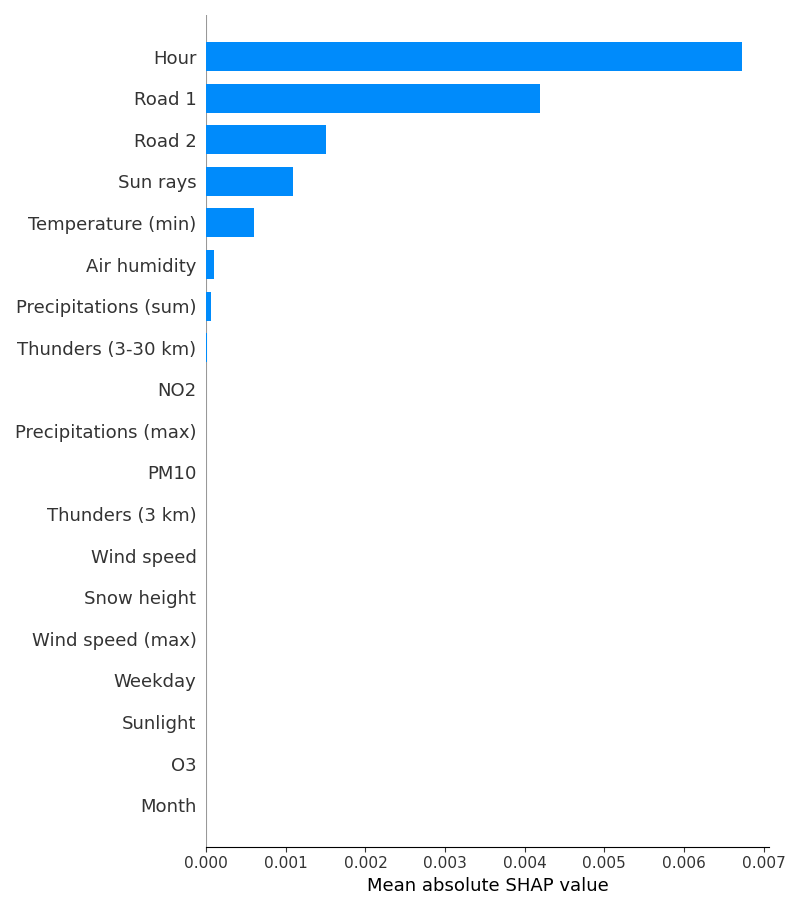}
    \caption{SHAP values, XGBoost, Pompales}
\end{minipage}
\quad
\begin{minipage}{.45\linewidth}
    \centering
    \includegraphics[width = 7cm]{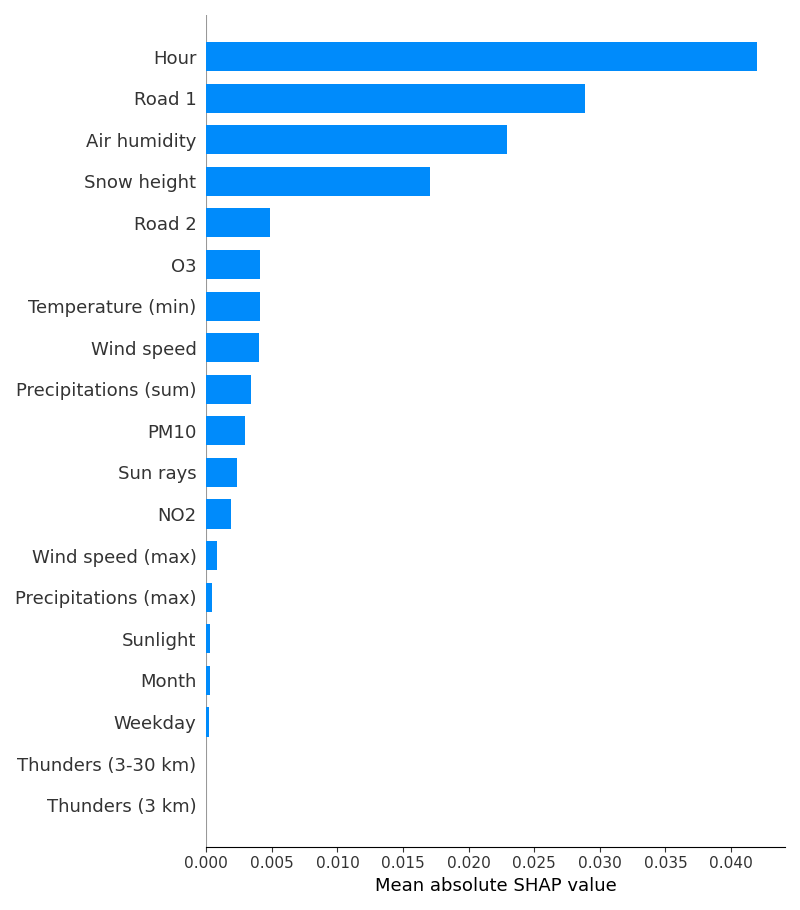}
    \caption{SHAP values, MLP, Pompales}
\end{minipage}
\end{figure}

\FloatBarrier

\vspace{1pts}

\subsection{Sainte-Croix}
\begin{figure*}[!h]
    \centering
    \includegraphics[width =0.9\textwidth, height=4in]{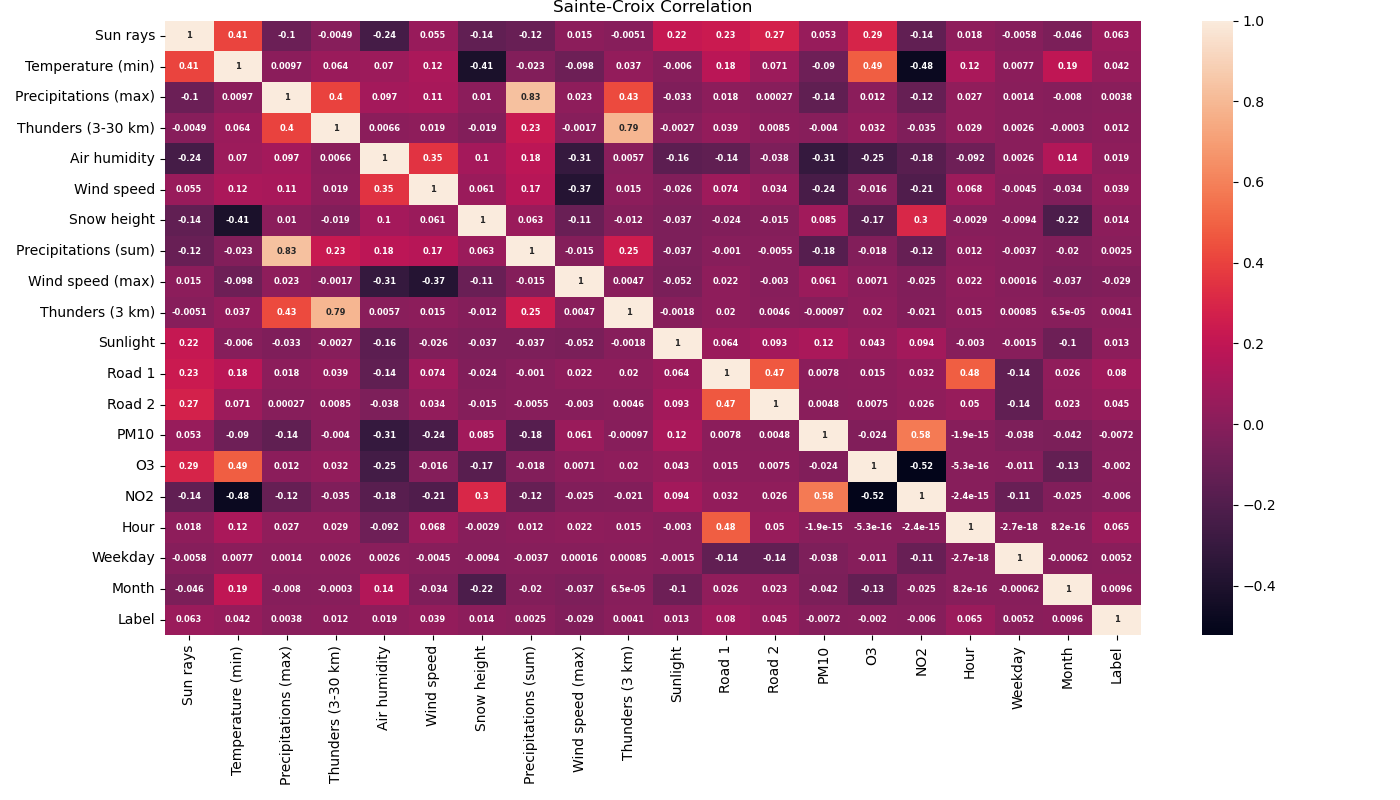}
    \caption{Correlation matrix, Sainte-Croix}
    \label{sainte-croix_corr}
\end{figure*}

\begin{figure}[!h]
\begin{minipage}{.45\linewidth}
    \centering
    \includegraphics[width =7cm]{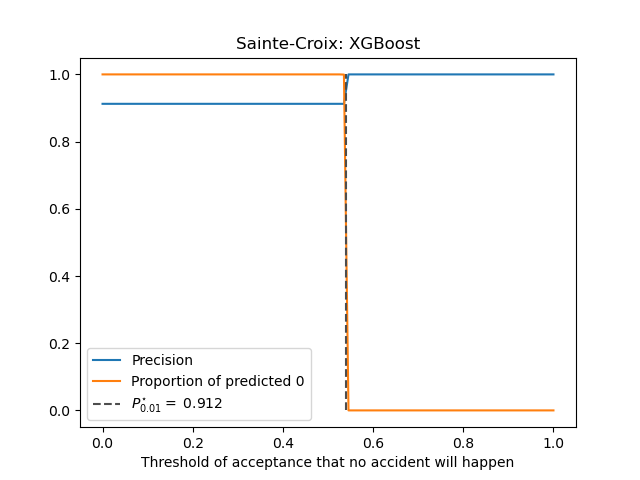}
    \caption{Precision curve, XGBoost, Sainte-Croix}
\end{minipage}
\quad
\begin{minipage}{.45\linewidth}
    \centering
    \includegraphics[width = 7cm]{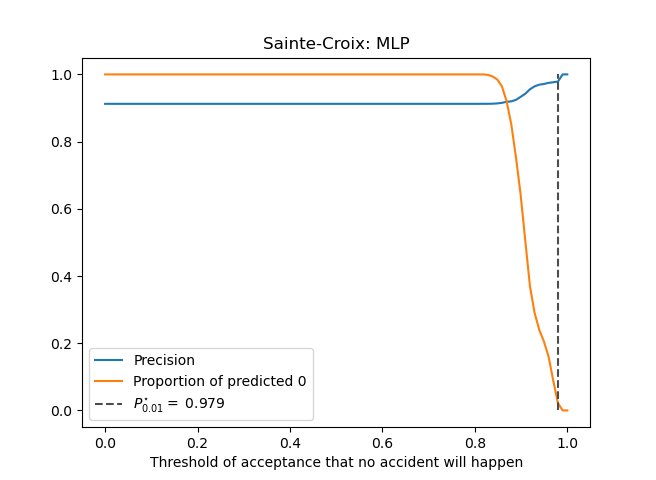}
    \caption{Precision curve, MLP, Sainte-Croix}
\end{minipage}
\end{figure}

\begin{figure}[!h]
\begin{minipage}{.45\linewidth}
    \centering
    \includegraphics[width =7cm]{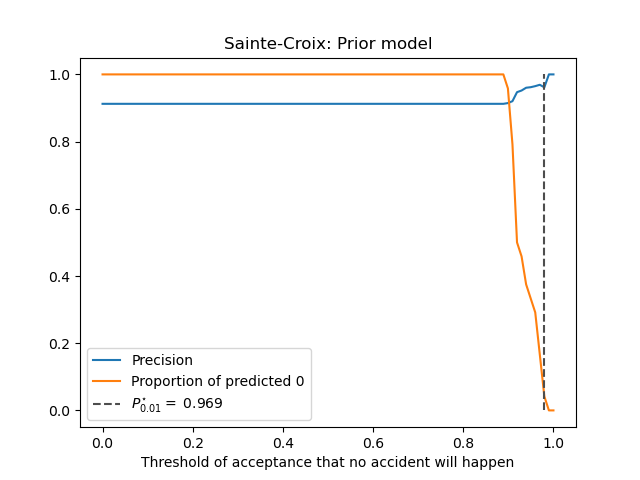}
    \caption{Precision curve, Prior model, Sainte-Croix}
\end{minipage}
\quad
\begin{minipage}{.45\linewidth}
    \centering
    \includegraphics[width = 7cm]{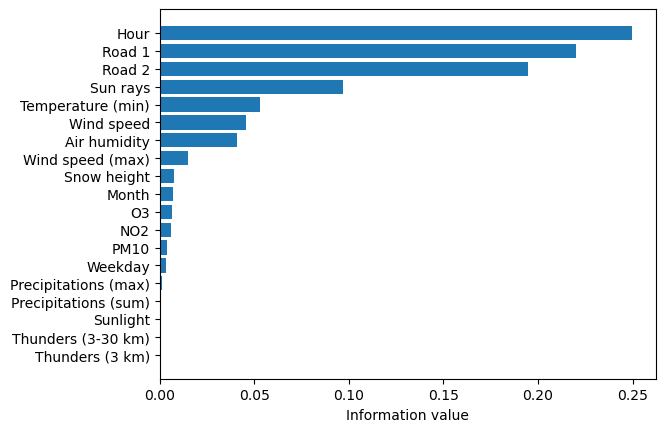}
    \caption{Information values, Sainte-Croix}
\end{minipage}
\end{figure}

\begin{figure}[!h]
\begin{minipage}{.45\linewidth}
    \centering
    \includegraphics[width =7cm]{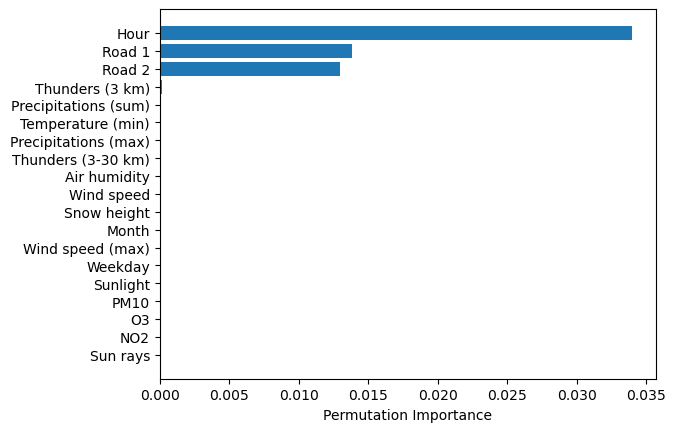}
    \caption{Permutation importance, XGBoost, Sainte-Croix}
\end{minipage}
\quad
\begin{minipage}{.45\linewidth}
    \centering
    \includegraphics[width = 7cm]{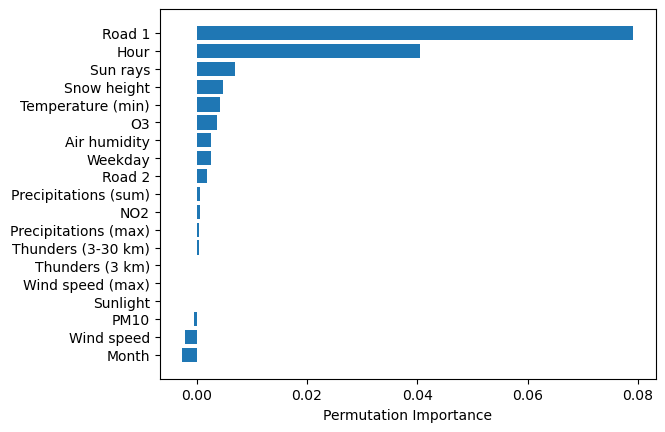}
    \caption{Permutation importance, MLP, Sainte-Croix}
\end{minipage}
\end{figure}

\begin{figure}[!h]
\begin{minipage}{.45\linewidth}
    \centering
    \includegraphics[width =7cm]{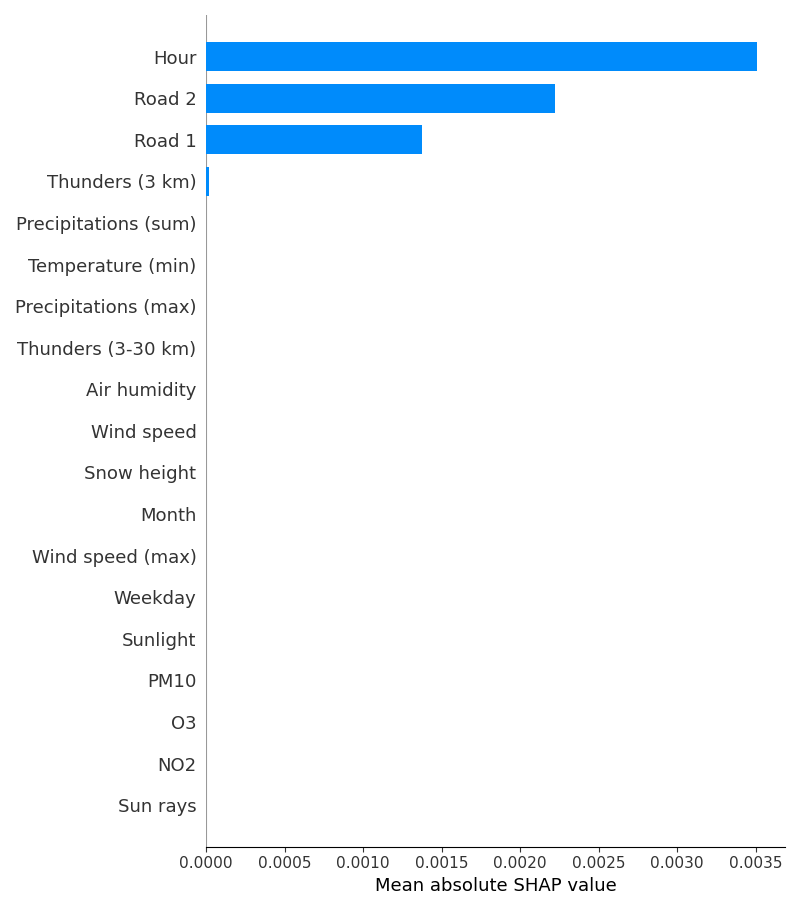}
    \caption{SHAP values, XGBoost, Sainte-Croix}
\end{minipage}
\quad
\begin{minipage}{.45\linewidth}
    \centering
    \includegraphics[width = 7cm]{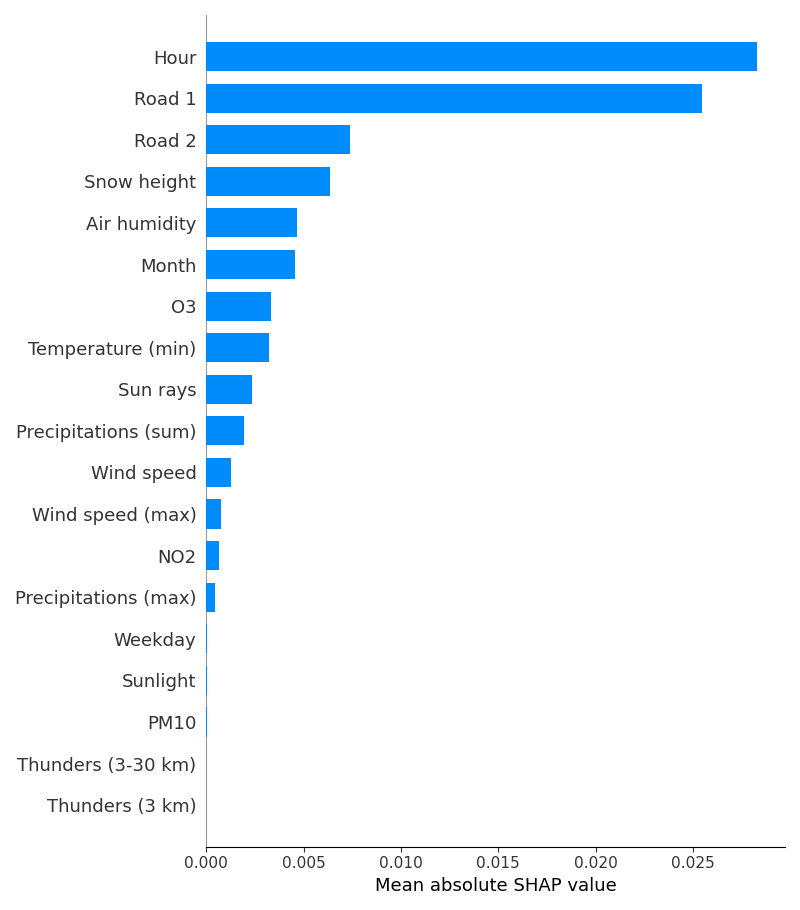}
    \caption{SHAP values, MLP, Sainte-Croix}
\end{minipage}
\end{figure}

\FloatBarrier

\vspace{1pts}

\subsection{Payerne}
\begin{figure*}[!h]
    \centering
    \includegraphics[width =0.9\textwidth, height=4in]{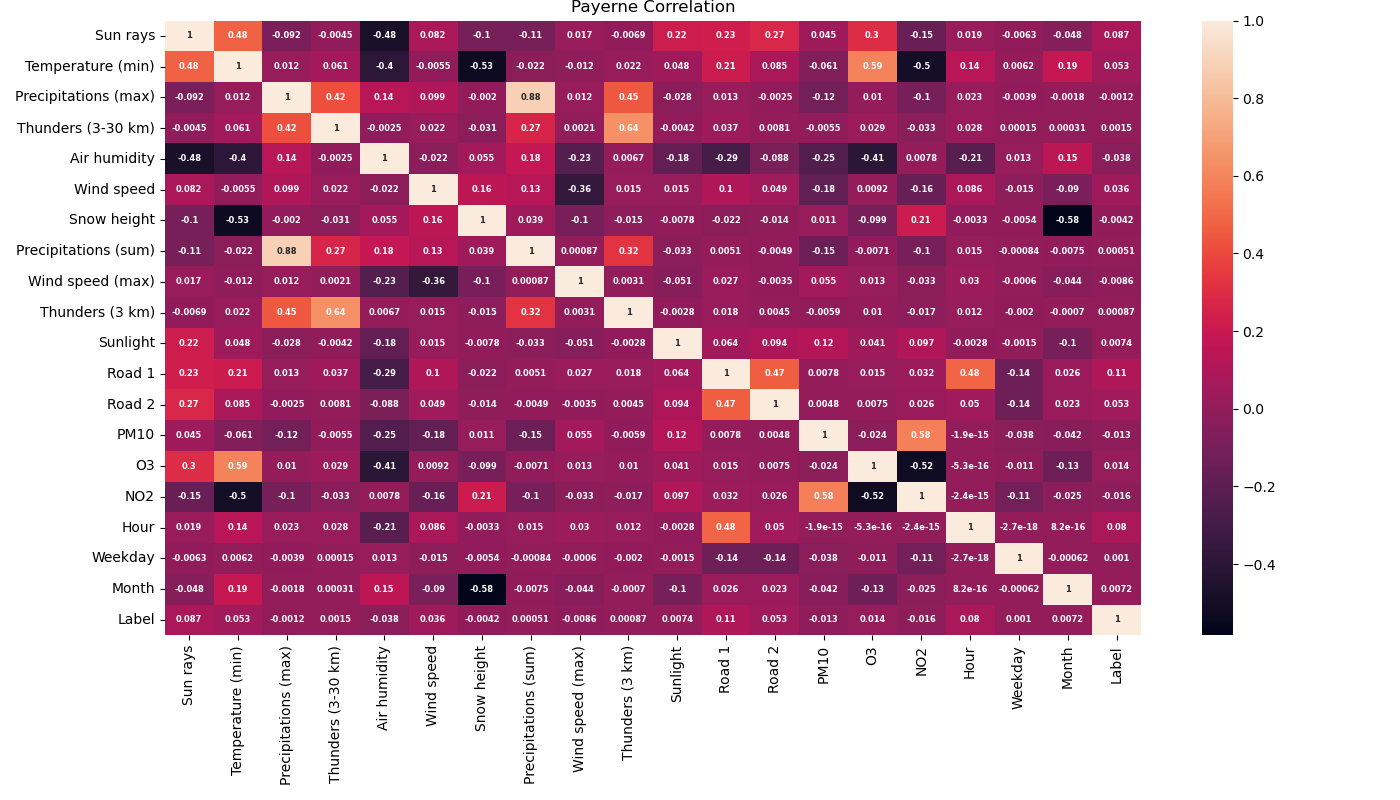}
    \caption{Correlation matrix, Payerne}
    \label{payerne_corr}
\end{figure*}

\begin{figure}[!h]
\begin{minipage}{.45\linewidth}
    \centering
    \includegraphics[width =7cm]{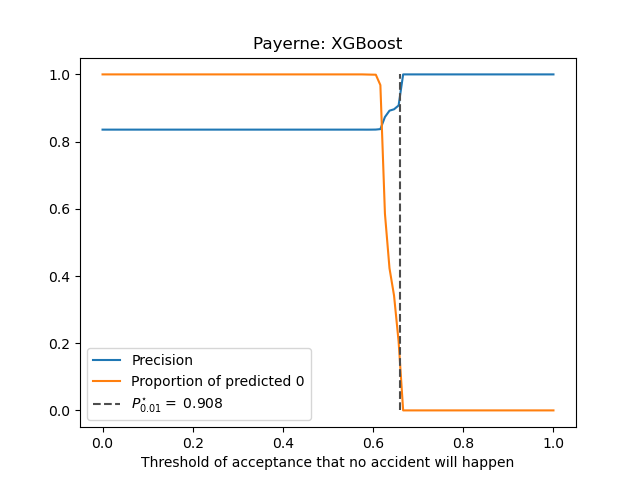}
    \caption{Precision curve, XGBoost, Payerne}
\end{minipage}
\quad
\begin{minipage}{.45\linewidth}
    \centering
    \includegraphics[width = 7cm]{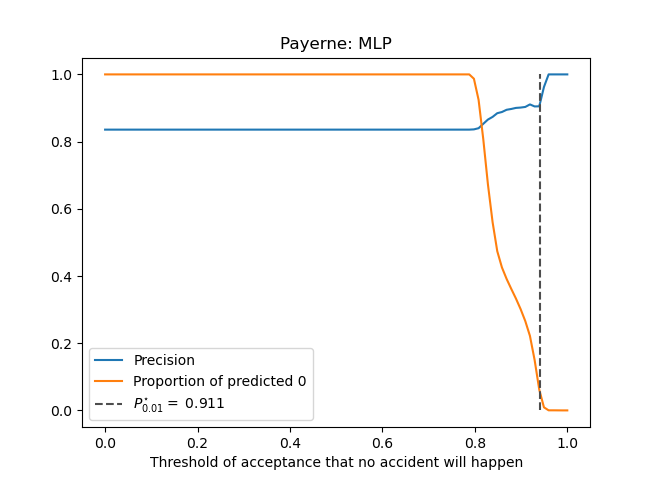}
    \caption{Precision curve, MLP, Payerne}
\end{minipage}
\end{figure}

\begin{figure}[!h]
\begin{minipage}{.45\linewidth}
    \centering
    \includegraphics[width =7cm]{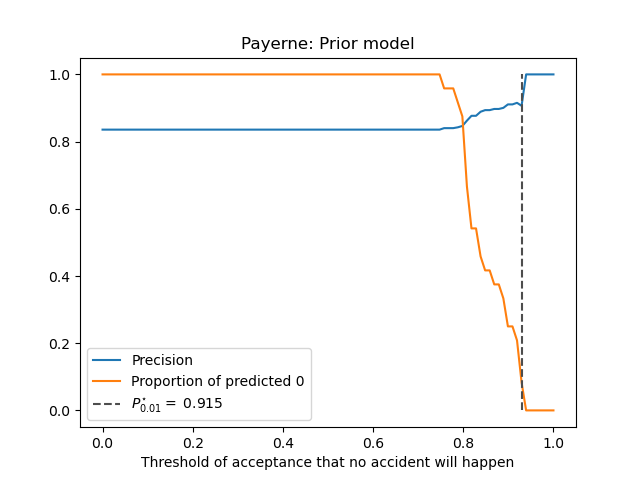}
    \caption{Precision curve, Prior model, Payerne}
\end{minipage}
\quad
\begin{minipage}{.45\linewidth}
    \centering
    \includegraphics[width = 7cm]{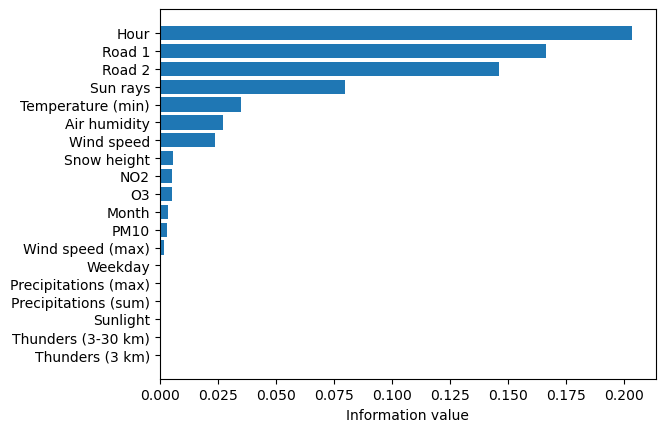}
    \caption{Information values, Payerne}
\end{minipage}
\end{figure}

\begin{figure}[!h]
\begin{minipage}{.45\linewidth}
    \centering
    \includegraphics[width =7cm]{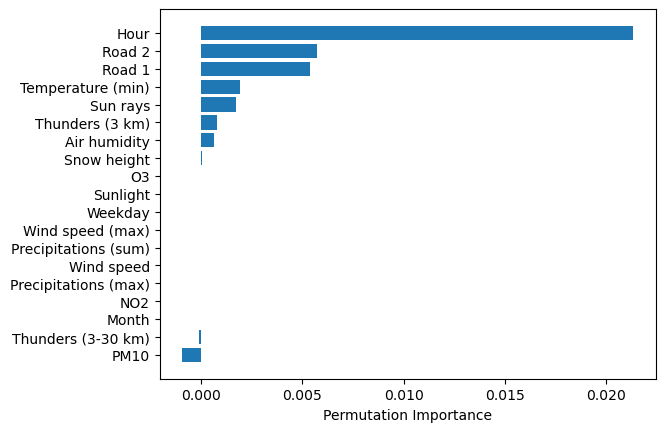}
    \caption{Permutation importance, XGBoost, Payerne}
\end{minipage}
\quad
\begin{minipage}{.45\linewidth}
    \centering
    \includegraphics[width = 7cm]{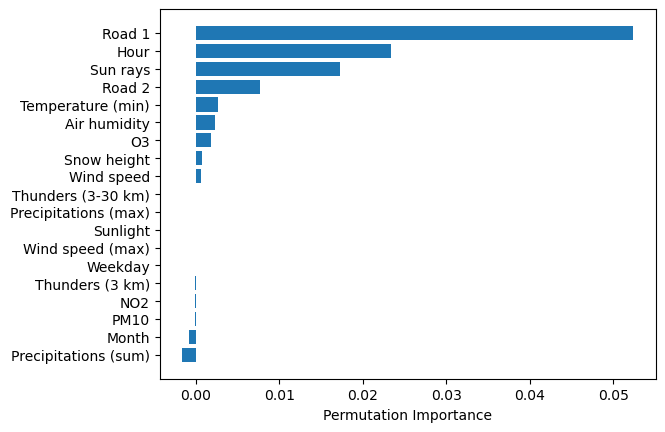}
    \caption{Permutation importance, MLP, Payerne}
\end{minipage}
\end{figure}

\begin{figure}[!h]
\begin{minipage}{.45\linewidth}
    \centering
    \includegraphics[width =7cm]{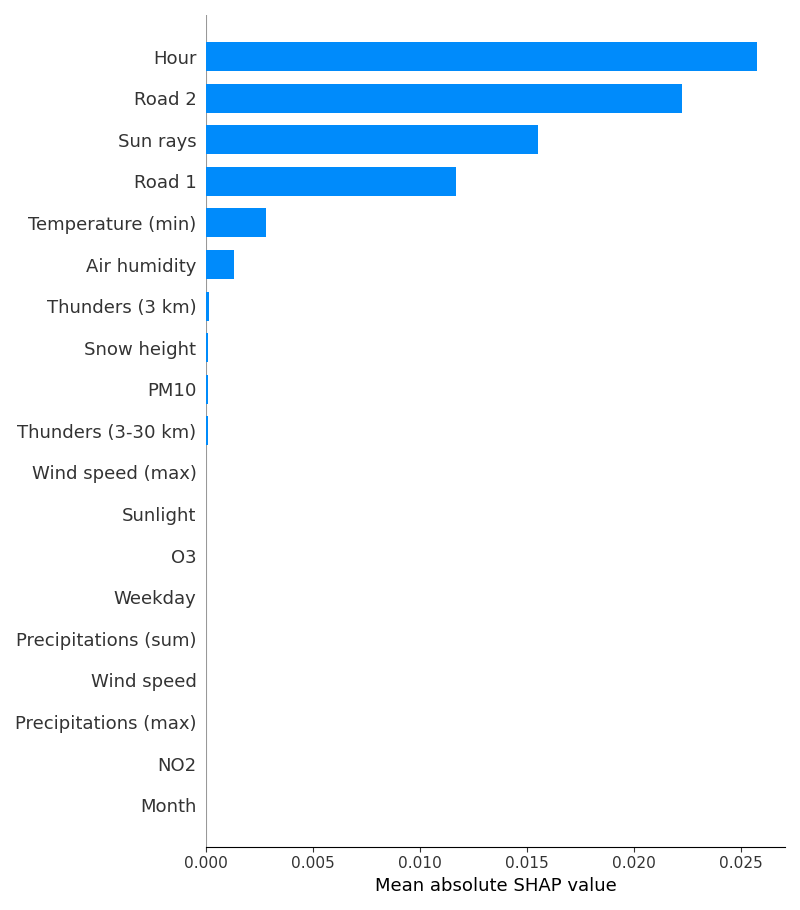}
    \caption{SHAP values, XGBoost, Payerne}
\end{minipage}
\quad
\begin{minipage}{.45\linewidth}
    \centering
    \includegraphics[width = 7cm]{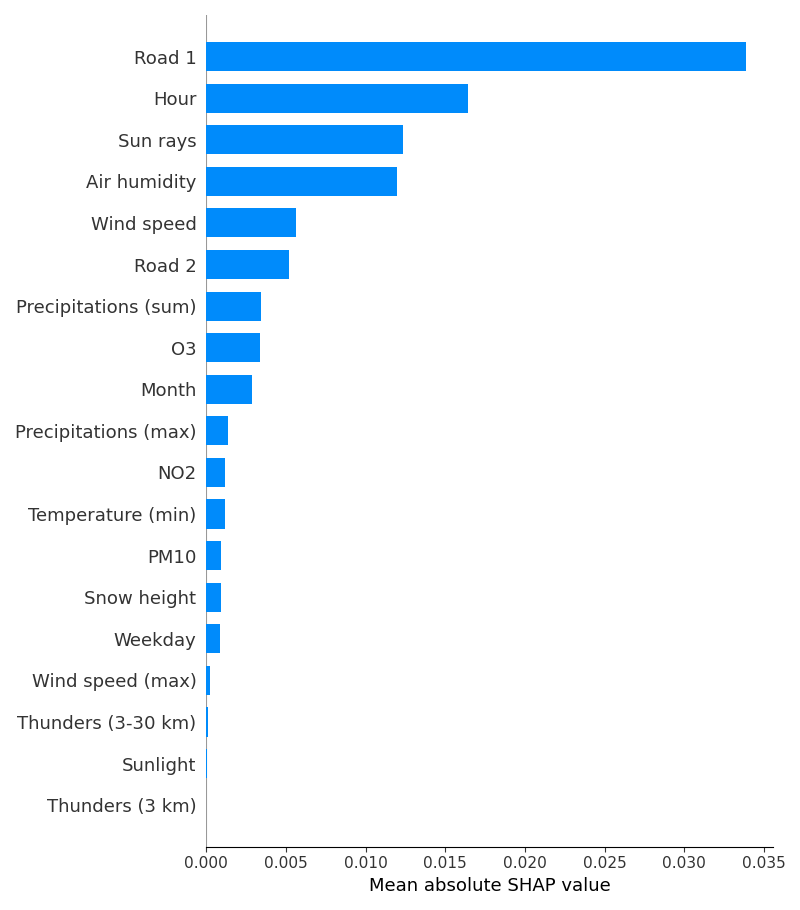}
    \caption{SHAP values, MLP, Payerne}
\end{minipage}
\end{figure}

\FloatBarrier

\vspace{1pts}

\subsection{Val-de-Travers}
\begin{figure*}[!h]
    \centering
    \includegraphics[width =0.9\textwidth, height=4in]{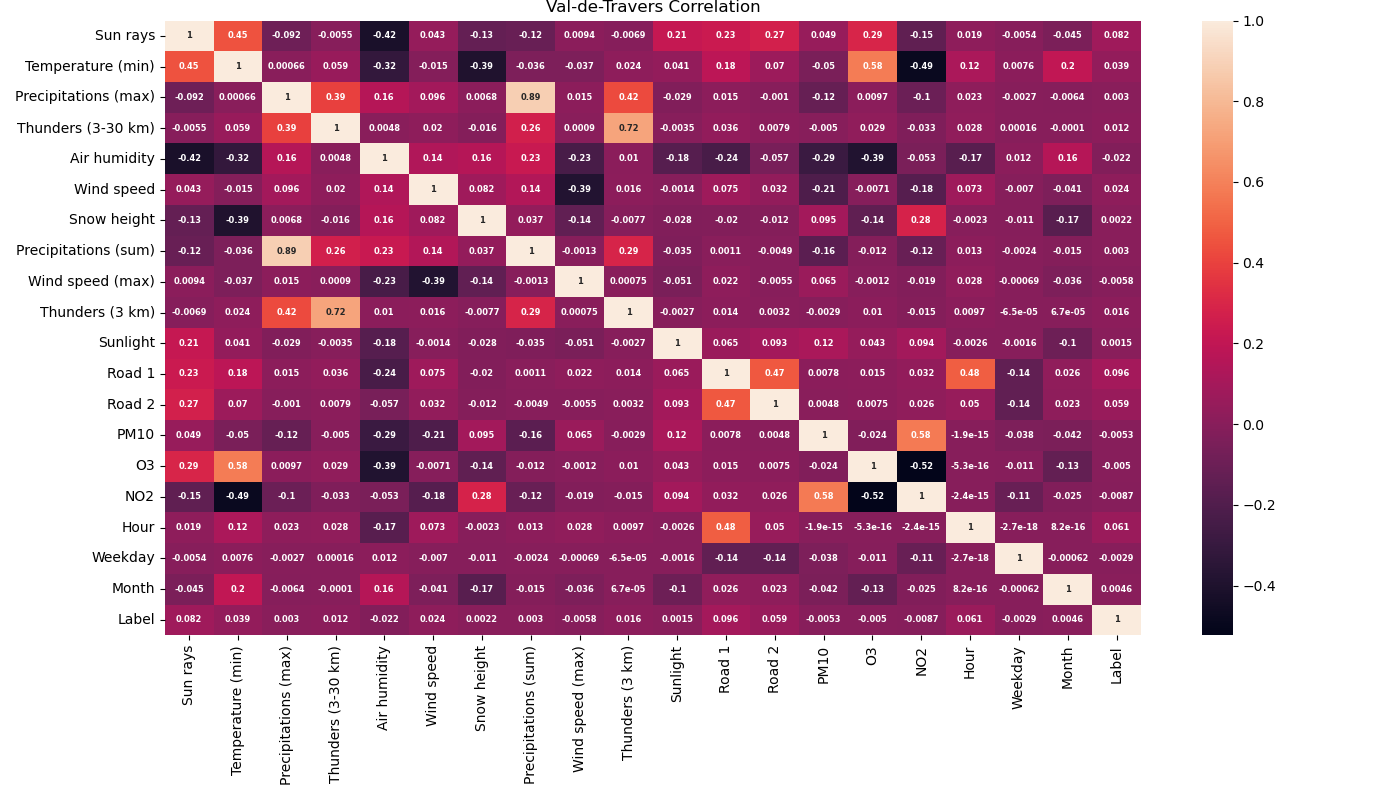}
    \caption{Correlation matrix, Val-de-Travers}
    \label{Figures/val-de-travers-correlation}
\end{figure*}

\begin{figure}[!h]
\begin{minipage}{.45\linewidth}
    \centering
    \includegraphics[width =7cm]{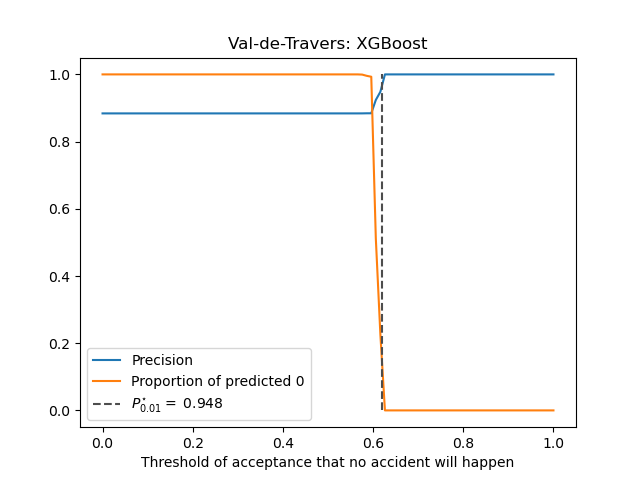}
    \caption{Precision curve, XGBoost, Val-de-Travers}
\end{minipage}
\quad
\begin{minipage}{.45\linewidth}
    \centering
    \includegraphics[width = 7cm]{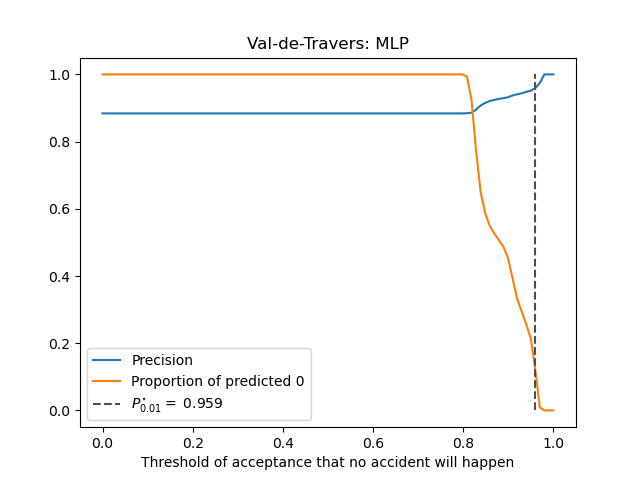}
    \caption{Precision curve, MLP, Val-de-Travers}
\end{minipage}
\end{figure}

\begin{figure}[!h]
\begin{minipage}{.45\linewidth}
    \centering
    \includegraphics[width =7cm]{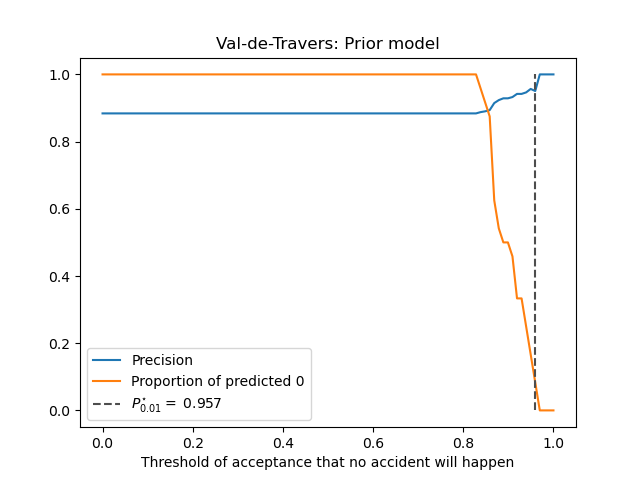}
    \caption{Precision curve, Prior model, Val-de-Travers}
\end{minipage}
\quad
\begin{minipage}{.45\linewidth}
    \centering
    \includegraphics[width = 7cm]{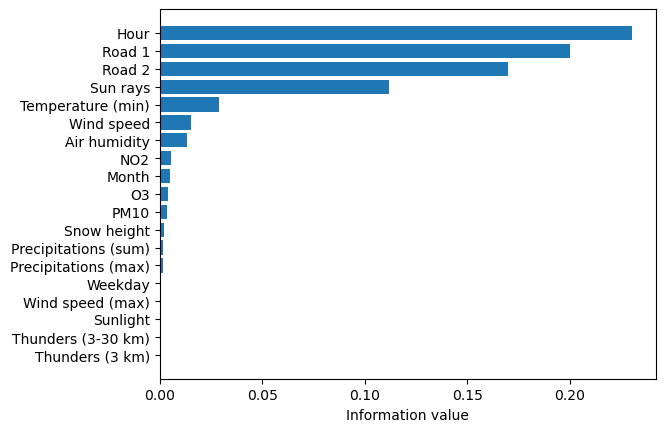}
    \caption{Information values, Val-de-Travers}
\end{minipage}
\end{figure}

\begin{figure}[!h]
\begin{minipage}{.45\linewidth}
    \centering
    \includegraphics[width =7cm]{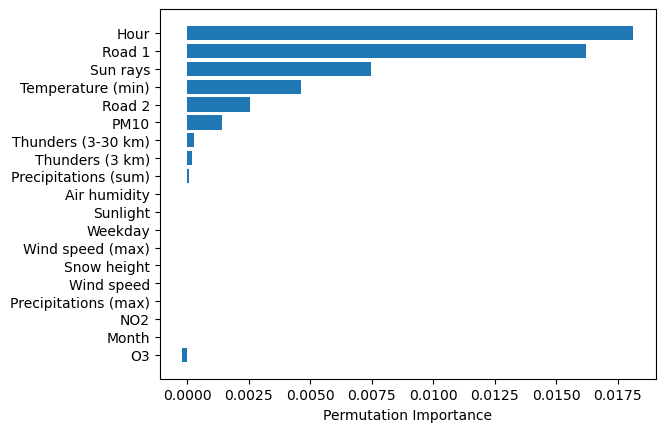}
    \caption{Permutation importance, XGBoost, Val-de-Travers}
\end{minipage}
\quad
\begin{minipage}{.45\linewidth}
    \centering
    \includegraphics[width = 7cm]{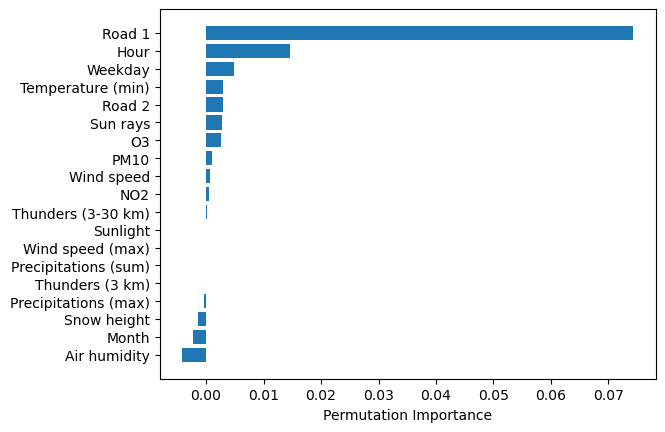}
    \caption{Permutation importance, MLP, Val-de-Travers}
\end{minipage}
\end{figure}

\begin{figure}[!h]
\begin{minipage}{.45\linewidth}
    \centering
    \includegraphics[width =7cm]{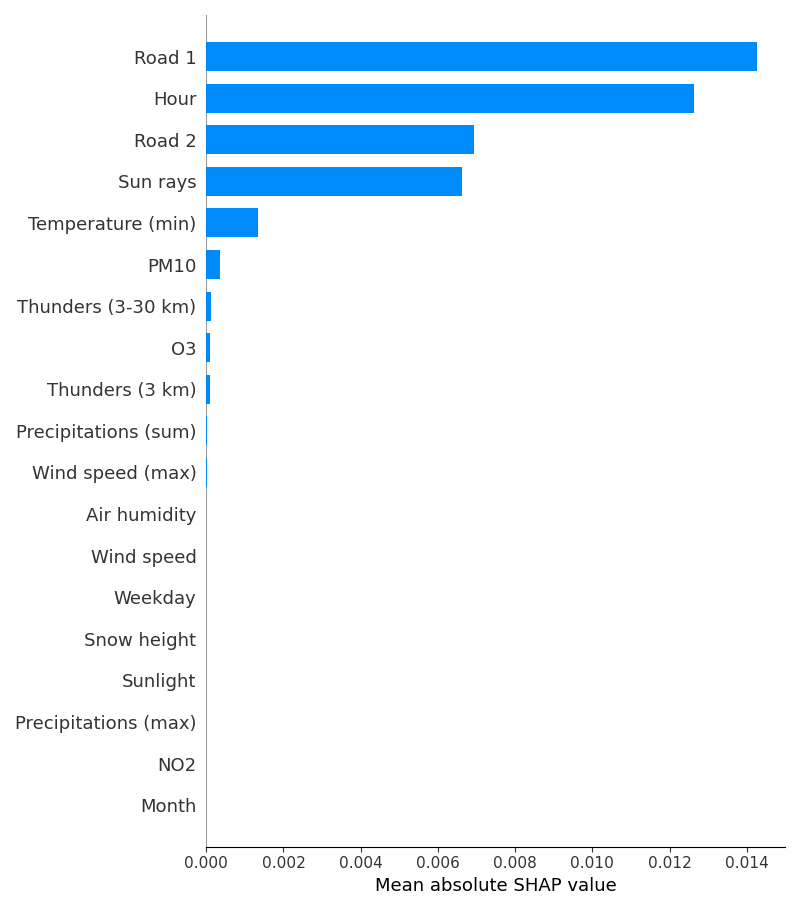}
    \caption{SHAP values, XGBoost, Val-de-Travers}
\end{minipage}
\quad
\begin{minipage}{.45\linewidth}
    \centering
    \includegraphics[width = 7cm]{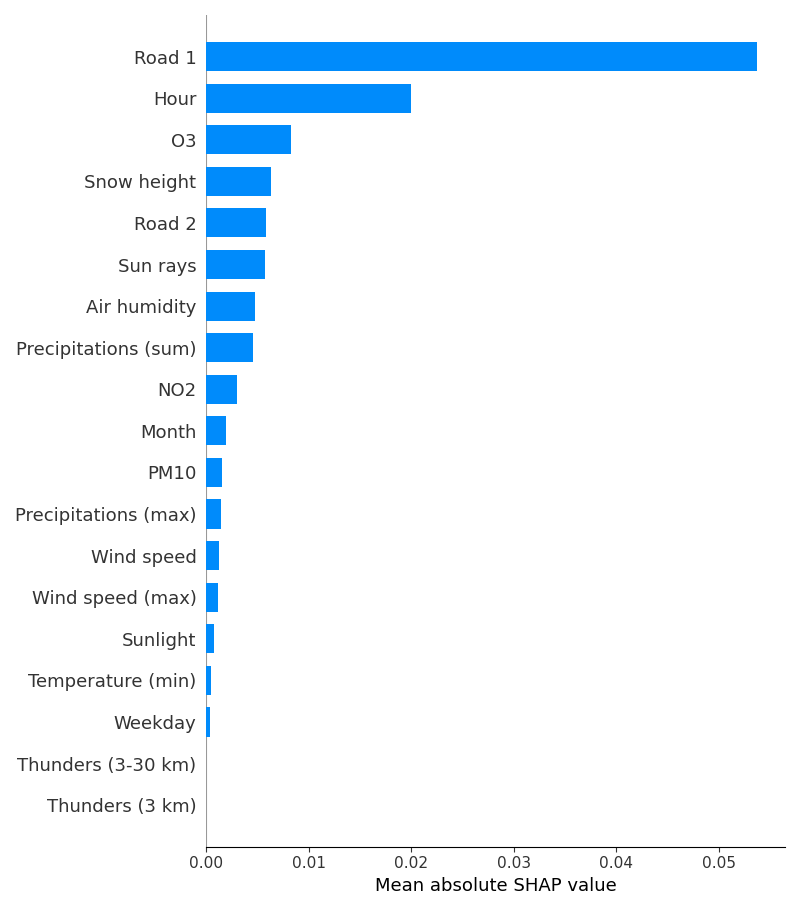}
    \caption{SHAP values, MLP, Val-de-Travers}
\end{minipage}
\end{figure}

\FloatBarrier

\vspace{1pts}

\subsection{Malviliers}
\begin{figure*}[!h]
    \centering
    \includegraphics[width =0.9\textwidth, height=4in]{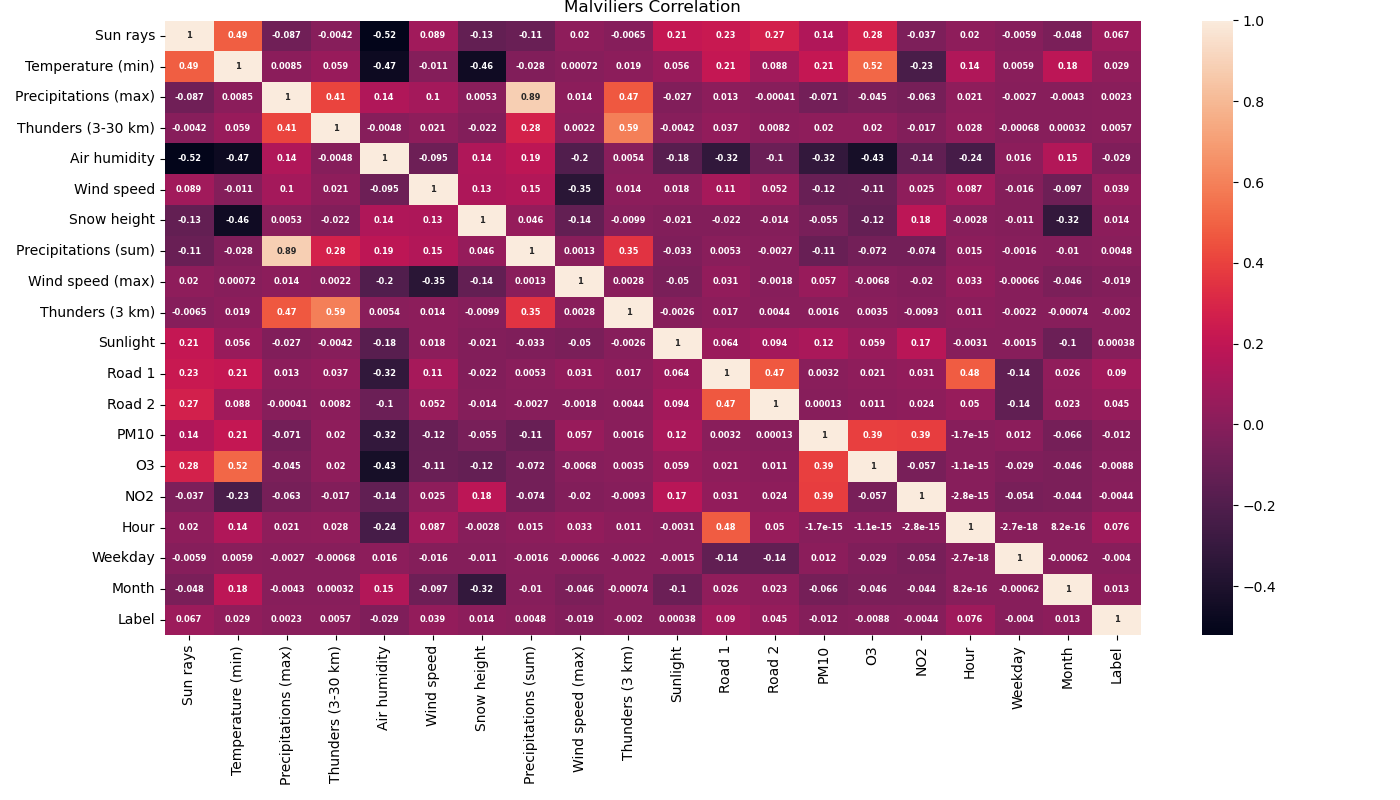}
    \caption{Correlation matrix, Malviliers}
    \label{malviliers_corr}
\end{figure*}

\begin{figure}[!h]
\begin{minipage}{.45\linewidth}
    \centering
    \includegraphics[width =7cm]{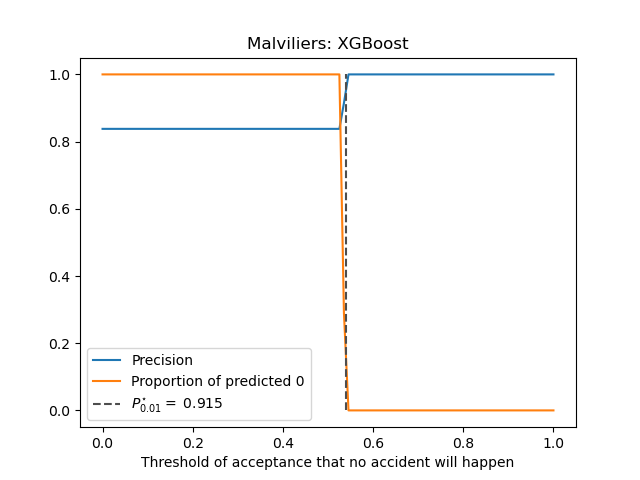}
    \caption{Precision curve, XGBoost, Malviliers}
\end{minipage}
\quad
\begin{minipage}{.45\linewidth}
    \centering
    \includegraphics[width = 7cm]{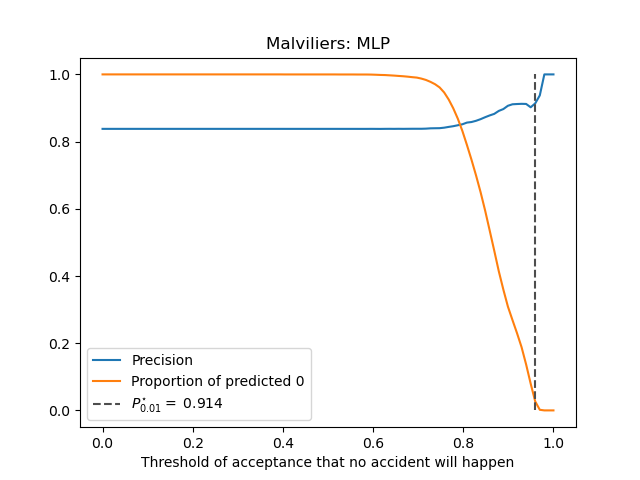}
    \caption{Precision curve, MLP, Malviliers}
\end{minipage}
\end{figure}

\begin{figure}[!h]
\begin{minipage}{.45\linewidth}
    \centering
    \includegraphics[width =7cm]{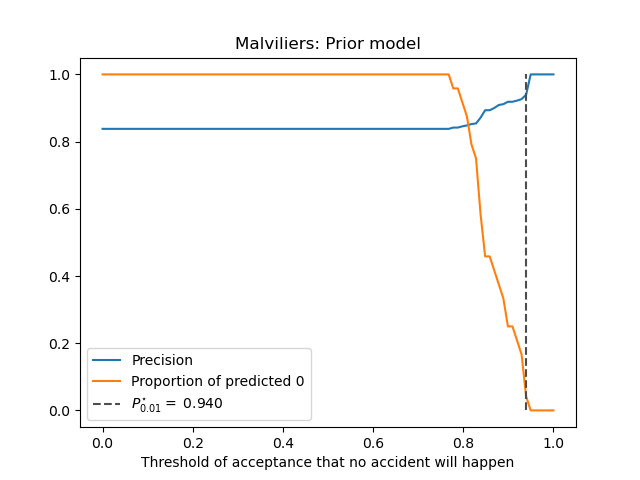}
    \caption{Precision curve, Prior model, Malviliers}
\end{minipage}
\quad
\begin{minipage}{.45\linewidth}
    \centering
    \includegraphics[width = 7cm]{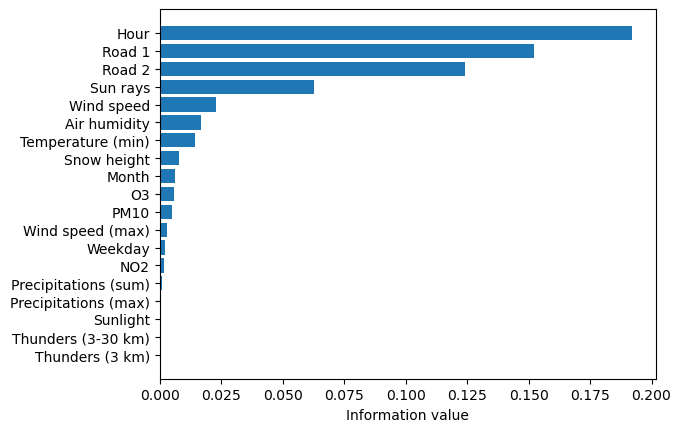}
    \caption{Information values, Malviliers}
\end{minipage}
\end{figure}

\begin{figure}[!h]
\begin{minipage}{.45\linewidth}
    \centering
    \includegraphics[width =7cm]{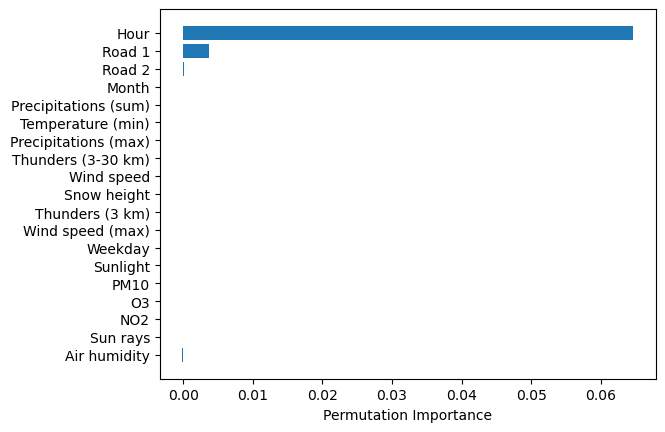}
    \caption{Permutation importance, XGBoost, Malviliers}
\end{minipage}
\quad
\begin{minipage}{.45\linewidth}
    \centering
    \includegraphics[width = 7cm]{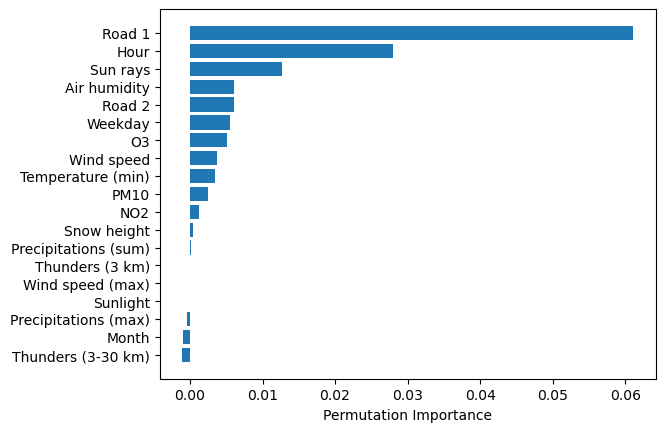}
    \caption{Permutation importance, MLP, Malviliers}
\end{minipage}
\end{figure}

\begin{figure}[!h]
\begin{minipage}{.45\linewidth}
    \centering
    \includegraphics[width =7cm]{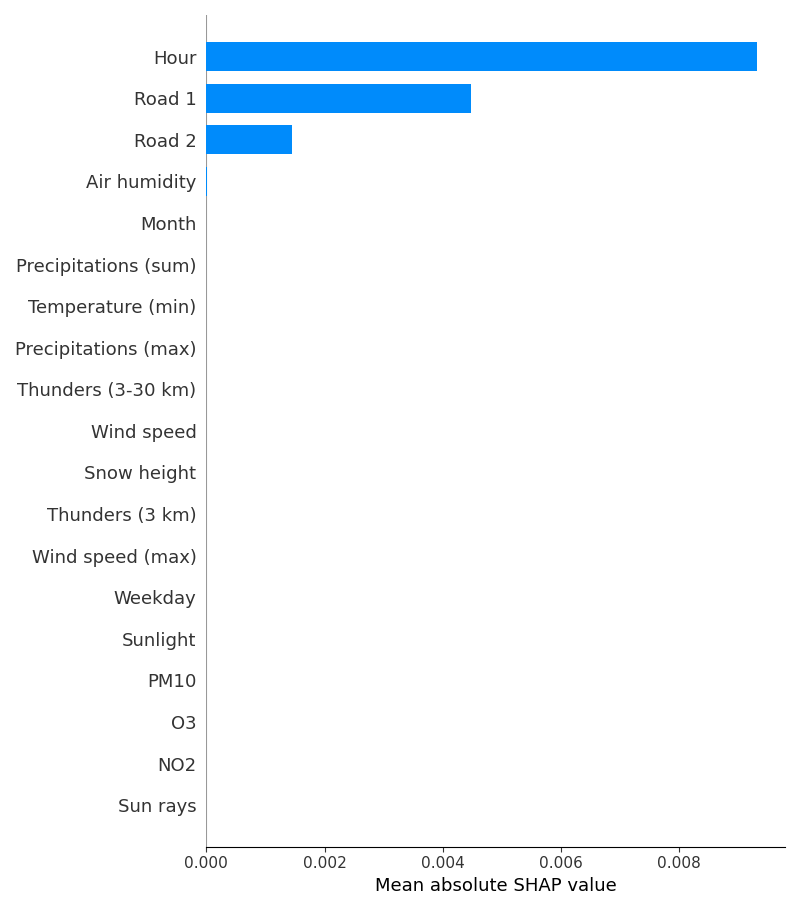}
    \caption{SHAP values, XGBoost, Malviliers}
\end{minipage}
\quad
\begin{minipage}{.45\linewidth}
    \centering
    \includegraphics[width = 7cm]{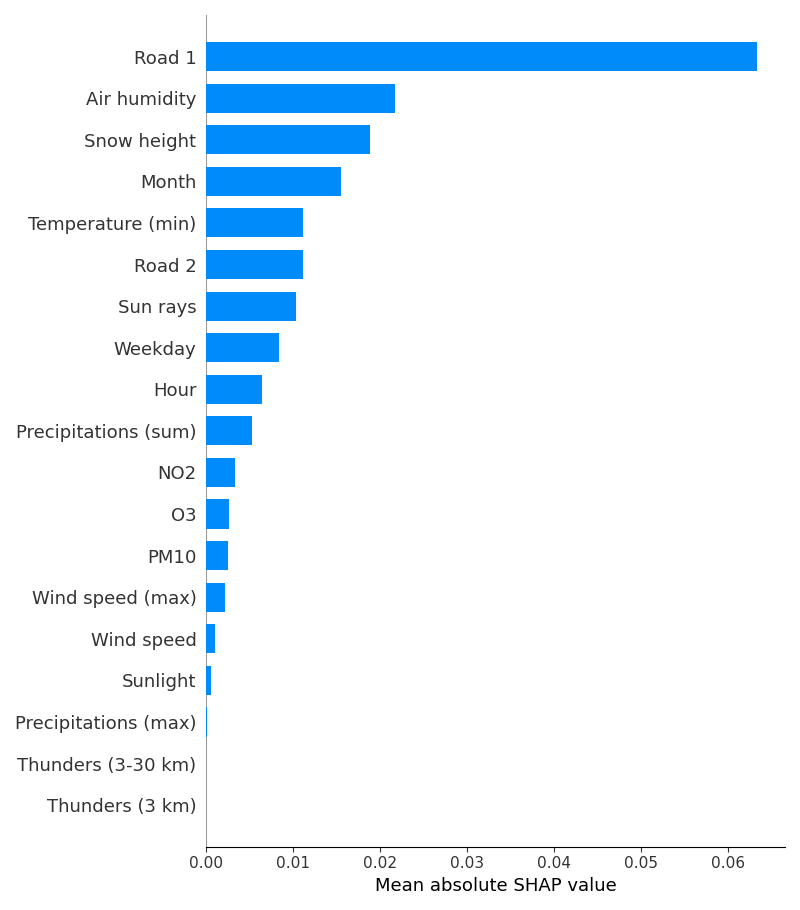}
    \caption{SHAP values, MLP, Malviliers}
\end{minipage}
\end{figure}

\end{document}